%% file: ss11.tex
\documentclass[preprint,showpacs,preprintnumbers,floatfix,superscriptaddress,nofootinbib,showkeys]{revtex4}

\pdfoutput=0

\usepackage[reqno]{amsmath}
\usepackage{amsthm}
\usepackage{amssymb}

\usepackage{graphicx}
\usepackage{lscape}
\usepackage{longtable}
\usepackage{afterpage}

\newcommand{\etal}{{\em et~al.}}                
\newcommand{\beq}{\begin{equation}}
\newcommand{\eeq}{\end{equation}}
\newcommand{\bea}{\begin{eqnarray}}
\newcommand{\eea}{\end{eqnarray}}
\begin{document}


\title{
Symmetry Energy II: Isobaric Analog States
}

\author{Pawe\l~Danielewicz\email{danielewicz@nscl.msu.edu}}

\affiliation{National Superconducting Cyclotron Laboratory and\\
Department of Physics and Astronomy, Michigan State University,
\\
East Lansing, Michigan 48824, USA\\
}

\author{Jenny Lee\email{jennylee@ribf.riken.jp}}

\affiliation{RIKEN Nishina Center for Accelerator-Based Science\\
Wako, Saitama 351-0198, Japan\\
}

\begin{abstract}
Using excitation energies to isobaric analog states (IAS) and charge invariance, we extract nuclear symmetry coefficients, from a mass formula, on a nucleus-by-nucleus basis.   Consistently with charge invariance, the coefficients vary weakly across an isobaric chain.  However, they change strongly with nuclear mass and range from $a_a \sim 10\, \text{MeV}$ at mass $A \sim 10$ to $a_a \sim 22 \, \text{MeV}$ at $A \sim 240$.  Variation with mass can be understood in terms of dependence of nuclear symmetry energy on density and the rise in importance of low densities within nuclear surface in smaller systems.  At $A \gtrsim 30$, the dependence of coefficients on mass can be well described in terms of a~macroscopic volume-surface competition formula with $a_a^V \simeq 33.2 \, \text{MeV}$ and $a_a^S \simeq 10.7 \, \text{MeV}$.  Our further investigation shows, though, that the fitted surface symmetry coefficient likely significantly underestimates that for the limit of half-infinite matter.  Following the considerations of a~Hohenberg-Kohn functional for nuclear systems, we determine how to find in practice the symmetry coefficient using neutron and proton densities, even when those densities are simultaneously affected by significant symmetry-energy and Coulomb effects.  These results facilitate extracting the symmetry coefficients from Skyrme-Hartree-Fock (SHF) calculations, that we carry out using a variety of Skyrme parametrizations in the literature.  For the parametrizations, we catalog novel short-wavelength instabilities.  In our further analysis, we retain only those parametrizations which yield systems that are adequately stable both in the long- and short-wavelength limits.  In comparing the SHF and IAS results for the symmetry coefficients, we arrive at narrow ($\pm 2.4 \, \text{MeV}$) constraints on the symmetry energy values $S(\rho)$ at $0.04 \lesssim \rho \lesssim 0.13 \, \text{fm}^{-3}$.  Towards normal density the constraints significantly widen, but the normal value of energy $a_a^V$ and the slope parameter $L$ are found to be strongly correlated.  To narrow the constraints, we reach for the measurements of asymmetry skins and arrive at $a_a^V = (30.2\,$--$\,33.7) \, \text{MeV}$ and $L=(35\,$--$\,70)\,\text{MeV}$, with those values being again strongly positively correlated along the diagonal of their combined region.  Inclusion of the skin constraints allows to narrow the constraints on $S(\rho)$, at
$0.04 \lesssim \rho \lesssim 0.13 \, \text{fm}^{-3}$, down to $\pm 1.1 \, \text{MeV}$.  Several microscopic calculations, including variational, Bruckner-Hartree-Fock and Dirac-Bruckner-Hartree-Fock, are consistent with our constraint region on $S(\rho)$.

\end{abstract}

\pacs{21.65.-f, 21.10.Dr, 21.10.Gv, 21.60.Jz}


\keywords{
symmetry energy, isobaric analog state, nuclear matter, Hohenberg-Kohn functional, Skyrme-Hartree-Fock model, symmetry coefficient, binding formula
}

\maketitle


\section{Introduction}
Evolution of bulk nuclear properties with changing neutron-proton asymmetry is obviously not as well known as are the properties characteristic for commonly encountered combinations of neutron and proton numbers.  Within nuclear structure, modest changes in asymmetry imply gradual displacement of the Fermi levels and make the effects of the average evolution with asymmetry compete with microscopic, shell and pairing, effects.  Within reactions, nuclear systems undergo complicated changes, that need to be carefully modeled when analyzing data, with any average effects of asymmetry on the reactions competing with a variety of physical effects subject to modeling uncertainties.  In~this paper we attempt to learn about the average effects of neutron-proton asymmetry on nuclear energies, exploiting excitation energies to isobaric analog states and to reach the conclusions in as model-independent manner as possible.  Our early efforts in this direction have been reported in~\cite{Danielewicz:2004et,Danielewicz:2007pf,Danielewicz200936}.  As we progress, we find that we need to reassess our strategy.

Nuclear structure can principally provide information about impact of asymmetry on bulk properties at subnormal and normal densities only.  Reactions and astrophysical observations, on the other hand, can provide information about supranormal densities and about subnormal as well, though potentially with a lesser precision than structure can.  Within the structure, observables  investigated historically in this context included systematics of nuclear binding energies~\cite{Moller:1993ed,Danielewicz:2003dd,Diep07} as well as of fission barriers \cite{Moller:1993ed}.  To emphasize the effects of asymmetry, energy differences have been employed \cite{PhysRevC.81.054302,PhysRevC.81.067302}, including differences between neutron and proton separation energies \cite{Danielewicz:2003dd,PhysRevC.81.024324}.  Moreover, implications of measurements of asymmetry skins have been explored~\cite{Centelles:2010qh,PhysRevC.82.024321,Diep07,Horowitz:2000xj,PhysRevC.67.034305,Danielewicz:2003dd}.  Further, systematics of the strength distribution for nuclear collective excitations for nuclear systems have been utilized \cite{Li:2010kfa,Li:2007bp,Carbone:2010az,Klimkiewicz:2007zz}.  In the context of asymmetry dependence of the bulk nuclear properties, a strategy has been proposed to analyze informational content of different observables \cite{PhysRevC.81.051303}.  Within reactions, such observables as neutron-to-proton yield ratios \cite{PhysRevLett.102.122701,PhysRevLett.78.1644} and differences between neutron and proton flows~\cite{Russotto2011471} have been employed.  Among other reaction observables, charged pion ratios, in particular, rise hope of learning about average effects of asymmetry at supranormal densities~\cite{Zhang:2009qe}.
Within astrophysics, mass-radius relation and maximal mass for neutron stars have been linked to the nuclear symmetry energy~\cite{Horowitz:2000xj,0004-637X-722-1-33}.

Analyzing excitation energies to the lowest states for a given isospin, such as pursued here, has some history, with the excitation energies used to access both the systematics of pairing energy and of symmetry energy \cite{deShalitTalmi,Janecke196597,Zeldes197612,PhysRevC.61.041303,Vogel:1998km}.  As a substitute for the excitation energies, combinations of masses of neighboring isobars have been further exploited \cite{Janecke200323,Janecke2007317}.  From the perspective of a mass formula, though, the excitation energies give the most direct access to the systematic of symmetry energy, while still affected by the microscopic effects.  With our interest in extrapolating to an infinite system, we study a wide range of nuclear masses, including those light -- not always associated with a mass formula.  The advantage in pursuing a wide range of masses is in getting different contributions to the symmetry energy from surface areas where the density changes, potentially testing the dependence of symmetry energy on density for uniform matter.  The compilation \cite{Antony:1997}, actually aiming at the Coulomb displacement energies, provides a particularly large set of energies for isobaric analog states (IAS) for different nuclei, suitable for our purposes.  To extract smoothly varying bulk contribution to the excitation energies, we need to apply shell corrections, preferably extending to low mass and charge numbers.  A relatively recent set of those corrections is by Koura {\em et al.} \cite{Koura2000,Koura:2005}.  In the more remote past, von Groote {\em et al.} put forward corrections \cite{Groote1976418} extending to low masses, of utility to us.  For completeness, we also employ the corrections by Moller {\em et al.} \cite{Moller:1993ed}.

For relating the nucleus-by-nucleus results on symmetry energy to universal properties of bulk nuclear matter, the most desirable would be a model-independent extrapolation such as based on the surface-volume decomposition for those properties.  In the first paper~\cite{Danielewicz:2007pf} of this series, referred further to as I, we have investigated the surface and volume contributions to the energy, within the Skyrme-Hartree-Fock (SHF) description of half-infinite nuclear matter.  In this, we built upon earlier efforts by Kohler~\cite{Kohler:1969,Kohler:1976} and by C$\hat{\text{o}}$te {\em et al.}~\cite{Cote:1978,Farine:1980,PhysRevC.24.303}.  On their own, the SHF calculations for spherical nuclei~\cite{Reinhard:1991,reinhard:014309} can serve as a test of extrapolation procedures applied to the results from IAS data.  In trying to understand the progress of volume-surface separation with change in nuclear mass, it is particularly useful to calculate also nuclei of masses much larger than occurring in nature \cite{reinhard:014309}.  An~obstacle that we have encountered in the analysis, of the evolution of properties with mass, is that of instability of many of the Skyrme parameterizatizations utilized in the literature.  The~Skyrme parameterizations have been compiled with the assistance of Stone, see Ref.~\cite{PhysRevC.68.034324} and~I.  Energies for the unstable interactions tend to deviate from the expectations based on half-infinite matter of~I, even for very large mass numbers~$A$.  The unstabilities that may emerge in the long-wavelength limit are normally identified in terms of Landau parameters~\cite{PinesNoziers,Backman1975209,vuong07}.  However, we find also short-wavelength instabilities for the Skyrme interactions and we develop formal criteria for the emergence of these instabilities.  Specifics of those criteria parallel to some degree the expectations of Lesinski \etal~\cite{Lesinski:2006cu}.

When using the Skyrme interactions as a reference, we find that expectations from half-infinite matter are approached at a slower pace for the portion of the nuclear energy associated with asymmetry than for the portion that represents the energy of symmetric matter.  One reason for this is that the energy of symmetric matter is quadratic in density variations while the symmetry energy is linear.  Here the variations are relative to the density for symmetric matter, that minimizes energy.  The slow pace of approach to bulk limit hampers the ability to learn about the symmetry energy of bulk matter from IAS in a model-independent manner, which forces us to employ the outcomes from the plethora of Skyrme parameterizations, as representing possible variants of reality, and exploit them to narrow conclusions on the symmetry energy of bulk matter.  The resulting limits on the values of symmetry energy end up being fairly narrow at densities below $0.13 \, \text{fm}^{-3}$, but fan out in the vicinity of normal density.  To cope with the lack of resolution in the latter region, we reach for the measurements of asymmetry skins.  The skins correlate strongly with dimensionless slope of the symmetry energy, that we term stiffness.  Upon combining the IAS and skin constraints, we arrive at fairly narrow constraints on the values of symmetry energy as a~function of density, both in the normal density region and at more subnormal densities.

In the next section, we analyze the excitation energies to IAS, in order to obtain symmetry-energy coefficients on a nucleus-by-nucleus basis.

\section{Symmetry-Energy Coefficients from Excitation Energies}
\label{sec:SymIAS}

\subsection{General Considerations}
\label{subs:General}

The starting point for the analysis of symmetry energy in nuclei is the premise that the ground-state energy of a nucleus may be represented as
\beq
E=E_\text{nuc} + E_\text{Cou} + E_\text{mic} \, ,
\eeq
where $E_\text{nuc}$ is the bulk nuclear contribution to the energy, smoothly changing with neutron~$N$ and proton~$Z$ numbers, $E_\text{Cou}$ is the Coulomb contribution, also smoothly changing and $E_\text{mic}$ is the microscopic contribution, including shell and pairing contributions, changing abruptly with $N$ and $Z$.  The nuclear contribution can be next split into the energy $E_0(A)$ for a nucleus with equal number of neutrons and protons, $A=N+Z$, and a correction $E_1$ associated with the neutron-proton asymmetry, $N - Z$,
\beq
\label{eq:Enuc}
E_\text{nuc} = E_0 + E_1 \, .
\eeq
Under the charge symmetry of nuclear interactions, the nuclear contribution to the energy should be symmetric under neutron-proton interchange, i.e.\ an even function of asymmetry or isoscalar.  If we were to expand the energy in powers of relative asymmetry, $\eta=(N-Z)/A$, we should get even powers only:
\beq
\label{eq:E1}
E_1 = a_a(A) \, \frac{(N-Z)^2}{A} + \ldots \, .
\eeq
In the practice of the analysis of nuclei, it has not been necessary to include terms higher than quadratic.  With \eqref{eq:Enuc} and \eqref{eq:E1}, the bulk nuclear contribution to net energy acquires the form
\beq
E_\text{nuc} = E_0(A) + a_a(A) \, \frac{(N-Z)^2}{A} \, .
\eeq
Here, $a_a(A)$ is the generalized mass-dependent (a)symmetry coefficient.
The energy of a symmetric nucleus is further expanded in powers of $A$:
\beq
\label{eq:E0}
E_0(A) = -a_V \, A + a_S \, A^{2/3} + \ldots \, .
\eeq
With the radius of the nucleus changing as $A^{1/3}$, the leading term in \eqref{eq:E0} may be interpreted as associated with the volume contribution to the energy and the next -- with the surface contribution.  The subsequent term, proportional to $A^{1/3}$, may be tied with the contribution of the surface curvature \cite{Myers:1969}.  From the microscopic side, the volume contribution to the energy may be calculated within a consideration of infinite nuclear matter.

The premise above is successful already at the very basic level, where only volume and surface terms are retained in $E_0$, $a_a$ is taken as $A$-independent and the Coulomb term is taken in the form such as for a uniform sphere of charge,
\beq
\label{eq:ECou}
E_{Cou} \simeq a_C \, \frac{Z^2}{A^{1/3}} \, ,
\eeq
with $\displaystyle a_C \approx \frac{3}{5} \frac{e^2}{4 \pi \epsilon_0} \frac{1}{r_0} $ and
$\displaystyle
\frac{4 \pi}{3}\, r_0^3 = \frac{1}{\rho_0}
$,
where $\rho_0$ is normal density.  The success of the basic energy formula is discussed in virtually every introductory nuclear textbook.

The above quick outline glosses though over some obvious issues pertinent to the central topic of the present work.  Thus, the charge invariance of strong interactions constraints the form of the $E_1$ energy-component more strongly than does the charge symmetry -- the energy needs to be an isoscalar in a broader sense than in \eqref{eq:E1}, invariant under rotations in the isospin space, not just under neutron-proton interchange.  The operator of net nuclear isospin is ${\pmb T} = \sum_{i=1}^A {\pmb t}_i$, where ${\pmb t}_i$ are operators for individual nucleons.  The tradition in nuclear physics is opposite to that in high-energy physics, of attributing the positive $z$-component of isospin to the neutron, rather than to proton, equivalent to attributing the positive component to the down rather than up quark.  If we rewrite the energy $E_1$ in the form \eqref{eq:E1}, in terms of isospin, with
\beq
N-Z = 2 \sum_{i=1}^A t_{zi} = 2 T_z \, ,
\eeq
we get
\beq
E_1 =  \frac{4 \, a_a(A) }{A} \, T_z^{2} \, ,
\eeq
observed to be invariant under reflections in the isospin space.
Under charge invariance, though, different coordinate components of the ${\pmb T}$-operator need to be treated democratically.  Correspondingly, the expression for $E_1$ needs to be generalized to
\beq
\label{eq:E1T}
E_1 = \frac{4 \, a_a(A) }{A} \, {\pmb T}^{2} = \frac{4 \, a_a(A) }{A} \, \big(T_z^{2} + {\pmb T}_\perp^{2} \big) =   \frac{4 \, a_a(A) }{A} \, T \big(T + 1 \big) \, .
\eeq
In minimizing energy for the ground state, the net isospin will achieve it lowest possible value of $T= |T_z| = \frac{1}{2} |N-Z|$. In systems where the symmetry energy is sizeable, with large $T$, the distinction between \eqref{eq:E1T} and \eqref{eq:E1} is going to make relatively little impact.  On the other hand, the modification \eqref{eq:E1T} opens up the interesting possibility of considering the lowest-energy state for a nucleus, at a given $N$ and $Z$, constrained to a specific~$T$, including the possibility of $T > |T_z|$.  Notably, considering contributions to the minimal energy at $T \ge |T_z|$, those $N=Z$ properties that minimize nuclear energy, giving rise to the specific minimal $E_0(A)$ and $a_a(A)$, obviously do not depend on $T$ one starts from, i.e.\ are the same for $T > |T_z|$ as they are for the ground state with $T=|T_z|$.  Given the same $Z$ as for the ground state, the Coulomb energy should be the same for the excited minimal-energy state at $T > |T_z|$ and for the ground state.  A state of minimal energy, for a given $T> |T_z|$, is normally an IAS of the ground state of another nucleus within the same isobaric analog chain.  Following the reasoning above, the excitation energy to the IAS should be
\beq
\label{eq:EIAS}
E_\text{IAS}^* = E_\text{IAS} - E_\text{gs}  = \frac{4 \, a_a(A) }{A} \, \Delta {\pmb T}_\perp^2 + \Delta
E_\text{mic} \, .
\eeq
In the above, $\Delta {\pmb T}_\perp^2$ is the change in the transverse isospin squared, between the ground and the excited state,
\beq
\Delta {\pmb T}_\perp^2 = \Delta {\pmb T}^2=  T \, (T+1)
- |T_z| \, (|T_z| + 1) \equiv {\pmb T}_\perp^2 - |T_z| \, ,
\eeq
and $\Delta E_\text{mic}$ is the difference in the microscopic corrections to the energy, between the IAS and the ground state.

Aside from the above, another issue glossed over in the original discussion is that of the coupling between the nuclear and Coulomb terms in the energy minimization for the ground state.  The coupling may be examined e.g.\ within the Thomas-Fermi theory obviously consistent with the bulk limit.  The interplay between the nuclear and Coulomb terms is particularly important in the heavy high-$Z$ nuclei, with the interplay leading to a specific distribution of asymmetry within the nuclear volume and affecting the net density as well.  The above coupling complicates the dependence of the net energy on asymmetry at a given mass number~$A$.  This and other finer details, required in modeling the nuclear and Coulomb energies, make it quite difficult to arrive at unambiguous results on the symmetry energy, when analyzing ground-state masses \cite{Danielewicz:2003dd,Kir08,mendoza-2008}.  On the other hand, irrespectively of the coupling and the other details, the bulk portion of excitation energy to the IAS of a ground-state should be linear in the square of transverse isospin.  However, because of the coupling, the coefficient of proportionality between the energy and ${\pmb T}_\perp^2$ might depend on $Z$ in heavier systems:
\beq
\label{eq:Eaaz}
E_\text{IAS}^* =  \frac{4 \, a_a(A,Z) }{A} \, \Delta {\pmb T}_\perp^2 + \Delta
E_\text{mic}
= \frac{4 \, a_a(A,Z) }{A} \, \Delta \Big[T \big( T + 1 \big) \Big] + \Delta
E_\text{mic}
\, .
\eeq
A second-order term, proportional to $\big({\pmb T}_\perp^2 \big)^2$, might become important in $E_\text{IAS}^*$, when the net energy of the excited state crosses the particle-drip threshold.  The above dependence on~${\pmb T}_\perp^2$ can be contrasted with the situation of ground states, at a given $A$, crossing the drip threshold at different $T=|T_z|$ values for different signs of $T_z$, suggesting importance of odd powers of $T_z$ in the high-order expansion of nuclear energy.

Notwithstanding the above considerations, analyses have been carried out in the past leading to a strong dependence of $a_a$ on $Z$, in particular for light systems \cite{PhysRevC.81.067302} and for small $|N-Z|$ \cite{Janecke200323,Janecke2007317}, in no particular relation to the Coulomb-symmetry-energy interplay.  In addition, claims of proportionality of the symmetry energy to $T (T + b)$ have been made, with $b$ different from~1, such as $b \sim 4$ \cite{BlattWeisskopf,Janecke196597,Janecke200323,Janecke2007317}.  The latter claims would imply the excitation energy to an IAS of the general form
\beq
\label{eq:IASb}
E_\text{IAS}^* =  \frac{4 \, a_a(A,Z) }{A} \, \Delta \Big[T \big( T + b(A,Z) \big) \Big] + \Delta
E_\text{mic} \, .
\eeq

In what follows, we will exploit data on excitation energies to the IAS of ground states, to learn on the symmetry energy.  On one hand, we shall try to extract the dependence of symmetry coefficients on $A$.  On the other hand, we shall try to find out whether there is an evidence for an independent dependence of $a_a$ on $Z$.  Finally, we will examine whether there is an evidence for $b \ne 1$ in the symmetry energy.  Because of the limited amount of data that is available for the purpose of our investigation, we will not be able to fully relax the anticipated form of the symmetry energy, but rather have to constrain it in a manner suitable for the specific part of our investigation.  The latter is, in particular, important in that any changes in the value of~$b$, compared to~1, may be traded off, in an obvious manner, against the $Z$-dependence of~$a_a$ \cite{PhysRevC.81.067302}.

\subsection{Mass-dependent Symmetry Coefficients from Excitation Energies to IAS}

The primary source of data on the energies of IAS of ground states has been for us the compilation by Antony \etal\ \cite{Antony:1997}.  That compilation has been complemented by more recent results of measurements of IAS, in particular for $^\text{5}$He \cite{springerlink:10.1140/epjad/i2005-06-080-6}, $^\text{11}$Be~\cite{Shimoura1998387},  $^\text{14}$B \cite{Takeuchi2001255}, $^\text{37}$K \cite{PhysRevC.75.055503},
$^\text{40}$Sc~\cite{Cameron2004293}, $^\text{42}$K \cite{SINGH20011}, $^\text{60}$Zn \cite{Tuli2003347},
$^\text{91}$Nb \cite{Baglin19991}, $^\text{109}$In \cite{Blachot1999505}, $^\text{112}$Sb  \cite{Frenne1996639}, $^\text{128}$I \cite{KANBE2001227}, $^\text{130}$I \cite{SINGH200133}, $^\text{170}$Tm and $^\text{170}$Lu \cite{BAGLIN2002611}, and $^\text{192}$Au \cite{Baglin1998717}.

For some nuclei, an IAS of the ground state has not been identified with the spectrum of an adjacent isobar, as e.g.\ in the case of an IAS of the ground state of $^\text{79}$Se in $^\text{79}$Br, but an IAS of a low-lying excited state has been.  In that case, we {\em estimate} the energy of the ground-state IAS within $^\text{79}$Br, by assuming that the energy difference between the IAS of the specific excited state and that of the ground state is approximately the same as the energy difference between the original states in $^\text{79}$Se.  When all energies are known, the latter type of approximation is typically observed to be valid to within
\beq
\label{eq:errorA}
\sim \frac{1 \, \text{MeV}}{A^{0.85}} \, ,
\eeq
excellent for heavy nuclei and still reasonable for light.

As far as microscopic corrections are concerned, those of most utility for us in the literature are the corrections, $E_\text{mic}^\text{est} (A,Z)$, estimated by Koura {\em et al.}~\cite{Koura2000,Koura:2005}, as these extend to relatively low mass-numbers $A$.  Our interest in low numbers is due to the fact that we want to see as much variation in the symmetry-energy coefficients as possible, in order to improve our chances of correctly extrapolating our results in the limit of $A \rightarrow \infty$.  A much older set of corrections, also extending to low $A$, is that estimated by von Groote {\em et al.} \cite{Groote1976418}.  Finally, we employ the more commonly employed set of corrections by Moller {\em et al.} \cite{Moller:1993ed}, extending down to $N=Z=8$.  Besides pairing and shell effects, the corrections should compensate also for shape effects.  The latter are indeed intertwined with shell effects for heavy nuclei.   With this, the corrected energies refer to spherical nuclei.  The corrected energies will be marked in the paper with an apostrophe:
\beq
E' = E - E_\text{mic}^\text{est} \, .
\eeq
Upon application of the corrections, excitation energies to IAS should directly reflect the symmetry energy, with
\beq
\label{eq:EIASpb}
E_\text{IAS}^{* \prime } =  \frac{4 \, a_a(A,Z) }{A} \, \Delta \Big[T \big( T + b(A,Z) \big) \Big] + \Delta \delta  \, ,
\eeq
being the most general case.  Here, $\delta$ represents any remaining difference between the actual and estimated microscopic contribution to the energy, $\delta = E_\text{mic} - E_\text{mic}^\text{est}$.  The obvious issue that we encounter is that of the shell corrections being developed in the literature for ground states only, while we need corrections for the excited IAS, in order to assess the symmetry energy.  To solve this problem we assume that the microscopic corrections weakly depend on $T_z$ within an isospin multiplet:
\beq
\label{eq:Emicmult}
E_\text{mic}(A, T, T_z) \simeq E_\text{mic} (A, T, \pm T) \, .
\eeq
When both $T_z = \pm T$ members of the isospin multiplet are available, we use a linear combination of their corrections.  We expect that the error in the procedure to be governed by the finding in \eqref{eq:errorA}.  Whether the assumption above works should be verified by the smoothness of the results on symmetry energy.  In practice, we encounter stronger fluctuations in the results for $a_a$ than expected on the basis of \eqref{eq:errorA} alone, i.e.\ the uncertainty in \eqref{eq:Emicmult}, introduced by the transcription/interpolation procedure, turns out to be in the practical sense irrelevant.

\begin{figure}
\centerline{\includegraphics[width=.75\linewidth]{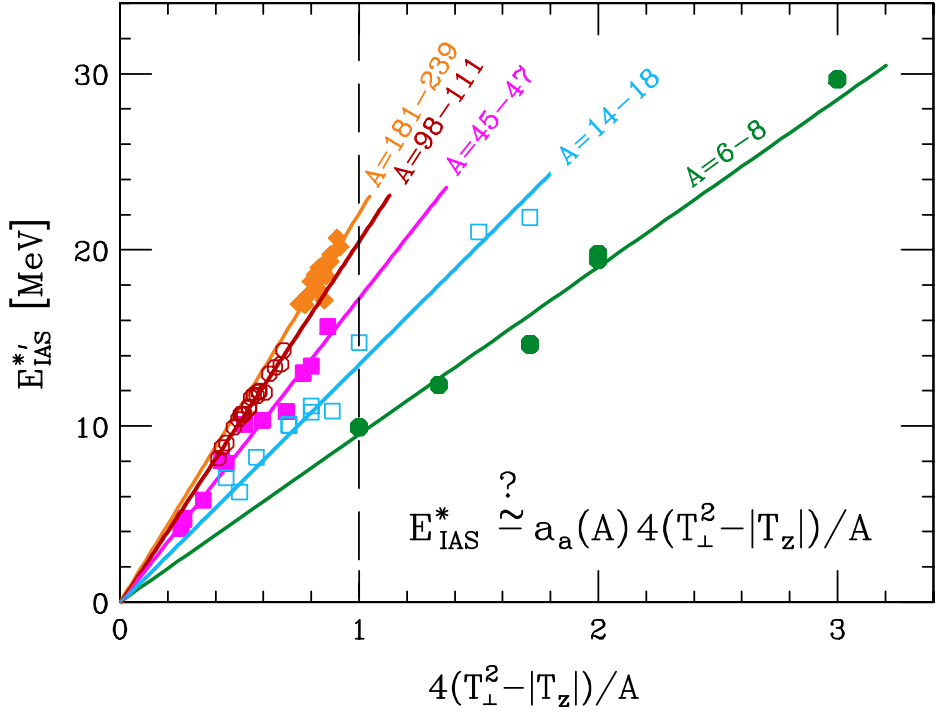}}
\caption{Excitation energies to the ground-state IAS, corrected for microscopic effects~\cite{Koura:2005}, plotted vs change in $T(T+1)$ scaled with~$A$, represented by different symbols for different indicated intervals in the mass number $A$.  Solid lines represent linear least-square fits to the represented energies.  The fits are forced to pass through the coordinate origin.  The dashed vertical line helps to read off the values of $a_a(A)$ for the specific $A$-intervals, from intersection of the fitted lines with the vertical line.
}
\label{fig:EIASTTA}
\end{figure}

Testing equations such as \eqref{eq:EIAS} on a nucleus-nucleus basis is hampered by the limited number of ground-state IAS known for individual nuclei.  However, in the bulk limit within which the symmetry energy is considered, nuclear properties are expected to change only gradually with mass and charge numbers.  Correspondingly, we can attempt to lump results from different nuclei, provided $A$ and $Z$ do not change much, in order to test various suppositions.  As a first step, we test Eq.~\eqref{eq:EIAS} by taking narrow intervals in $A$ and by plotting corrected excitation energies of IAS, vs the change in the isospin squared divided by $A$. This is done in Fig.~\ref{fig:EIASTTA}.  As evident in that figure, the excitation energies to the ground-state~IAS, within narrow intervals of~$A$, indeed rise approximately linearly with the change in $T(T+1)$ scaled with~$A$.  The slope, identified with the generalized symmetry coefficient, can be read off from the intersection of the fit to data with the dashed vertical line at $4\Delta [T(T+1)]/A =1$.

\begin{figure}
\centerline{\includegraphics[width=.85\linewidth]{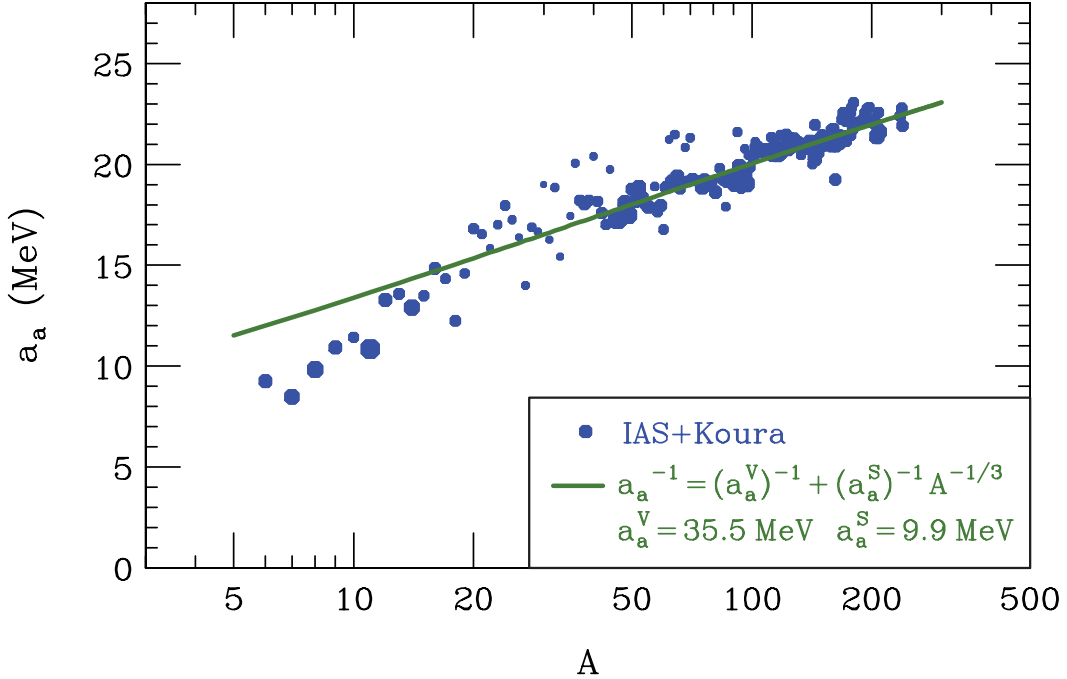}}
\caption{Generalized mass-dependent asymmetry coefficient $a_a(A)$ extracted from excitation energies to ground-state IAS within individual isobaric chains $A$, when applying shell corrections by Koura \etal\;   Symbol size reflects relative significance of a result.  The line represents a fit at $A \ge 30$, assuming a surface-volume competition in the asymmetry coefficient.
}
\label{fig:asyma}
\end{figure}

We next calculate $a_a(A)$ from the dependence of $E_\text{IAS}^{*\prime}$ on $4T(T+1)/A$ for individual values of~$A$.  The obtained parameter values are shown in Fig.~\ref{fig:asyma}.  Some caveats in the determination of $a_a(A)$ represented in Fig.~\ref{fig:asyma} should be mentioned.  Thus, for some isobaric chains~$A$, only a single ground-state IAS is known, likely making the deduced $a_a$-values fragile.  That is often the case for large~$A$ (that tend to be odd).  For low $A$, often many ground-state IAS are known, e.g.\ as many as $n=7$ IAS for $A=24$.  Other issues have to do with correlations underlying the fitted data, and with estimated uncertainties for deduced coefficient values.

One correlation issue is that of the different excitation energies within one nucleus getting correlated
through the subtraction of the single ground-state energy, from the energies of different excited states.
The issue is that, in spite of the applied corrections using estimated microscopic contributions, both the excited- and ground-state energies are bound to contain some residual microscopic contributions $\delta$.  These contributions will make the deduced~$a_a(A)$ fluctuate and it is desirable to minimize the anticipated fluctuations.  Assuming an independence in the residual microscopic contributions to the ground and excited states and inverting the respective correlation matrix, see Appendix \ref{apx:Stat}, one finds that the quantity which needs to be minimized, in seeking the parameters of the symmetry energy $E_1(A,Z,T)$ for an $(A,Z)$ nucleus, is
\beq
\label{eq:chimod}
\sum_{i=1}^n \left[ \left(E_\text{IAS}^{* \prime} \right)_i - \Delta E_1(A,Z,T_i) \right]^2 - \frac{1}{n+1} \left[  \sum_{i=1}^n \left[ \left(E_\text{IAS}^{* \prime} \right)_i - \Delta E_1(A,Z,T_i) \right]  \right]^2 \, ,
\eeq
rather the simple sum of squares in the first term of \eqref{eq:chimod}.  In the above expression,
$i$ is the index for ground-state IAS in the spectrum and $n$ is the number of such IAS.  In the expression~\eqref{eq:chimod}, the energies of the ground and excited states are treated democratically.  The~expression~\eqref{eq:chimod} can be contrasted with two other ones, the naive one arrived when the correlation due to common ground-state subtraction is ignored and another one arrived at when the information in the ground state is disregarded.  In the naive case, the second subtraction term that appears in~\eqref{eq:chimod} is missing.  When the ground state is disregarded, the prefactor in the subtraction term changes from $1/(n+1)$ to $1/n$.

If $E_1$ from the r.h.s.\ of \eqref{eq:Eaaz} is used in \eqref{eq:chimod}, the best-fit value for the asymmetry coefficient, from minimizing \eqref{eq:chimod}, is found to be
\beq
\label{eq:deltaaz}
a_a(A,Z) \simeq \frac{ \sum_{i}^{n} x_{i} \, \left(E_\text{IAS}^{* \prime} \right)_i - \frac{1}{n+1}  \sum_{i}^{n}  x_{i}  \sum_j^{n} \left(E_\text{IAS}^{* \prime} \right)_j }{\sum_{i}^{n} x_{i}^2 - \frac{1}{n+1} \left( \sum_{i}^{n}  x_{i}  \right)^2} \, .
\eeq
Here, $x_i$ are the abscissas in Fig.~\ref{fig:EIASTTA}, $x_i = 4 \left( \Delta {\pmb T}^2 \right)_i/A$.  The associated uncertainty for the coefficient \eqref{eq:deltaaz} is expected to be
\beq
\label{eq:azerror}
\delta a_a(A,Z) \simeq \frac{\delta_\text{rms}} {\sqrt{  \sum_{i}^{n} x_{i}^2 - \frac{1}{n+1} \left( \sum_{i}^{n}  x_{i}  \right)^2  } } \, ,
\eeq
where $\delta_\text{rms}$ on the r.h.s.\ represents the r.m.s.\ value for $(E_\text{mic} - E_\text{mic}^\text{est})$.

For Fig.~\ref{fig:asyma}, we combine information from different $Z$-values for a~given~$A$, as discussed below and in Appendix \ref{apx:Stat}, to arrive at~$a_a(A)$.  Sizes of the symbols in Fig.~\ref{fig:asyma} represent relative significance of the results  within a certain range of~$A$, tied to the inverse of the factor multiplying $\delta_\text{rms}$ in \eqref{eq:azerror}.  Principally, $\delta_\text{rms}$ could be $A$-dependent, which would have been the case if $\delta_\text{rms}$ were actually adequately represented by the observation \eqref{eq:errorA}.  Potential use of~\eqref{eq:errorA} for $\delta_\text{rms}$ presents, though, the following problems.  Thus, the use of Eq.~\eqref{eq:errorA} in~\eqref{eq:azerror} would predict a rapid shrinkage of the errors in the values of $a_a(A)$ with $A$, which is inconsistent with the scatter of values of $a_a(A)$ in Fig.~\ref{fig:asyma}, which scatter persists little changed up to high~$A$.  Second, even at low $A$, the errors predicted with \eqref{eq:errorA} would underestimate the scatter of the values in Fig.~\ref{fig:asyma} by a factor of~2 or more.  In fact, the scatter in Fig.~\ref{fig:asyma} is consistent with an approximately constant $\delta_\text{rms}$ as a function of~$A$, with the value of $\delta_\text{rms} \approx 0.5 \, \text{MeV}$.  For example, the values of $a_a(A)$ at $A \gtrsim 20$ appear consistent with a simple competition of surface and volume effects in the symmetry coefficient, cf.~I,
\beq
\label{eq:aaVS}
\frac{1}{a_a(A)} = \frac{1}{a_a^V} + \frac{A^{-1/3}}{a_a^S} \, .
\eeq
If we fit Eq.~\eqref{eq:aaVS} to the values of $a_a(A)$ from IAS at $A \ge 30$, we arrive at $\chi^2$ per degree of freedom of about~1.  The value of $\delta_\text{rms} \approx 0.5 \, \text{MeV}$ is close to the typical accuracy of mass formulas combining macroscopic and microscopic contributions, of about $0.65 \, \text{MeV}$~\cite{Moller:1993ed,Koura2000,Koura:2005}.  The accuracy of the mass formulas improves down to about $0.50 \, \text{MeV}$, when only heavier, $A \gtrsim 70$, nuclei are considered.  It is possible that focusing on just on one aspect of the nuclear energy allows for an improved quality of a macroscopic description, down to lower~$A$.  The specific values of parameters $a_a^V$ and $a_a^S$ are quite stable with respect to the applied cut-off in mass, i.e.\ about the same values are obtained for $A \ge 18$ and $A \ge 60$.

\begin{figure}
\centerline{\includegraphics[width=.85\linewidth]{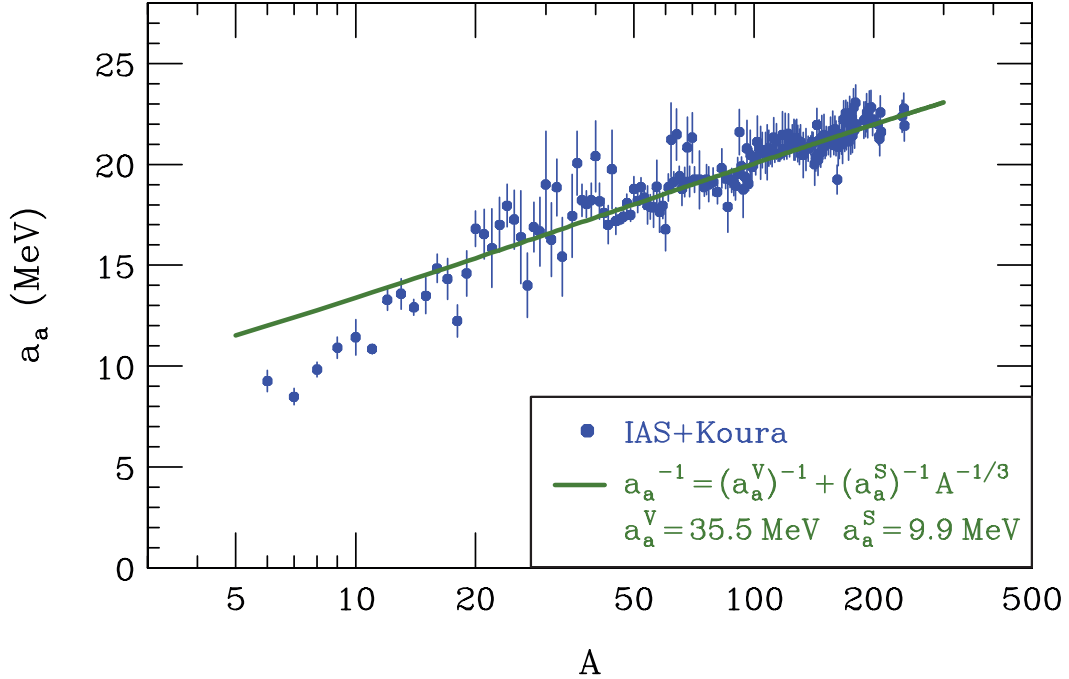}}
\caption{Same results as in Fig.~\ref{fig:asyma}, for the generalized asymmetry coefficient $a_a(A)$, now displayed with the estimated errors.  The line represents a fit at $A \ge 30$ assuming a combination of volume and surface symmetry terms.
}
\label{fig:asyme}
\end{figure}

Relatively small variations \eqref{eq:errorA} of energy differences between states of different isospin, within one isobaric chain, as compared to $\delta_\text{rms} \sim 0.5 \, \text{MeV}$, brings about another correlation in the analysis of IAS data: on the scale of $\delta$, the~isospin symmetry in excitation energies appears nearly exact, so that the spectra for different~$Z$, connected by isospin symmetry, do not bring independent information pertinent to~$a_a(A)$. To cope with this correlation problem, to the extent possible, when calculating $a_a(A)$ for Fig.~\ref{fig:asyma}, we lump together the spectra of ground-state IAS from different $Z$.  Other than that, in calculating $a_a$, we drop states for which we lack shell corrections.
For reason discussed in the next subsection, when we have enough many states for a given $N=Z$ nucleus, we drop the $T=0$ ground state from our analysis.  The results from Fig.~\ref{fig:asyma} are represented next again in Fig.~\ref{fig:asyme}, now with the errors estimated following Eq.~\eqref{eq:deltaaz} with $\delta_\text{rms} = 0.5 \, \text{MeV}$.

Within some narrow intervals of $A$ in Fig.~\ref{fig:asyme}, the values of $a_a(A)$ appear primarily displaced in one direction, away from an expected smooth behavior, revealing the final correlation issue in the IAS analysis.  Apparently, some systematic effects are not fully accounted for by the $E_\text{mic}^\text{est}$-subtraction and these correlate the $\delta$-values, with the correlation propagated onto the correlations between the $a_a(A)$-values.  This is actually common for analyses of nuclear energies in terms of a combination of bulk and microscopic contributions \cite{Moller:1993ed}.  The~consequence is that the results from the analysis over more narrow intervals of~$A$ may be biased.  Less bias is expected from analysis over wider regions, where displacements of different sign may balance out.

Next, following the previously laid out strategy and paying attention to the issues above, we test the robustness of the results so-far, on the generalized coefficient $a_a$.

\subsection{Robustness of the Results on Symmetry Coefficients}

In assessing the robustness, we first test the dependence of the generalized symmetry coefficients on $Z$.  Such a dependence was claimed in the literature, when these coefficients were obtained in a different manner~\cite{Janecke200323,Janecke2007317,PhysRevC.81.067302}.  In investigating separate $Z$, we obviously do not lump anymore the IAS spectra within one $A$ and we obtain the results directly from~\eqref{eq:deltaaz} and \eqref{eq:azerror}, with one caveat, though.  We exclude from the analysis the ground states of $N=Z$ nuclei, or $T=0$ states, if more than one IAS of a ground state is known in the excitation spectrum.  In practical terms, this exclusion (cf.~the discussion before) amounts to the replacement of the prefactors $1/(n+1)$ by $1/n$ in Eqs.~\eqref{eq:chimod}, \eqref{eq:deltaaz} and~\eqref{eq:azerror}.  Without such an exclusion, the coefficients for $N=Z$ nuclei only, of masses $A = (12$--$40)$, come out slightly, still within individual errors, but {\em systematically} depressed compared to the coefficients for other nuclei within the same isobaric chain.  When it is possible to exclude the $T=0$ states, no systematic discontinuity is seen anymore in the behavior of the coefficients with $Z$.  We suspect that the effect is associated with an interplay between the treatment of pairing in the microscopic corrections of Koura \etal~\cite{Koura:2005} and the competition between the $T=1$ and $T=0$ pairing in the $N=Z$ nuclei.

\begin{figure}
\centerline{\includegraphics[width=.60\linewidth]{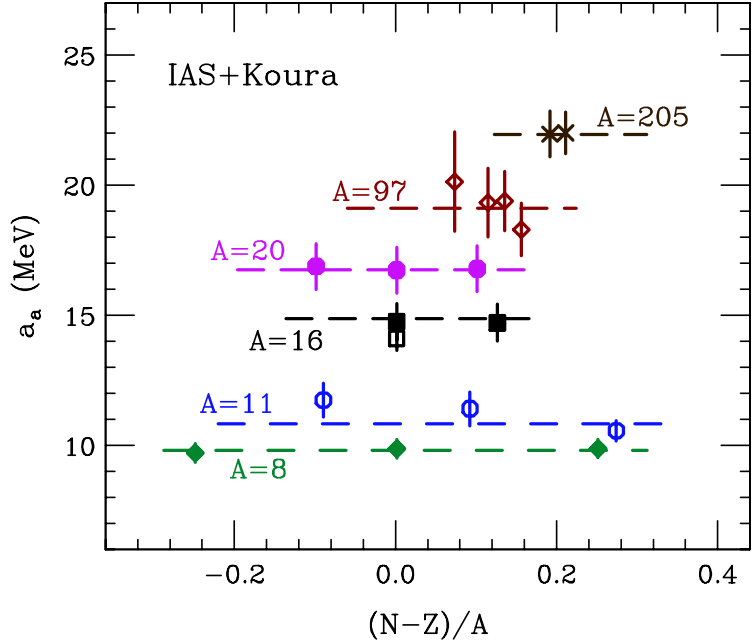}}
\caption{Generalized symmetry coefficient for individual nuclei, within selected isobar chains characterized by mass numbers~$A$, presented here as a function of asymmetry.  The horizontal dash lines represent the coefficient values for the isobaric chains as a whole.  For $^\text{16}$O, we show both the coefficient value when disregarding (filled square) and including (open square) the $T=0$ ground state in our analysis.  The observed type of downward shift is characteristic for the $N=Z$ nuclei in the region of $A=(12$--$40)$.
}
\label{fig:asymz}
\end{figure}

Exemplary results from our analysis, with the emphasis on a span of mass numbers and of coefficient values and on the cases with a larger number of isobars, are shown in Fig.~\ref{fig:asymz}.  Within our results, we find no evidence for a dependence of the coefficients on asymmetry.  In fact, the results between different $Z$ tend to agree better than is expected on the basis of individual errors, due to the correlations between different members of a chain introduced by strong effects of isospin symmetry, when portions of the spectra for the members overlap within the space of net isospin.  For mirror nuclei, with IAS of the same isospin known in the two spectra, we find cases where the discrepancy between the symmetry coefficients is of the order of 10~keV, i.e.~two orders of magnitude smaller than the individual errors for the coefficients!

In arriving at the symmetry coefficients so far, we assumed the symmetry-energy contribution to the excitation energies to IAS of the form represented in Eqs.~\eqref{eq:EIAS}--\eqref{eq:Eaaz}, i.e.\ $E_1 \propto T(T+1)$.  We will now relax that assumption and consider the form $E_1 \propto T(T + b)$, that underlies Eq.~\eqref{eq:IASb}, examining whether there is an evidence within the IAS energies for $b \ne 1$, when extrapolating in isospin of the assessed states towards $T=0$.  This question is important in the context of quartetting in the nuclei \cite{deShalitTalmi} and in the context of the principal origin of the symmetry energy~\cite{Satula2003152}.  We generally lack enough many states within individual nuclei and even within individual isobaric chains, to address the question of $b$ on a nucleus-by-nucleus or even on individual-$A$ basis.  However, if we consider relatively narrow intervals of~$A$, the parameters of the symmetry energy may vary little enough, so that results from different nuclei may be combined to address the validity of Eq.~\eqref{eq:IASb} and of the corresponding Eq.~\eqref{eq:EIASpb}.  In order to illuminate the behavior of the excitation energies to IAS extrapolated to $T=0$, for consecutive intervals in~$A$, we plot in Fig.~\ref{fig:eat} the energies scaled with $A/[4(T - |T_z|)]$, against $(T + |T_z|)$, where the isospin numbers pertain to IAS.  According to Eq.~\eqref{eq:IASb}, the scaled energies should align themselves with a line of which the slope is $a_a(A)$ and of which the intersection with the abscissa is $-b(A)$.  An alignment is indeed seen in Eq.~\eqref{eq:IASb} for individual mass regions.  Except for the highest mass interval, lines fitted to the scaled energies intersect the abscissa around -1, giving support to $b \sim 1$.

\begin{figure}
\includegraphics[width=.48\linewidth]{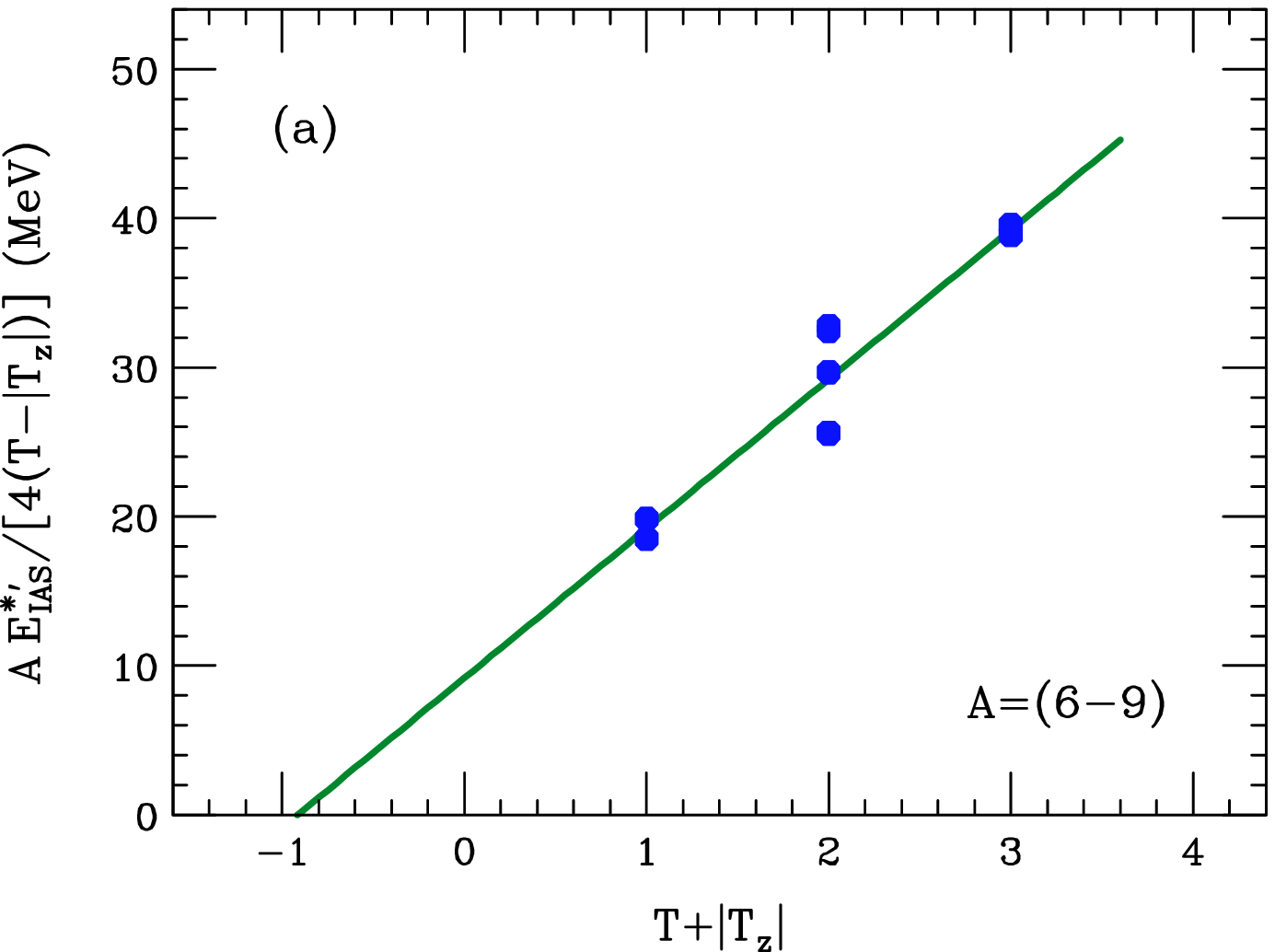}\hspace*{.04\linewidth}
\includegraphics[width=.48\linewidth]{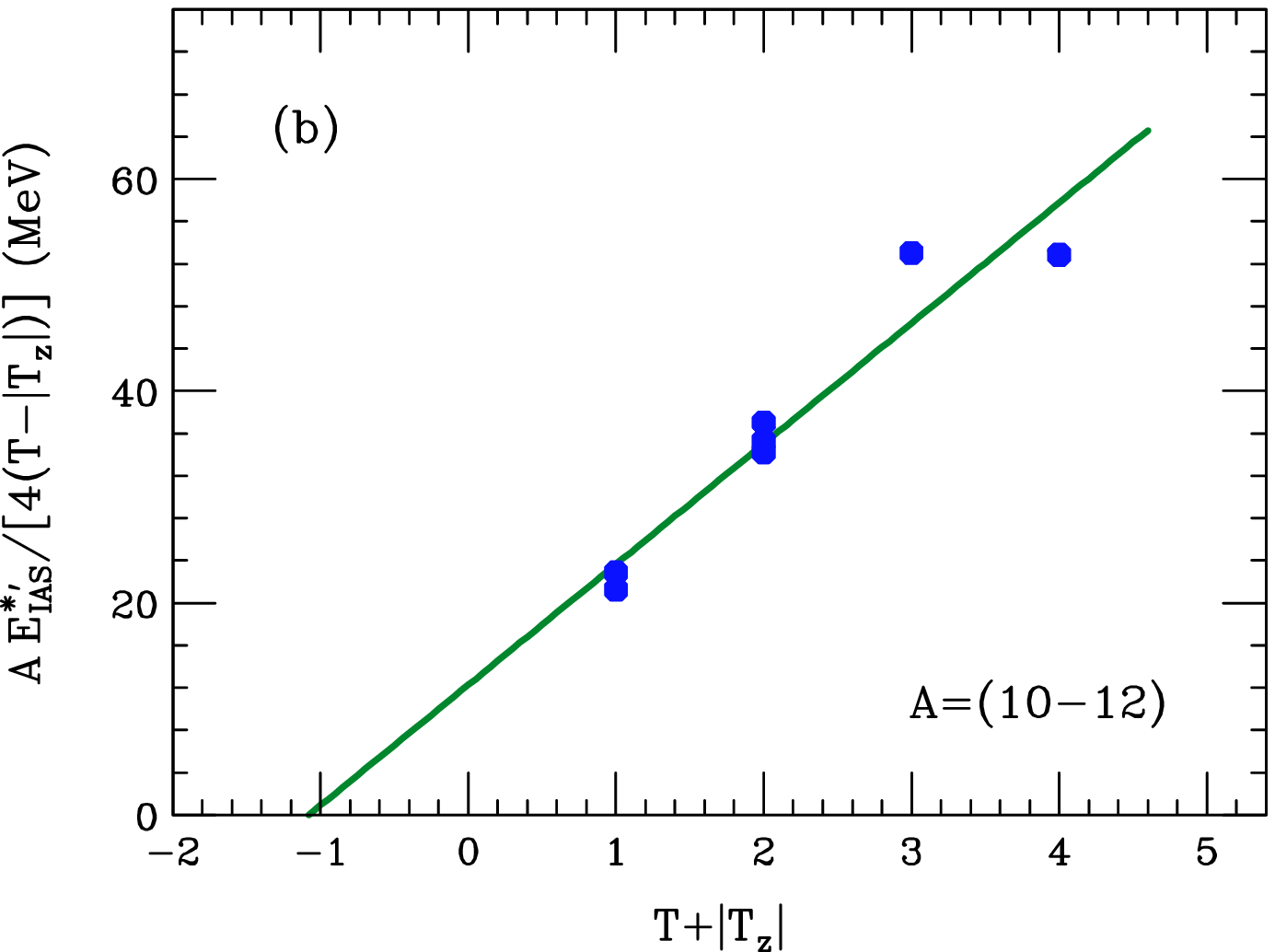}\\[2ex]
\includegraphics[width=.48\linewidth]{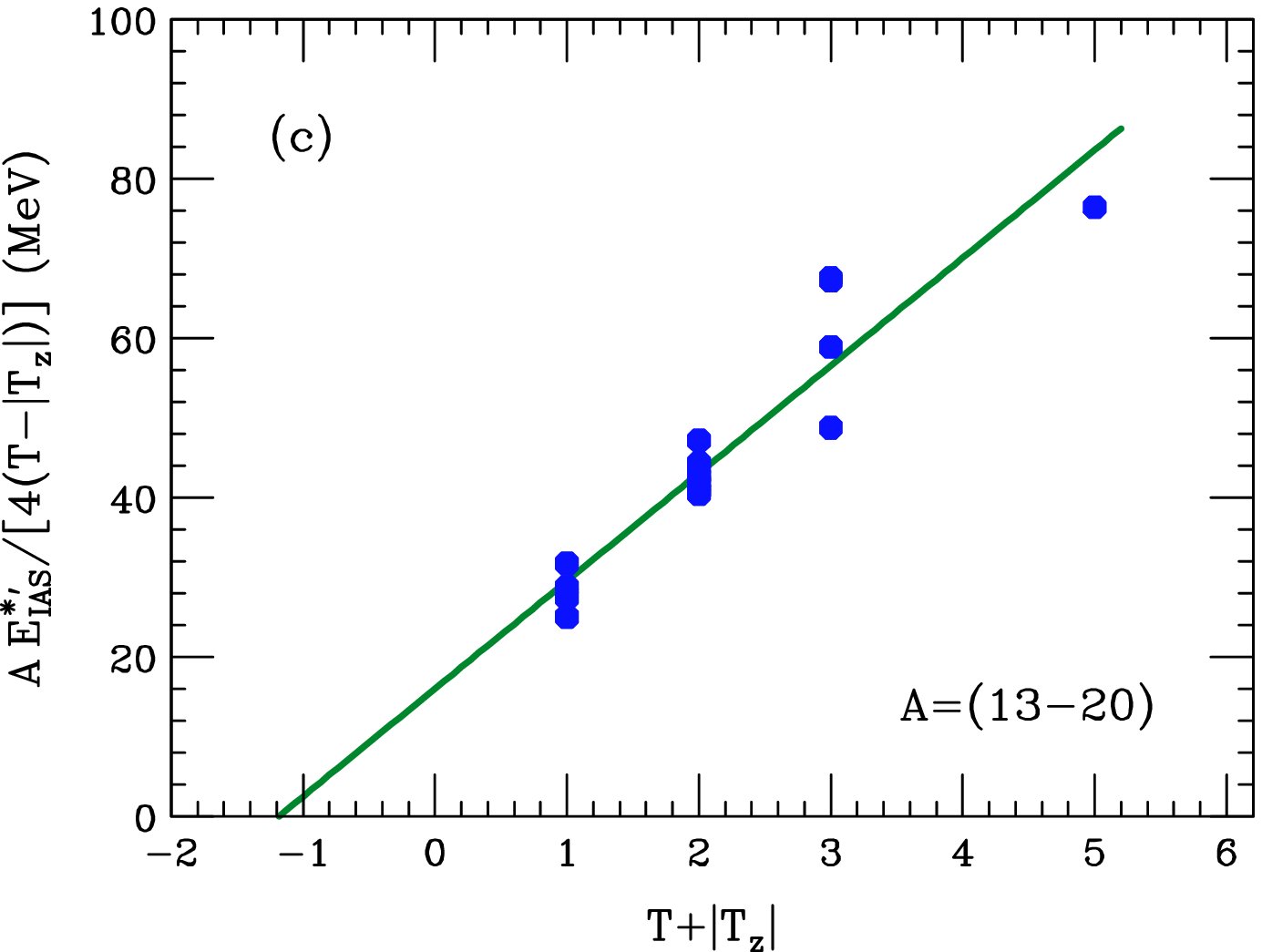}\hspace*{.04\linewidth}
\includegraphics[width=.48\linewidth]{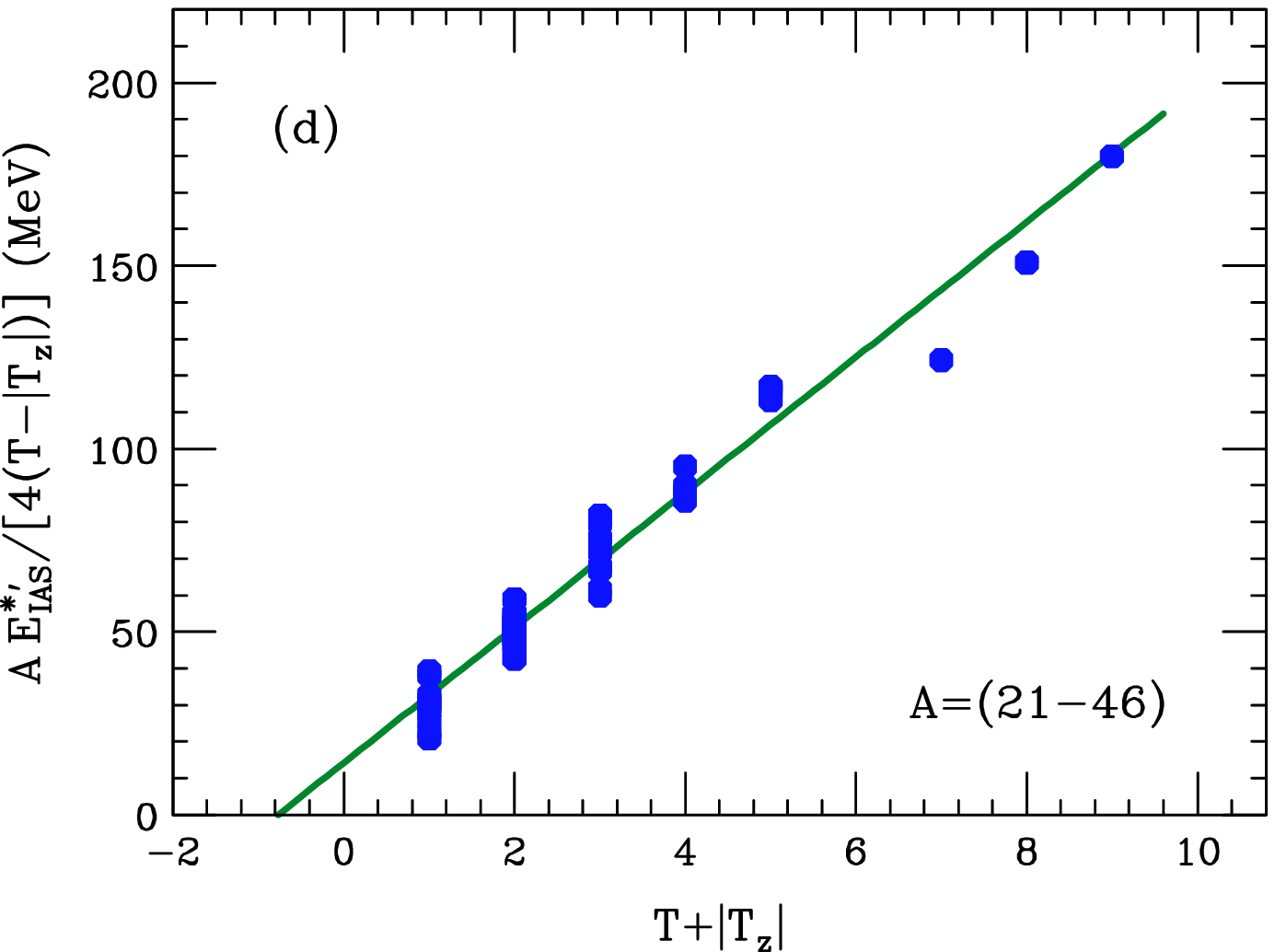}
\caption{
\label{fig:eat}
Excitation energies to ground-state IAS, with microscopic corrections \cite{Koura:2005} incorporated, scaled with mass number and with the difference of quantum numbers for the net and third components, are plotted as symbols vs the sum of the quantum numbers for the net and third isospin components, in the indicated consecutive mass intervals.  Lines indicate linear fits to the scaled excitation energies.  Line slope and its intersection with the abscissa are expected to represent, respectively, the generalized symmetry coefficient and negative of the parameter $b$ for the specific mass interval.  This figure continues to the next page\ldots}
\end{figure}
\addtocounter{figure}{-1}
\begin{figure}
\includegraphics[width=.48\linewidth]{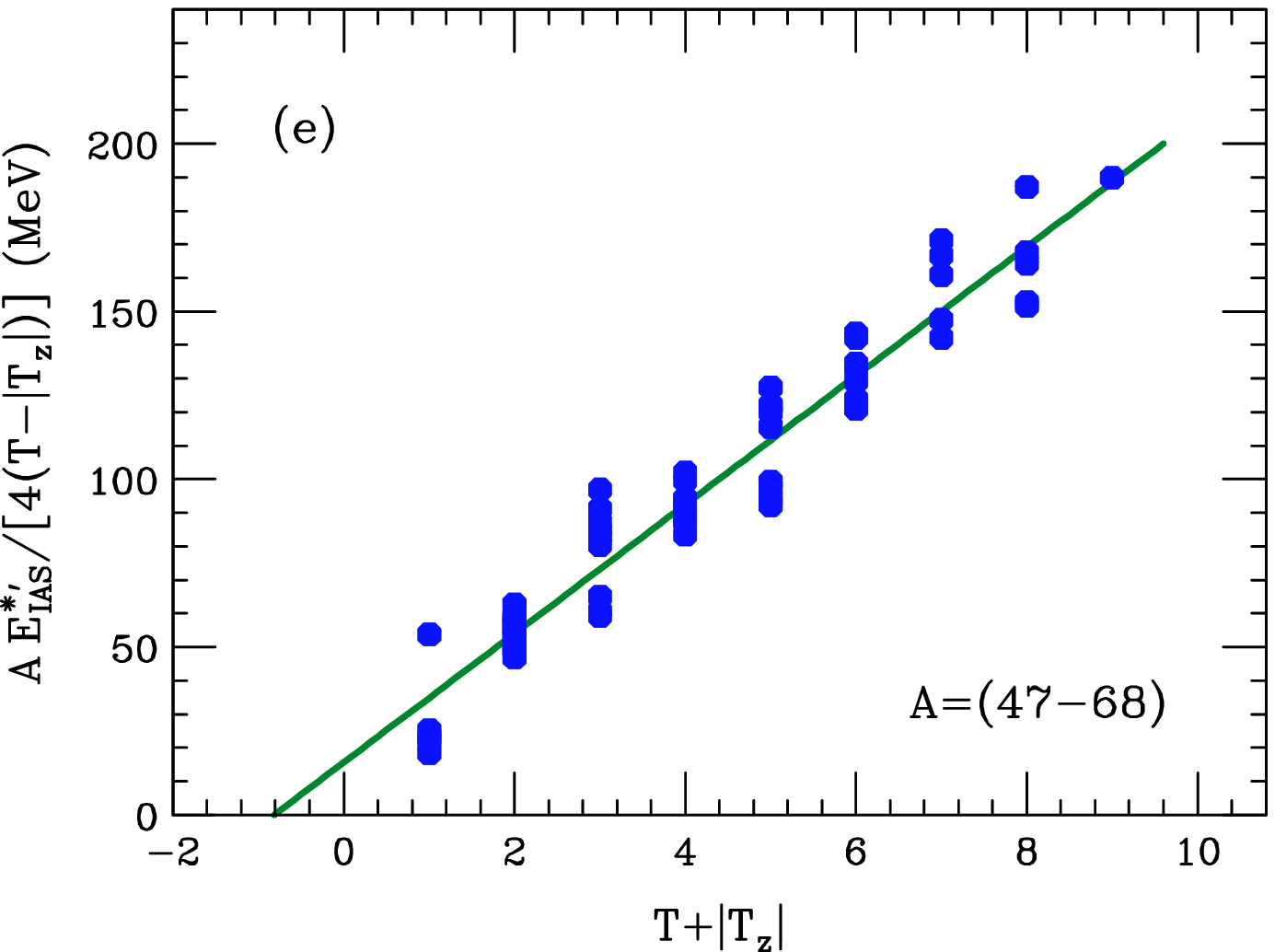}\hspace*{.04\linewidth}
\includegraphics[width=.48\linewidth]{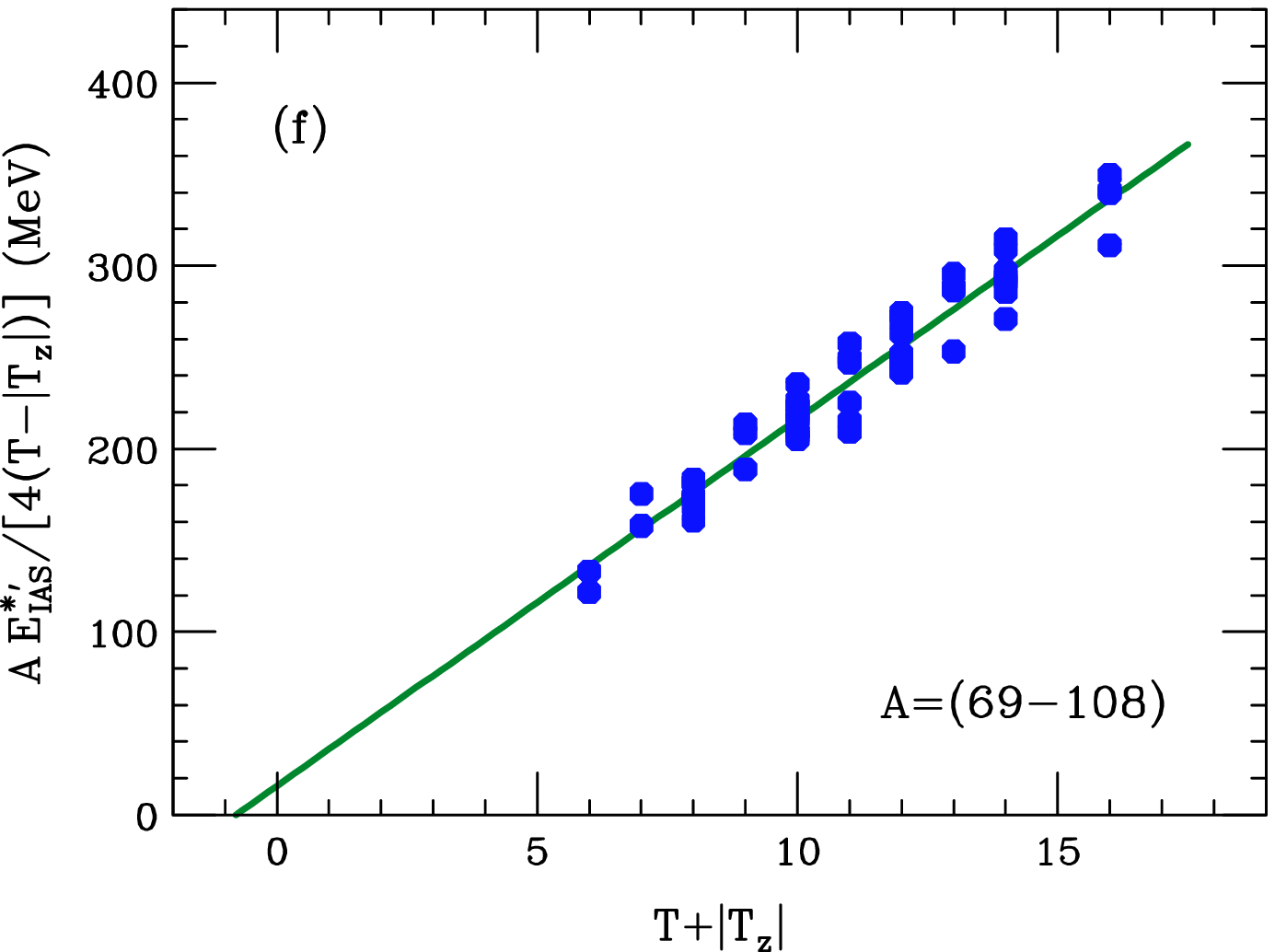}\\[2ex]
\includegraphics[width=.48\linewidth]{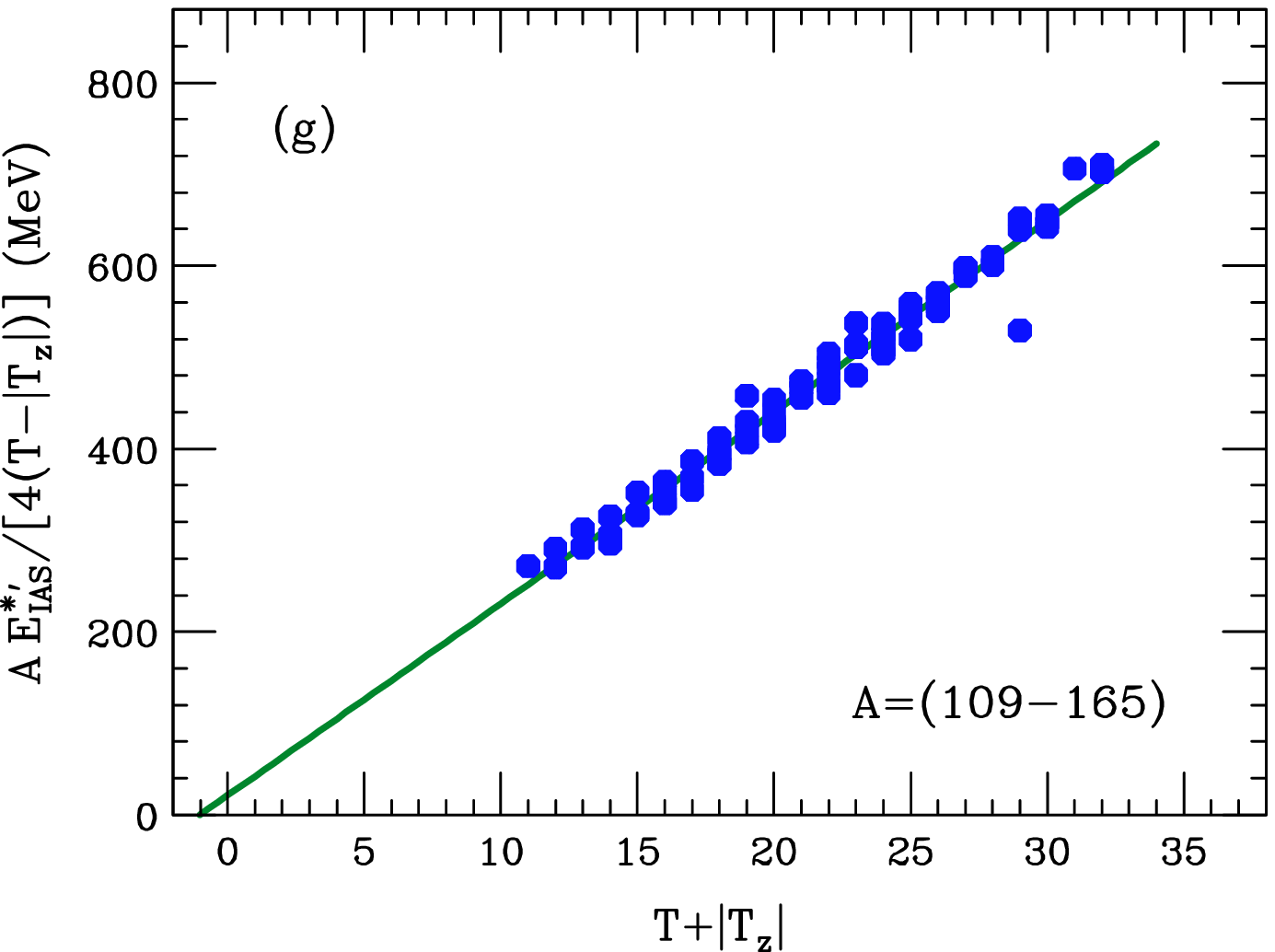}\hspace*{.04\linewidth}
\includegraphics[width=.48\linewidth]{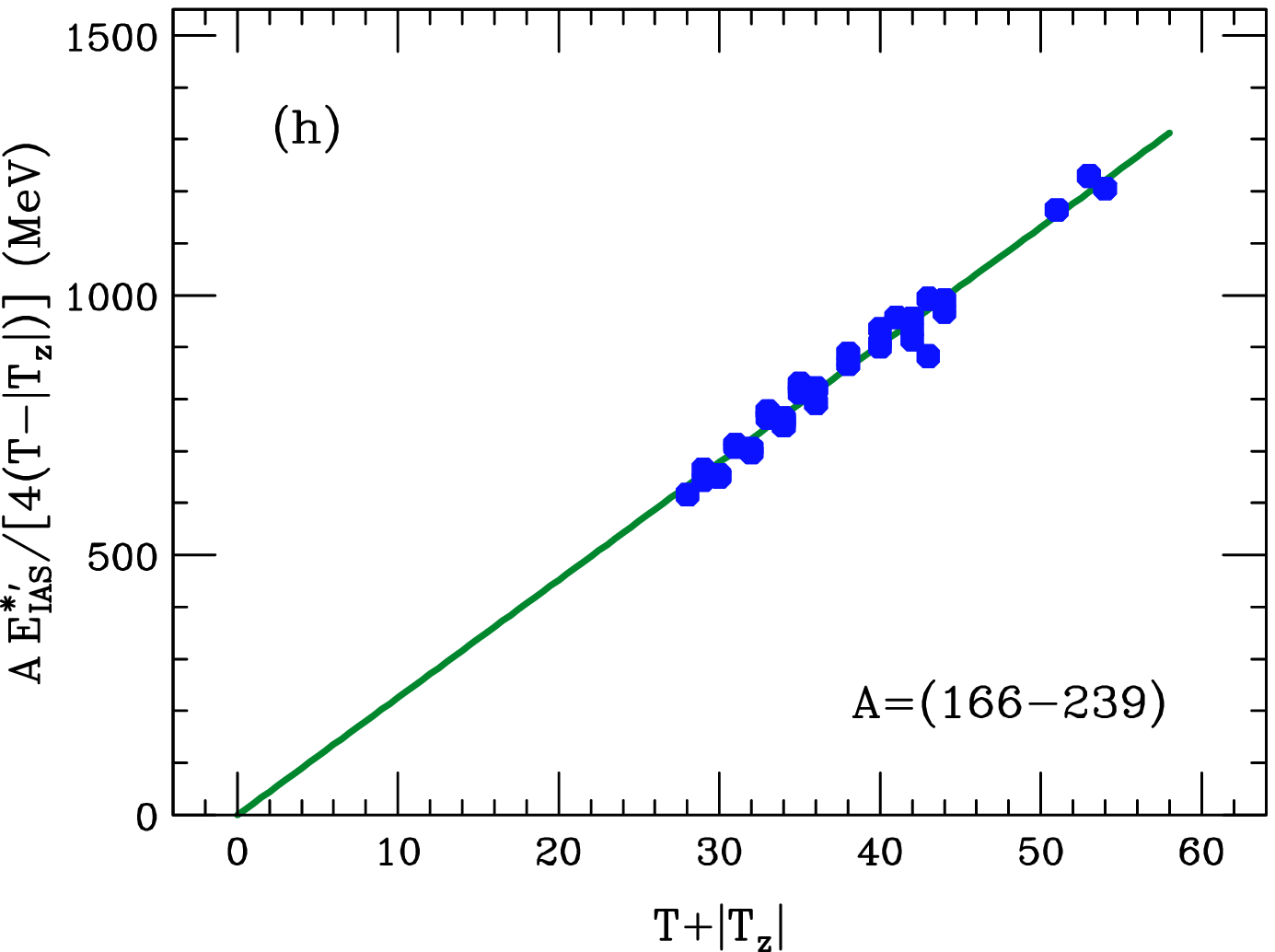}
\caption{-- Continued.}
\end{figure}

\begin{figure}
\centerline{\includegraphics[width=.60\linewidth]{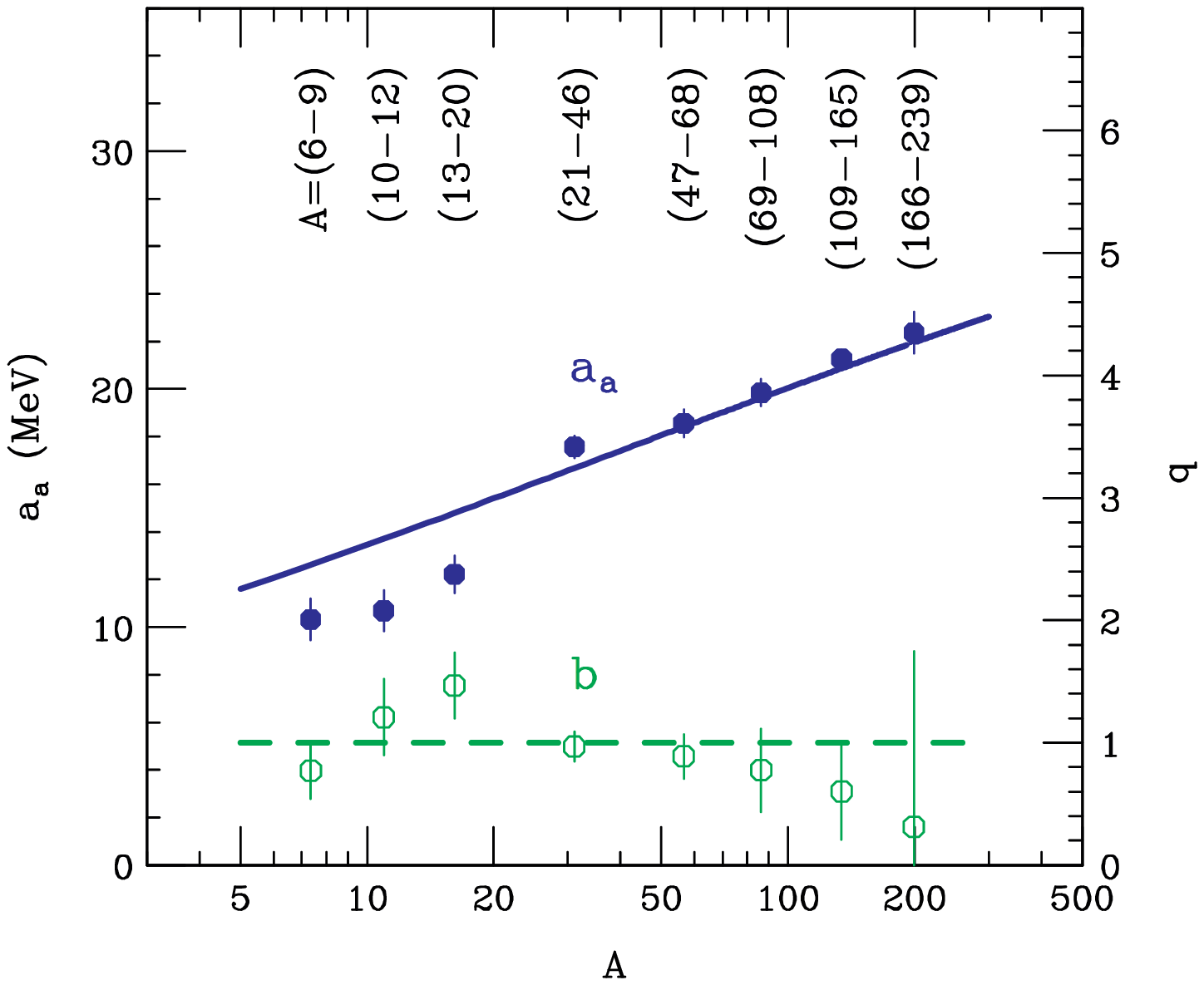}}
\caption{Generalized symmetry coefficient $a_a$ (filled symbols) and parameter $b$ (open symbols) from fitting Eq.~\eqref{eq:EIASpb} to IAS data, with shell corrections \cite{Koura:2005} applied, over indicated mass regions.  The~left scale in the figure is for $a_a$ and the right is for $b$.  The lines serve to guide the eye.  The~dashed line represents $b=1$ while the solid line reproduces the surface-volume fit to the mass dependence of~$a_a$ of Fig.~\ref{fig:asyme}.
}
\label{fig:asb}
\end{figure}

In place of fitting lines to the energies in Fig.~\ref{fig:eat}, to arrive at estimates of parameter~$b$, the~analysis of proper covariance matrix of Appendix \ref{apx:Stat} shows that more faithful results should result from using the result \eqref{eq:chimod} and, specifically, from minimizing terms such as in \eqref{eq:chimod}, summed over isotopes contributing to a given mass interval.  With the excitation energies represented in Eq.~\eqref{eq:EIASpb} as linear both in $b$ and $a_a$, analytic (however generally nested) expressions follow both for the optimal $b$ and $a_a$ values and for their errors.  The results from the discussed optimization, within the consecutive mass intervals, are next shown in Fig.~\ref{fig:asb}.  The fitted values of $b$ in the figure fluctuate around $b=1$ and the values of $a_a$ generally follow the mass-dependence obtained when assuming $b=1$.  The results from the mass intervals are somewhat fragile in that too narrow mass intervals result in unsatisfactorily large errors while wider mass intervals encompass $a_a$-values that are likely to evolve with~$A$.  However, we see no indications for significant deviations of $b$ from~1 and in the following generally rely on results obtained when enforcing $b=1$.

The next issue we address is that of biasing of the results by the employed microscopic corrections.  The general expectation is that, without such corrections, physical quantities fluctuate around those representing the bulk limit.  As a result, coarse fits to physical quantities, across broad ranges of $N$ and $Z$ assuming weak changes across shells, such as with Eq.~\eqref{eq:aaVS}, should yield largely unchanged results whether made with corrected or uncorrected data.  We should mention here that, when shell corrections are developed for net nuclear energy, there is a definite possibility that their application could produce a~relatively smooth net energy as a function of $N$ and $Z$, but {\em not} a smooth symmetry energy coefficient, following procedures such as here, given the low contributions of the symmetry energy to the net energy for relatively small changes in isospin.

\begin{figure}
\centerline{\includegraphics[width=.85\linewidth]{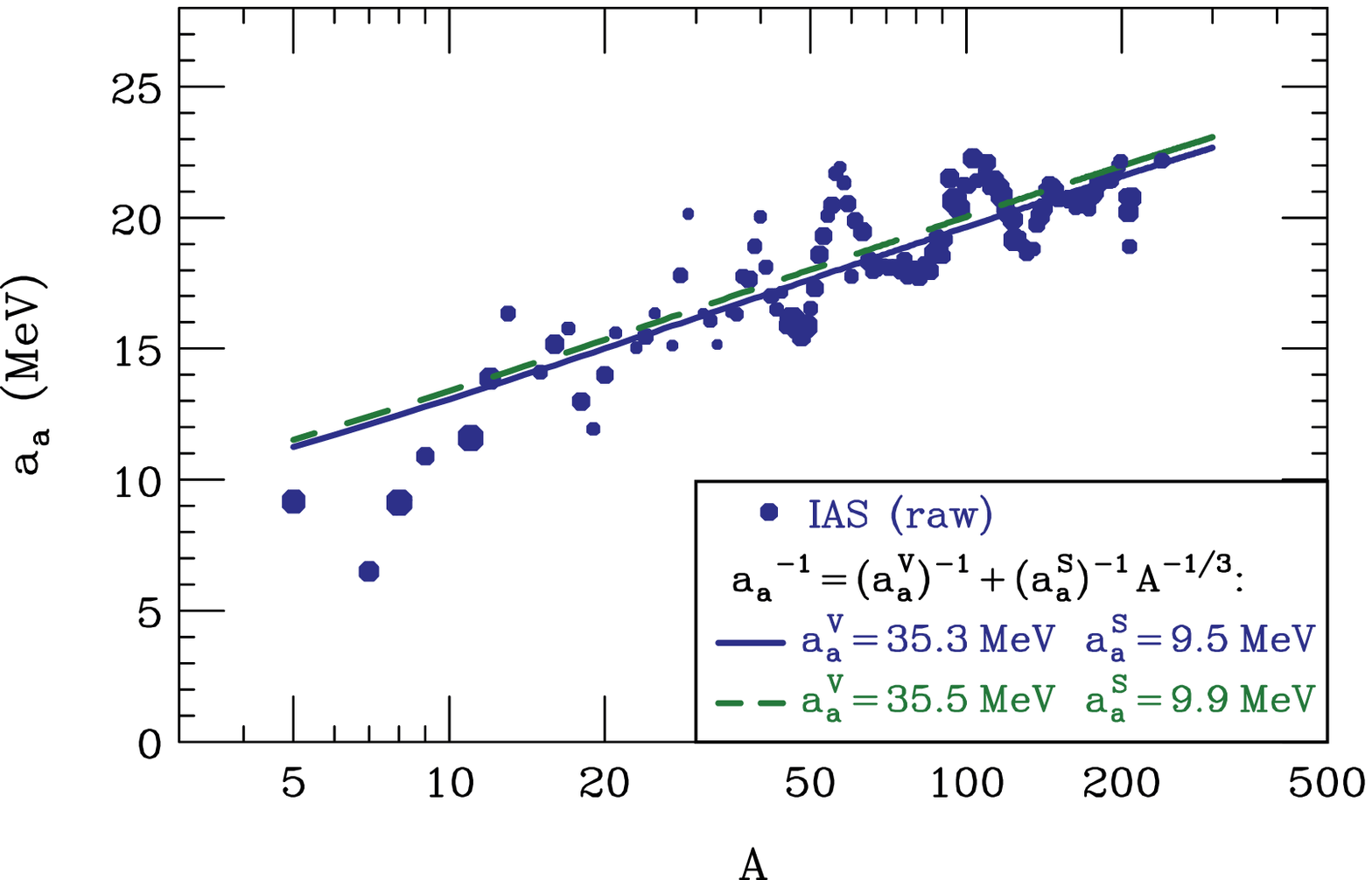}}
\caption{Generalized mass-dependent asymmetry coefficient $a_a(A)$ extracted from excitation energies to ground-state IAS within individual isobaric chains $A$, when no microscopic corrections are applied to the state energies.  Symbol size reflects relative significance of a result.  The solid line represents a fit at $A \ge 30$ to the displayed results, assuming a surface-volume competition in the asymmetry coefficient.  The dashed line represents a fit to the results with corrections by Koura \etal~\cite{Koura:2005}, reproduced here from Fig.~\ref{fig:asyma}.
}
\label{fig:asyma_ns}
\end{figure}

In Fig.~\ref{fig:asyma_ns}, we next show results from an analysis of IAS states carried out as before, but without application of any microscopic corrections to the energies of the states.  The~only other change we make, compared to the previous analysis, is that for $A$-even nuclei we analyze separately the portions of spectra with even and odd $T$.  That separation is to avoid an impact of the $T=1$ pairing on the results.  It follows from Fig.~\ref{fig:asyma_ns} that the raw results, without any microscopic corrections applied, indeed generally oscillate around the previous results with those corrections \cite{Koura:2005}: previous fit at $A \ge 30$ to the results with corrections is represented with a dashed line in Fig.~\ref{fig:asyma_ns}.  Parameters from an analogous surface-volume fit to the results without corrections, represented by a solid line in Fig.~\ref{fig:asyma_ns}, are surprisingly close to the parameters from the previous fit.  Parameters of the fit to the results from raw data are reasonably stable when the cut-off in $A$ is varied between 18 and 45.  Interestingly, $\chi^2$~per degree of freedom, for deviation of the last results from that fit, becomes equal to about~1, when $\delta \sim 4 \, \text{MeV}$ is assumed.

\begin{figure}
\centerline{\includegraphics[width=.85\linewidth]{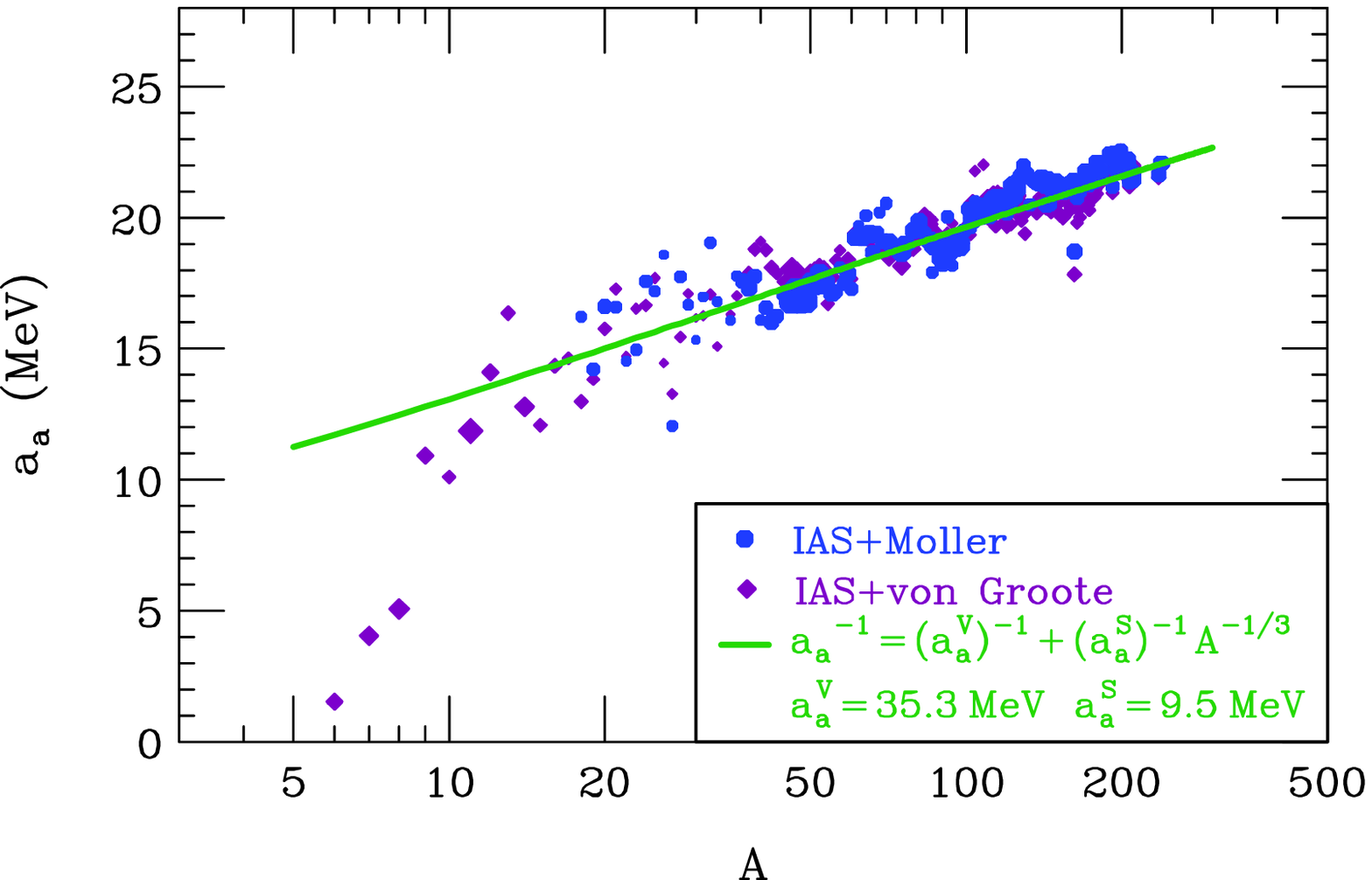}}
\caption{Generalized mass-dependent asymmetry coefficient $a_a(A)$ extracted from excitation energies to ground-state IAS within individual isobaric chains $A$, when microscopic corrections
by M\"{o}ller \etal~\cite{Moller:1993ed} (filled circles) and by von Groote \etal~\cite{Groote1976418} (filled diamonds) are employed.  The~solid line represents the surface-volume fit to the results obtained when no microscopic corrections are applied, reproduced here from Fig.~\ref{fig:asyma_ns}.
}
\label{fig:asyma_mh}
\end{figure}

Impact of alternate microscopic corrections is next illustrated in Fig.~\ref{fig:asyma_mh}, where we show results for the asymmetry coefficient obtained from the excitation energies to IAS when applying, respectively, the microscopic corrections to the energies by M\"{o}ller \etal~\cite{Moller:1993ed} and by von Groote \etal~\cite{Groote1976418}.  As corrections by M\"{o}ller \etal\  have been constructed only for nuclei with $N$, $Z \ge 8$, the corresponding results for light nuclei are missing.  In comparing Figs.~\ref{fig:asyma_mh}, \ref{fig:asyma_ns} and \ref{fig:asyma}, one can see that the corrections by M\"{o}ller \etal\ leave somewhat stronger systematic effects of uncompensated shell effects.  Otherwise the results are fairly similar.  The corrections by von Groote \etal\  produce similar results those by Koura \etal\  down to $A \sim 9$ and then the coefficients sink to lower values than in the case of Koura \etal\ Overall, for $A \gtrsim 20$ there is a good degree of agreement between the results obtained as a function of mass with the three sets of corrections, and with the line representing the coarse expectation when applying no corrections.  There are some differences, though, if one looks into finer details, as discussed below.  In particular, when relieving the condition $b=1$, the results for~$a_a$ with M\"{o}ller \etal\ corrections  tend to support $b \sim 1$, while the results with von Groote \etal\  are noisy with that regard.

Significant differences emerge for the values of $a_a^V$ and $a_a^S$, though, when the $a_a(A)$-results for different corrections are fitted, at $A \ge 30$, with the volume-surface formula \eqref{eq:aaVS}.  Thus, the corrections of M\"{o}ller \etal\  yield $a_a^V = 39.73 \, \text{MeV}$ and $a_a^S = 8.48 \, \text{MeV}$, while those of von Groote \etal\  yield $a_a^V = 31.74 \, \text{MeV}$ and $a_a^S = 11.27 \, \text{MeV}$,
to be compared with $a_a^V = 35.51 \, \text{MeV}$ and $a_a^S = 9.89 \, \text{MeV}$ for Koura \etal\ \  In spite of the large differences in the coefficient values, the volume-surface formulas, with the different parameter sets for Koura \etal , M\"{o}ller \etal\ and von Groote \etal\ corrections, are consistent with each other in the region of $30 \le A \le 240$, to within $0.4 \, \text{MeV}$ reached at the edges of the mass interval.  This will turn out to be important when drawing conclusions regarding nuclear matter.  Averaging the asymmetry capacitances (cf.~I) over the three parametrizations, at  fixed $A$ values, yields a compromise volume-surface formula with the parameter values of $a_a^V = 35.34 \, \text{MeV}$ and $a_a^S = 9.67 \, \text{MeV}$, not far from Koura \etal\ alone.  We adopt the formula with those parameters as a smooth representation of the $a_a(A)$ results in the region of $30 \ge A \ge 240$, with an rms systematic error across $A$, on account of uncertainty in the microscopic corrections, of $\sim 0.3 \, \text{MeV}$.  When comparing those results to coefficients from theoretic calculations of nuclei, an additional error will be associated with the extraction of coefficients representing those calculations.

\section{Symmetry Energy within Functional Theory}
\label{sec:HohenbergKohn}

We now turn to a discussion of the symmetry energy within the Hohenberg-Kohn functional theory~\cite{hohenberg-1964}.  We carried out such a discussion in I, but there we suppressed the Coulomb energy, which is not the case here.  The purpose of our discussion is fewfold.  Thus, we we are going to obtain functional expressions for the terms of interest within an energy formula.  We will explicitly demonstrate the coupling between the symmetry energy term and the Coulomb term within an energy formula, for the third component of isospin, and the lack of such a coupling for the transverse components of isospin.  Before, a coupling for the third component was discussed by one of us within a macroscopic model~\cite{Danielewicz:2003dd}.  We will further establish here a relation between the densities of neutrons and protons in a nucleus and the symmetry coefficient, in the presence of Coulomb effects.  That relation will facilitate an~extraction of the symmetry coefficient from results of Skyrme calculations of nuclear ground states, while suppressing shell effects.  Finally, our considerations will allow us to explore the fragility of volume-surface decomposition for the symmetry coefficient.

\subsection{Hohenberg-Kohn Functional}

Our starting point here will be a Hohenberg-Kohn (HK) energy functional~\cite{hohenberg-1964} of the local densities of baryon number and isospin components.  The energies in the functional are obtained for the state that minimizes net energy, while meeting the constraint of specific densities as a function of position.  We will assume that the functional is smoothed out, in particular across discrete values of baryon number and isospin, making it defined for continuous values of the latter and making it analytic in the densities, cf.~\cite{hohenberg-1964} and~I.  That functional will be partitioned into nuclear and Coulomb parts, according to the contributions to the hamiltonian.  Given the long-range nature of the Coulomb interaction, reducing sensitivity to finer details, we will assume the basic folding form in terms of proton density and, at this stage, we will not be concerned with the exchange term.  Furthermore, we~will partition the nuclear part into the functional for isospin-symmetric matter and the reminder:
\beq
\label{eq:HK}
\overline{E}(\rho, {\pmb \rho}) = E_\text{nuc} (\rho, {\pmb \rho}) + E_\text{Cou} (\rho, \rho_3) =
E_0 (\rho) + E_1 (\rho, {\pmb \rho}) + E_\text{Cou} (\rho, \rho_3) \, .
\eeq
Here the densities are
\beq
{\pmb \rho} ({\pmb r}) = 2 \langle \sum_{i=1}^A {\pmb t} \, \delta({\pmb r} - \pmb{r}_i)  \rangle \, ,
\eeq
and in this context
\beq
\rho_3 \equiv \rho_n - \rho_p \equiv \rho_{np} \, .
\eeq
One more symbol, $\rho_{\underline{1}}$, is utilized for the neutron-proton difference
in the context of the stability of Skyrme interactions in Appendix \ref{Appendix:SkyrmeStability}, for notational convenience there.
As was discussed, the Coulomb energy is in the form
\beq
\label{eq:ECoufun}
E_\text{Cou} (\rho, \rho_3) = \frac{1}{8} \frac{e^2}{4 \pi \epsilon_0} \int \text{d} {\pmb r}_1 \, \text{d} {\pmb r}_2 \, \big(\rho - \rho_3 \big)({\pmb r}_1) \, \frac{1}{| {\pmb r}_1 - {\pmb r}_2 |} \,  \big(\rho - \rho_3 \big)({\pmb r}_2) \, .
\eeq
Given the charge invariance of nuclear interactions, the symmetry term $E_1$ needs to be of the form
\beq
\label{eq:EaS}
E_1 (\rho , {\pmb \rho}) = \int \text{d} {\pmb r}_1 \, \text{d} {\pmb r}_2 \, {\pmb \rho} ({\pmb r}_1) \, {\mathcal S}( \rho, {\pmb r}_1, {\pmb r}_2) \,
{\pmb \rho} ({\pmb r}_2) + {\mathcal O}({\pmb \rho}^4) \, ,
\eeq
where ${\mathcal S}$ is a symmetric bilinear positive definite operator, cf.~I, generally nonlocal.  Given the short-range nature of nuclear interactions, for weakly nonuniform matter the operator~${\mathcal S}$ may be approximated in the local form
\beq
\label{eq:Sloc}
{\mathcal S}( \rho, {\pmb r}_1, {\pmb r}_2 ) \simeq \frac{S(\rho)}{\rho} \big( {\pmb r}_1 \big)  \, \delta ({\pmb r}_1 - {\pmb r}_2) \, ,
\eeq
where $S(\rho)$ is the symmetry energy in uniform matter.  The finding, in the SHF calculations of I, was that for ground-state nuclei such an approximation works at densities $\rho \gtrsim \rho_0/4$, where $\rho_0$ is the normal density.

In the following, we shall consider the minimal energy $\overline{E}$ of Eq.~\eqref{eq:HK}, under the constraints of a fixed nucleon number and vector isospin,
\beq
\int  \text{d} {\pmb r} \, \rho( {\pmb r} ) = A \, \hspace*{2em} \text{and}  \hspace*{2em}
\int  \text{d} {\pmb r} \, {\pmb \rho} ( {\pmb r} ) = 2 {\pmb T} \, .
\eeq
Minimization of the energy, using Lagrange multipliers to incorporate the constraints, yields coupled equations for the densities
\beq
\label{eq:muset}
\frac{\delta \overline{E}}{\delta \rho({\pmb r})} = \mu \, \hspace*{2em} \text{and} \hspace*{2em}
\frac{\delta \overline{E}}{\delta {\pmb \rho}({\pmb r})} = {\pmb \mu} \, .
\eeq
Solving for those densities should yield, on one hand, average characteristics of nuclear densities.  On the other hand, inserting those densities into \eqref{eq:HK} should yield energy as a~function of nucleon number and isospin $\overline{E} (A, {\pmb T})$.  With the latter, underpinnings of the symmetry energy in the \emph{energy formula} should emerge, as well as possible interplay of the symmetry term with other within the formula.

The Lagrange multipliers $\mu$ and ${\pmb \mu}$ represent chemical potentials associated with the constrained quantities.  Thus, e.g.\ if we take
\beq
\delta \rho ( {\pmb r} ) = \rho(A + {\text d} A, {\pmb T}, {\pmb r}) - \rho(A , {\pmb T}, {\pmb r}) \, ,
\eeq
where the r.h.s.\ densities belong to the solutions of the set \eqref{eq:muset}, and we multiply by that particular $\delta \rho$ both sides of the first equation of the set \eqref{eq:muset} and integrate over space, we arrive at
\beq
\mu = \frac{\partial \overline{E}}{\partial A} \, .
\eeq
Similarly we find
\beq
\label{eq:muV}
{\pmb \mu} = \frac{1}{2} \, \frac{\partial \overline{E}}{\partial {\pmb T}} \, .
\eeq

In discussions, we shall utilize net densities, indicated with a $0$ subscript, obtained when minimizing energy with the Coulomb energy put to zero.  Distinction relative to the normal density $\rho_0$ will be made by the presence of a spatial argument and/or by explicit naming of the density.  The energy of symmetric matter may be expanded in deviations $\Delta \rho$ of density from that minimizing the energy at fixed $A$ and ${\pmb T} = 0$, $\rho_0 (A, ${\pmb T} = 0$, {\pmb r})$:
\beq
E_0(\rho) = E_0(\rho_0) + \mu_0 \int \text{d}{\pmb r} \, \Delta \rho  ({\pmb r}) + \int \text{d} {\pmb r}_1 \, \text{d} {\pmb r}_2 \,
\Delta \rho ({\pmb r}_1) \, {\mathcal K} (\rho_0, {\pmb r}_1 , {\pmb r}_2) \, \Delta \rho ({\pmb r}_2) + {\mathcal O}\left((\Delta \rho)^3\right)   \, ,
\eeq
where $\Delta \rho = \rho - \rho_0$ and ${\mathcal K}$ is a symmetric bilinear positive-definite operator that is generally nonlocal.  For densities that satisfy the constraint of a fixed nucleon number, the linear term in the expansion vanishes.  Of particular interest is the case of a system approaching the limit of half-infinite nuclear matter, cf.~I, with $A \rightarrow \infty$.  In the regions of $\rho_0({\pmb r}) \simeq \rho_0$, for $\Delta \rho$ weakly varying with position, the operator ${\mathcal K}$ may be approximated in the local form
\beq
{\mathcal K} (\rho_0, {\pmb r}_1 , {\pmb r}_2) \simeq \frac{K}{18 \rho_0} \, \delta ({\pmb r}_1 - {\pmb r}_2) \, ,
\eeq
where $K$ is the incompressibility.

With only the symmetry energy dependent on densities of transverse isospin, ${\pmb T}_\perp = (T_1, T_2) $, the densities that minimize the energy and their contributions to that energy are relatively easy to find.  Thus,
on carrying the differentiation in Eq.~\eqref{eq:muV}, we find
\beq
\label{eq:muperp}
{\pmb \mu}_\perp = 2 \int \text{d} {\pmb r}_1 \,  {\mathcal S}( \rho, {\pmb r}, {\pmb r}_1) \,
{\pmb \rho}_\perp ({\pmb r}_1) \, .
\eeq
Upon applying an operator inverse to ${\mathcal S}$ to both sides of \eqref{eq:muperp}, we find
\beq
\label{eq:rhot}
{\pmb \rho}_\perp ({\pmb r}) = \frac{{\pmb \mu}_\perp}{2} \int \text{d} {\pmb r}_1 \,  {\mathcal S}^{-1}( \rho, {\pmb r}, {\pmb r}_1) \, .
\eeq
Integration of both sides of the above relation yields
\beq
\label{eq:Tpmu}
2 \, {\pmb T}_\perp = \frac{{\pmb \mu}_\perp}{2} \int \text{d} {\pmb r} \,\text{d} {\pmb r}_1 \,  {\mathcal S}^{-1}( \rho, {\pmb r}, {\pmb r}_1) = \frac{{\pmb \mu}_\perp}{2} \, \frac{A}{a_a} \, ,
\eeq
where we introduced a mass- and potentially charge-dependent symmetry coefficient given by
\beq
\label{eq:aadf}
\frac{A}{a_a} = \int \text{d} {\pmb r} \,\text{d} {\pmb r}_1 \,  {\mathcal S}^{-1}( \rho, {\pmb r}, {\pmb r}_1) \, ,
\eeq
see also~I.

Given the anticipation of a limited nonlocality range for the operator ${\mathcal S}^{-1}$, the~factoring out of $A$ in \eqref{eq:aadf} should act to limit the $A$-dependence for $a_a$.  The remaining dependence on $A$ may be both due to nonlocality of ${\mathcal S}^{-1}$ and due to nonlinear dependence on a~local $\rho$ changing within the surface region.  The dependence on~$Z$ may result from a dependence of $\rho$ on $Z$, combined with a nonlinearity in ${\mathcal S}^{-1}$.  From \eqref{eq:Tpmu}, we get for the transverse components of the isospin chemical potential
\beq
\label{eq:muperpT}
{\pmb \mu}_\perp = \frac{4 a_a \, {\pmb T}_\perp}{A} \, .
\eeq
Upon integrating both sides of the above equation over transverse components of isospin, we get a contribution of those components to the energy
\beq
\label{eq:Eperp}
E_{a \perp} = 2 \int_0^{{\pmb T}_\perp} \text{d} {\pmb T}_\perp \, {\pmb \mu}_\perp = \frac{4 a_a \, {\pmb T}_\perp^2}{A} \, .
\eeq

While, with the above, we have principally already arrived at a functional justification
for the energy function behind our IAS analysis,
continuation of the functional analysis may facilitate getting insights into physical effects behind the deduced symmetry coefficients.  Of~general interest are also any differences in the coupling of the Coulomb term in the energy functional to the
contributions of different components of isospin to the symmetry energy.  Beyond insights,
continuing the analysis will result in practical procedures employed later in the paper.

In Eq.~\eqref{eq:rhot}, we see that the inverse operator ${\mathcal S}^{-1}$, integrated over one of its arguments, plays the role of a profile function for ${\pmb \rho}_\perp$.  In fact, from \eqref{eq:rhot} and \eqref{eq:Tpmu}, we find
\beq
{\pmb \rho}_\perp ({\pmb r}) = 2 \, {\pmb T}_\perp \, \frac{\int \text{d} {\pmb r}_1 \,  {\mathcal S}^{-1}( \rho, {\pmb r}, {\pmb r}_1)}{\int \text{d} {\pmb r} \,\text{d} {\pmb r}_1 \,  {\mathcal S}^{-1}( \rho, {\pmb r}, {\pmb r}_1)} \, .
\eeq
From \eqref{eq:Sloc} it follows that in weakly nonuniform matter
\beq
\label{eq:Siloc}
{\mathcal S}^{-1}( \rho, {\pmb r}, {\pmb r}_1 ) \simeq \frac{\rho }{S(\rho)} \big( {\pmb r} \big)  \, \delta ({\pmb r} - {\pmb r}_1) \, .
\eeq
In this context we introduce an isovector profile function, also termed asymmetric density (see~I), normalized in such a manner that its value approaches normal density when the isoscalar density approaches the normal density:
\beq
\label{eq:rhoaloc}
\rho_a ({\pmb r}) = a_a^V \int \text{d} {\pmb r}_1 \,  {\mathcal S}^{-1}( \rho, {\pmb r}, {\pmb r}_1) \, .
\eeq
Here, $a_a^V$ is the symmetry energy at normal density in uniform matter, $a_a^V = S(\rho_0)$.
In~weakly nonuniform matter then
\beq
\label{eq:rhoax}
\rho_a ({\pmb r}) \approx \frac{a_a^V \, \rho}{S(\rho)} \, ({\pmb r}) \, .
\eeq
The symmetry coefficient of Eq.~\eqref{eq:aadf} can be expressed in terms of $\rho_a$ as
\beq
\label{eq:Aaa}
\frac{A}{a_a} = \frac{1}{a_a^V} \int \text{d} {\pmb r} \, \rho_a ({\pmb r}) \, .
\eeq
The density of transverse isospin is further given by, in terms of $\rho_a$,
\beq
{\pmb \rho}_\perp ({\pmb r}) = 2 \, {\pmb T}_\perp \, \frac{\rho_a ({\pmb r})}{\int \text{d} {\pmb r} \, \rho_a ({\pmb r})} =  \frac{2 \, {\pmb T}_\perp}{A} \, \frac{a_a}{a_a^V} \, \rho_a ({\pmb r}) = \frac{{\pmb \mu}_\perp}{2a_a^V} \, \rho_a ({\pmb r}) \, .
\eeq

The nuclear energy depending quadratically on isospin in Eq.~\eqref{eq:Eperp} is analogous to the energy of a charged capacitor in electrostatics, depending quadratically on charge.  With this, the inverse of the factor multiplying the square of isospin in the nuclear energy, i.e.\ $A/a_a$, may be interpreted (up to a factor of~2) in terms of a~capacitance for asymmetry.  The integral \eqref{eq:Aaa} for the capacitance may broken up into volume and surface contributions
\beq
\begin{split}
\label{eq:CVS}
\frac{A}{a_a}  & = \frac{1}{a_a^V} \int \text{d} {\pmb r} \, \rho ({\pmb r}) + \frac{1}{a_a^V} \int \text{d} {\pmb r} \, (\rho_a - \rho) ({\pmb r})  \\
&  = \frac{A}{a_a^V} + \frac{1}{a_a^V} \int \text{d} {\pmb r} \, (\rho_a - \rho) ({\pmb r}) \approx  \frac{A}{a_a^V} + \frac{A^{2/3}}{a_a^S}   \, ,
\end{split}
\eeq
where
\beq
\label{eq:aaS}
\frac{1}{a_a^S} = \frac{4 \pi r_0^2}{a_a^V} \int \text{d} z \, (\rho_a^{\infty/2} - \rho^{\infty/2}) (z) \, .
\eeq
The densities on the r.h.s.\ of Eq.~\eqref{eq:aaS}, with superscript $\infty/2$, pertain to half-infinite matter.  The r.h.s.\ of Eq.~\eqref{eq:CVS}, obtained already in~I, is the basis of the volume-surface fits to the IAS results from the preceding section.  Within the volume-surface separation, the density dependence of the symmetry energy is contained in the surface coefficient $a_a^S$.  The faster the fall of symmetry energy with density in \eqref{eq:rhoax}, at subnormal densities in the surface, the more enhanced is the asymmetric density compared to isoscalar density.  The more extended is $\rho_a$ beyond the isoscalar surface, the greater is the contribution of the surface to the capacitance for asymmetry in~\eqref{eq:CVS} and \eqref{eq:aaS}.  That larger surface contribution should produce a stronger mass-dependence of the symmetry coefficient, in particular exhibited in a lower surface symmetry coefficient $a_a^S$ in the volume-surface breakdown in~\eqref{eq:CVS}.

In I we carried out SHF calculations of half-infinite nuclear-matter, arriving at the $\rho$- and $\rho_a$-profiles for that matter, following different Skyrme parameterizations.  We calculated the surface coefficients both for the energy of symmetric matter and symmetry energy (Eq.~\eqref{eq:aaS}) for the variety of those parameterizations.  Consistently with the qualitative considerations above, we observed a strong correlation between the values of $a_a^S$ and the slope parameter~$L$ of the symmetry energy of uniform matter with respect to density at $\rho_0$, defined in
\beq
\label{eq:SaL}
S(\rho) = a_a^V + \frac{L}{3 \rho_0} \big(\rho - \rho_0 \big) + \ldots \, ,
\eeq
see also Refs.~\cite{Danielewicz:2003dd,Steiner:2004fi}.

In the absence of Coulomb interactions, the results for the third component of isospin would have just paralleled those for the transverse components in Eqs.~\eqref{eq:Eperp} and \eqref{eq:aadf}.  Indeed, to zeroth-order in $e^2$, the density of the third component is
\beq
\label{eq:rho30}
\rho_3 ({\pmb r}) \equiv
(\rho_n - \rho_p)({\pmb r}) \approx \rho_{30} ({\pmb r}) = \frac{N - Z}{A} \, \frac{a_a^V}{a_a} \, \rho_{a0} ({\pmb r}) \, .
\eeq
With the Coulomb interactions there in the energy functional, though, the results for the third component cannot be generally put into such a compact and simultaneously useful form as for the transverse components.  Nonetheless, useful results may be arrived by perturbation.  The latter is principally already employed in the basic textbook formula used for the ground state energies, that was referred to in subsection \ref{subs:General}.  Thus, within the nuclear energy the largest contribution comes from the volume term of symmetric matter.  The other terms are added successfully independently from each other and in such a manner as if the associated physics had no impact on the interior represented by the volume term.  E.g.\ the Coulomb term is successfully implemented treating the nuclear interior as rigid, with no redistribution of protons on account of the Coulomb interactions.   At some level, though, the terms in the energy functional need to interplay as far as their contributions to the energy formula are concerned.  However, the apparent success of the most naive implementation suggests that most of the physics of the interplay could be grasped by considering that interplay at its lowest level, at least for nuclei that dominate the energy formula fit.  In the end, in carrying out the volume-surface breakdown for the symmetry coefficient, the effect of the surface appears incorporated perturbatively into the symmetry term with the breakdown expected to properly hold for heavier nuclei.

In the case of the Coulomb term interplaying with other in the energy formula, the~perturbation terms can be conveniently kept track of in terms of powers of $e^2$.  To the first order in~$e^2$, the Coulomb interactions polarize the $n$-$p$ density difference.  In analyzing the results of SHF calculations later in the paper, we need to correct for that polarization, in order to access $\rho_a$ and $a_a$ of importance for the transverse components of isospin.  To first order in~$e^2$, further, the Coulomb interactions stretch the nucleus, i.e.~affect the net density~$\rho$.  That effect we want to keep in the results of calculations, since it impacts $a_a$ for the transverse components and it is necessarily present in the data we want to compare to.  Finally, to the first order in $e^2$, coupling is there in the energy, between the Coulomb and symmetry terms.  That latter somewhat surprising result is associated with the fact that $\rho$ and $\rho_a$ have different dependencies on position.  An~unexpected aspect of that coupling is that its leading power in asymmetry is $(N - Z)$ rather than $(N-Z)^2$.  We now progress to details.

On carrying out the differentiation of the energy with respect to density of third isospin component in \eqref{eq:muset}, we find
\beq
\label{eq:mu3}
\mu_3 = 2 \int \text{d} {\pmb r}_1 \,  {\mathcal S}( \rho, {\pmb r}, {\pmb r}_1) \,
{ \rho}_3 ({\pmb r}_1) - \frac{1}{2} \, \Phi ({\pmb r}) \, ,
\eeq
where $\Phi$ is the Coulomb potential,
\beq
 \Phi ({\pmb r}) = \frac{1}{2} \, \frac{e^2}{4 \pi \epsilon_0} \int \text{d} {\pmb r}_1 \, \frac{1}{| {\pmb r} - {\pmb r}_1 |} \,  \big(\rho - \rho_3 \big)({\pmb r}_1) \, .
\eeq
Upon applying the inverse operator ${\mathcal S}^{-1}$ to both sides of Eq.~\eqref{eq:mu3}, we find
\beq
\label{eq:rho3}
\begin{split}
{ \rho}_3 ({\pmb r}) & = \frac{{\mu}_3}{2} \int \text{d} {\pmb r}_1 \,  {\mathcal S}^{-1}( \rho, {\pmb r}, {\pmb r}_1)
+ \frac{1}{4} \int \text{d} {\pmb r}_1 \,  {\mathcal S}^{-1}( \rho, {\pmb r}, {\pmb r}_1) \, \Phi ({\pmb r}_1) \\
& = \frac{\mu_3}{2 a_a^V} \, \rho_a ( {\pmb r} ) + \frac{1}{4} \int \text{d} {\pmb r}_1 \,  {\mathcal S}^{-1}( \rho, {\pmb r}, {\pmb r}_1) \, \Phi ({\pmb r}_1)
 \, .
\end{split}
\eeq
Integration of both sides over space yields
\beq
\label{eq:NZ}
N - Z = \frac{\mu_3}{2 a_a} \, A + \frac{1}{4 a_a^V}  \int \text{d} {\pmb r} \, \rho_a ( {\pmb r} ) \, \Phi ({\pmb r}) \, ,
\eeq
which replaces \eqref{eq:muperpT}.  An important aspect of ${\mathcal S}$ and its inverse is that the nonlocality range of those operators is expected to be short compared to the scale of variation of the potential~$\Phi$ due to long-range Coulomb interactions.  With this, one can approximate Eq.~\eqref{eq:rho3} with
\beq
\label{eq:rho3loc}
{ \rho}_3 ({\pmb r}) \simeq \frac{1}{2} \Big[ \mu_3 + \frac{1}{2} \, \Phi({\pmb r}) \Big] \int \text{d} {\pmb r} \,
  {\mathcal S}^{-1}( \rho, {\pmb r}, {\pmb r}_1) = \frac{1}{2 a_a^V} \Big[ \mu_3 + \frac{1}{2} \, \Phi({\pmb r}) \Big] \, \rho_a ( {\pmb r} )  \, .
\eeq
The importance of Eq.~\eqref{eq:rho3loc} is in relating, in a local manner, the neutron-proton density difference polarized by the Coulomb interactions, to the density profile for transverse isospin, unaffected by that polarization.  The validity of Eq.~\eqref{eq:rho3loc} is going to be enhanced when the two sides get integrated over some region, with the integration erasing any residual effects of the nonlocality of ${\mathcal S}^{-1}$.  In fact, upon integration over whole space, Eq.~\eqref{eq:rho3loc} produces the same result \eqref{eq:NZ} as the original equation \eqref{eq:rho3} with no range approximation.  We will use Eq.~\eqref{eq:rho3loc}, partially integrated over sides, in combination with \eqref{eq:rhoax} for nuclear interior, to extract $a_a$ from the outcomes of SHF calculations.

The mentioned interplay of the symmetry and Coulomb contributions to the energy functional, resulting in a term linear in $(N-Z)$ within an energy formula, beyond the terms present in \eqref{eq:E1} and \eqref{eq:ECou}, arises from the fact that the profile of proton density is expected to evolve with $(N-Z)$ at a constant $A$, even the under suppression of the Coulomb interactions, on account of the density dependence of the symmetry energy.  Specifically, from \eqref{eq:rho30}, the proton density from the HK functional \eqref{eq:HK}, minimized without the Coulomb interactions, is
\beq
\label{eq:rhop0}
\rho_{p0} ({\pmb r}) = \frac{1}{2} \, \Big[ \rho_0 - \frac{N-Z}{A} \frac{a_a^V}{a_a} \, \rho_{a0}    \Big]\big( {\pmb r}   \big) \, .
\eeq
Upon inserting \eqref{eq:rhop0} into the Coulomb energy $E_{Cou}$ \eqref{eq:ECoufun}, we can isolate a contribution not accounted for in \eqref{eq:ECou}, in the form
\beq
\label{eq:DECou}
\begin{split}
\Delta E_{Cou}^{iso} & = \frac{1}{2} \, \frac{e^2}{4 \pi \epsilon_0} \int  \text{d} {\pmb r} \, \text{d} {\pmb r}_1  \,
 \frac{1}{| {\pmb r} - {\pmb r}_1 |} \,  \bigg[ \rho_{p0} ( {\pmb r} ) \, \rho_{p0} ( {\pmb r}_1 )  - \Big(\frac{Z}{A}\Big)^2 \, \rho_{0} ( {\pmb r} ) \, \rho_{0} ( {\pmb r}_1 ) \bigg]\\
 & =  \frac{N-Z}{2A} \, \frac{e^2}{4 \pi \epsilon_0} \int  \text{d} {\pmb r} \, \text{d} {\pmb r}_1  \,
 \frac{1}{| {\pmb r} - {\pmb r}_1 |} \,  \bigg[ \rho_{p0} ( {\pmb r} ) +  \frac{Z}{A} \, \rho_{0} ( {\pmb r} ) \bigg] \, \bigg[ \rho_{0} ( {\pmb r}_1 )  - \frac{a_a^V}{a_a} \, \rho_{a0} ( {\pmb r}_1 ) \bigg] \\
& \approx   \frac{(N-Z) \, Z}{A^2} \int \text{d} {\pmb r} \, \text{d} {\pmb r}_1  \,
 \frac{1}{| {\pmb r} - {\pmb r}_1 |} \,   \rho_{0} ( {\pmb r} )  \, \bigg[ \rho_{0} ( {\pmb r}_1 )  - \frac{a_a^V}{a_a} \, \rho_{a0} ( {\pmb r}_1 ) \bigg] \, .
\end{split}
\eeq
The approximation in the last step is in dropping a term of second order in the difference between the densities $\rho_{0}$ and $(a_a^V/a_a) \, \rho_{a0}$, of dropping importance when the size of the nuclear system increases and Coulomb effects become important.  In the latter case the dropped contribution is by a factor of the order of $A^{-1/3}$ lower than that retained.  Upon further manipulations, valid consistently to the order of the dropped contribution, the last result in~\eqref{eq:DECou} can be incorporated into an energy formula so as to modify the charge radius in the denominator of Coulomb energy in Eq.~\eqref{eq:ECou}, Eqs.~(20) and~(21) of Ref.~\cite{Danielewicz:2003dd}.  In~a~similar manner, a contribution from the surface diffuseness in the Coulomb energy within the functional may be isolated:
\beq
\label{eq:DECod}
\begin{split}
\Delta E_{Cou}^{dif} =  & \frac{1}{2} \, \frac{e^2}{4 \pi \epsilon_0} \, \Big(\frac{Z}{A} \Big)^2 \int  \text{d} {\pmb r} \, \text{d} {\pmb r}_1  \,
 \frac{1}{| {\pmb r} - {\pmb r}_1 |} \, \\
  & \times \bigg[  \rho_{0} ( {\pmb r} ) \, \rho_{0} ( {\pmb r}_1 ) - \rho_0^2 \, \theta(r_0 \, A^{1/3} - r  ) \, \theta(r_0 \, A^{1/3} - r_1) \bigg] \, ,
 \end{split}
\eeq
where $\theta$ is the step function.
Following similar manipulations as with \eqref{eq:DECou}, dropping terms nominally smaller by $A^{-1/3}$ than the leading term retained, the contribution from \eqref{eq:DECod} can be incorporated into the effective radius for the Coulomb energy term in the energy formula, in parallel to the contribution from~\eqref{eq:DECou}, cf.~Eqs.~(20) and~(21) of Ref.~\cite{Danielewicz:2003dd}.

When an energy formula is fitted to the measured nuclear ground-state energies, it is found that changes to the contributions from \eqref{eq:DECou} and \eqref{eq:DECod} seriously affect conclusions on the symmetry energy, even when just the 2-parameter volume-surface fit is undertaken for the symmetry coefficient, and no attempt is made to determine the coefficient on a nucleus-by-nucleus basis.  Given that the valley of stability gives rise to a~correlation between mass number and typical isospin values, one can anticipate a similar sensitivity of conclusions on the symmetry energy, to details in the isoscalar terms in the energy formula.  In this context, indeed attractive becomes the ability to analyze the symmetry energy on a nucleus-by-nucleus or mass-by-mass basis, using IAS, without the need to refer to other terms in an energy formula than the symmetry term.

\subsection{Vulnerability of the Surface-Volume Decomposition for Symmetry Energy}

The ability to assess nuclear symmetry coefficients $a_a$ on a mass-by-mass basis, using IAS, followed by the volume surface decomposition for the deduced capacitance for asymmetry, appears to open up a possibility of accessing aspects of the symmetry energy for infinite matter, without a direct reference to a model or to any model-dependent nuclear characteristics.  Unfortunately, qualitative considerations and practical experimentation with models, for which bulk aspects of symmetry energy are known, demonstrate that such model-free extraction of symmetry-energy parameters cannot be relied upon, particularly as far as the surface symmetry parameter is concerned.  This contrasts the situation with parameters of symmetric matter, as we shall illustrate.  In part at least, this can be attributed to the fact that the energy of symmetric matter minimizes at normal density while the symmetry energy changes linearly with density.

As far as a qualitative consideration is concerned, let us consider schematically the energy of a symmetric nucleus with density largely uniform across the interior and then rapidly changing over the surface, so that the associated energy contribution can be treated in terms of a surface coefficient, in the spirit of the simple mass formula, first without Coulomb interactions:
\beq
\label{eq:E0AR}
E_0(A,R) = - a_V \, A + \frac{A K}{2} \, \Big(\frac{R - R_0}{R_0}\Big)^2 + \frac{a_S}{r_0^2} \, R^2 \, .
\eeq
Here, the energy per nucleon of uniform matter is expanded around normal density corresponding to the radius $R_0 = r_0 \, A^{1/3}$.  The surface energy coefficient might be expanded as well as a function of internal density (or $R$), but within the accuracy we will be working, the~dependence of $a_S$ on~$R$ will be of no relevance.  If we minimize $E_0$ of Eq.~\eqref{eq:E0AR} with respect to~$R$, we find to leading order, in the limit of $A \rightarrow \infty$,
\beq
\label{DR}
\Delta R = R - R_0 \simeq - \frac{2 \, a_S}{r_0^2 \, A K} \, R_0^3 = - \frac{2 \, a_S}{K} \, r_0  \approx - 0.18 \, \text{fm}  \, ,
\eeq
where the r.h.s.\ value is for typical values of $a_S$, $K$ and $r_0$.
The surface tension shrinks the radius by an approximately constant amount.  The change $\Delta \rho = \rho - \rho_0$ in density $\rho$ of uniform interior on account of \eqref{DR} follows from
\beq
\frac{\Delta \rho}{\rho_0} \simeq - 3 \, \frac{\Delta R}{R_0} \, ,
\eeq
yielding
\beq
\Delta \rho \simeq \frac{6 \, a_S}{K A^{1/3}} \, \rho_0 \, .
\eeq

The important aspect of the energy \eqref{eq:E0AR} is that, to the order of $A^{2/3}$, it is not affected by the deviation of the radius from $R_0$ in \eqref{DR} or of the density from normal.  Indeed, on account of the quadratic dependence of the energy of uniform matter on radius or density, we find from \eqref{eq:E0AR}
\beq
\label{eq:E0A}
E_0(A) \simeq - a_V \, A + a_S \, A^{2/3} \, ,
\eeq
upon inserting \eqref{DR} there.  However, there is no similar invariance there for the capacitance for asymmetry.  From Eqs.~\eqref{eq:Aaa}, \eqref{eq:CVS} and \eqref{eq:rhoax}, we find
\beq
\label{eq:Aaas}
\frac{A}{a_a} \simeq \frac{A}{a_a^V - L \, \Delta R /R_0} + \frac{R^2}{r_0^2 \, a_a^S}
\simeq \frac{A}{a_a^V} + \frac{A^{2/3}}{a_a^{S\prime}} \, ,
\eeq
where
\beq
\label{eq:aaSp}
\frac{1}{a_a^{S\prime}} = \frac{1}{a_a^S} - \frac{2 a_S \, L}{(a_a^V)^2 K} \, .
\eeq
We see that the compression effect due to surface tension changes the surface symmetry coefficient that would be deduced from the mass dependence of capacitance for asymmetry, or of symmetry energy, as compared to the expectation for normal interior density.  The~effect on the symmetry-energy characteristic~\eqref{eq:Aaas}, arising in surface part, can be contrasted with that on energy of symmetric matter where it is relegated to the curvature part of the energy~\eqref{eq:E0A}.  The difference stems from the difference, quadratic vs.\ linear, dependence of these two energies on density around normal.

Equation \eqref{eq:aaSp} suggests an increase in the surface symmetry coefficient deduced from mass dependence, for $L>0$, compared to the coefficient expected for half-infinite matter.  Physically, $L$ and $a_a^S$ are expected to be correlated.  For typical combinations of the values for Skyrme interactions, cf.\ I, $a_a^{S\prime}$ from \eqref{eq:aaSp} may be larger than $a_a^S$ by $\sim 15\%$.  The consideration so-far, however, ignored Coulomb interactions.  They would act to expand the nucleus, against the effects of surface tension.  With the interactions, the energy becomes
\beq
\label{eq:EAZR}
E(A,Z,R) = - a_V \, A + \frac{A K}{2} \, \Big(\frac{R - R_0}{R_0}\Big)^2 + \frac{a_S}{r_0^2} \, R^2 + a_C \, r_0 \, \frac{Z^2}{R} \, .
\eeq
Minimizing $E$ with respect to $R$, we now arrive at
\beq
\label{eq:DRc}
\Delta R \simeq - \frac{r_0}{K} \Big( 2 a_S - a_C \, \frac{Z^2}{A} \Big) = - r_0 \, \frac{a_C}{K} \Big[ \Big( \frac{Z^2}{A} \Big)_c  - \frac{Z^2}{A} \Big] \, .
\eeq
The ratio $x = Z^2/A$ in \eqref{eq:DRc} is the fissility parameter \cite{nix_studies_1965} and its critical value is
\beq
x_c = \Big( \frac{Z^2}{A} \Big)_c = \frac{2 a_S}{a_C} \approx 50 \, .
\eeq
According to \eqref{eq:DRc}, evolution of the fissility parameter towards critical erases the effects of tension on density and consequentially on the apparent surface symmetry coefficient.
Note that if we included a symmetry energy term in \eqref{eq:EAZR}, we would have arrived at a contribution to~$\Delta R$ being of second order in asymmetry.  If we next incorporated that contribution in the symmetry coefficient we would be actually turning to a quartic contribution in asymmetry to the net energy, which we want to refrain from at present.

The result \eqref{eq:DRc} suggests the representation of the symmetry coefficient in the form
\beq
\label{eq:aaL}
\frac{1}{a_a} = \frac{1}{a_a^V} + A^{-1/3} \bigg[ \frac{1}{a_a^S} - \frac{2 a_S \, L}{(a_a^V)^2 K} \Big(1 - \frac{x}{x_c}  \Big)     \bigg]
\, .
\eeq
The problem, with fitting a formula such as \eqref{eq:aaL} to the coefficients extracted from data, is that the variation generated by the data, combined with errors, allows in practice for fitting a~2- but not a 3-parameter formula and~$L$ is not known a priori.   To cope with that issue, one might exploit the correlation between~$L$ and $a_a^S$ from SHF calculations of half-infinite nuclear matter of~I.  On the other hand, one might carry out outright spherical SHF calculations~\cite{Reinhard:1991,reinhard:014309}, which we next pursue, using Skyrme interactions with different forms of symmetry energy and compare the calculated symmetry coefficients with those deduced from data.  With the latter one can additionally test the volume-surface decompositions both for the energy of symmetric matter and for the symmetry energy, assessing the meaning of fits to data based on such decompositions.  The tests, to be discussed, show larger deviations between the fitted and actual parameter values $a_a^V$ and $a_a^S$, than suggested by the analysis above.  Much better agreement than for $a_a^V$ and $a_a^S$ is found for $a_V$ and~$a_S$ characterizing the energy of a symmetric system.    The latter finding is suggested, indeed, by the analysis above.  Before we can discuss the SHF results, though, we need to address the nucleus-by-nucleus extraction of the symmetry-energy coefficients from the SHF calculations.

\section{Spherical Skyrme-Hartree-Fock Calculations}
\label{sec:SHF}

In trying to connect the symmetry energy extracted on a mass-by-mass basis from data, to the symmetry energy for uniform matter, one obvious choice is to use a theory that can predict both the properties of individual nuclei and of the uniform matter, which has enough flexibility to explore different physics scenarios.  The SHF approach \cite{sky56,PhysRevC.5.626} is the most straightforward and common of the possibilities.  We follow that approach in this Section and exploit in our numerical calculations the codes assuming spherical symmetry by P.-G.~Reinhard \cite{Reinhard:1991,reinhard:014309}.  The form of the Skyrme energy functional is provided in Appendix~\ref{Appendix:SkyrmeStability}.  In I, we obtained surface energy coefficients for most of the Skyrme parameterizations in the literature; for selected few parameterizations the coefficients were also obtained by other authors~\cite{Kohler:1976,PhysRevC.24.303}.  One issue that we need to face in the context of SHF calculations is of the extraction of symmetry coefficients from the results of those calculations.  Another is that of the choice of the Skyrme parameterizations for confrontations with data.  As discussed in Appendix~\ref{Appendix:SkyrmeStability}, reliance on zero-range density-dependent forces in the parametrization leads to potential instabilities for the predictions in the short-wavelength limit, in addition to those familiar from the long-wavelength limit.  Development of associated uncontrolled oscillations for the calculated systems may skew the conclusions on symmetry energy drawn from the data.  We start out with the extraction of the coefficients from the SHF calculations and next discuss the SHF parametrizations and results that represent them, on their own, before confronting the results for symmetry coefficients with those extracted from data.

\subsection{Symmetry Coefficients from SHF Calculations}
\label{subsec:aaSHF}

In trying to deduce the symmetry coefficients for individual nuclei from the SHF calculations, one needs to recognize that the Hartree-Fock (HF) approach violates isospin symmetry.  In consequence the same approach as for the data cannot be employed.  If one wanted to switch off the Coulomb interactions in the calculations, that impair extraction of symmetry energy from the ground states within an isobaric chain, one would be affecting the very coefficients one would want to extract and compare to data.  That Coulomb effect on coefficients, necessarily there in the data, will be, in fact, illustrated later in this Section.  An~additional issue with the SHF calculations is that of shell effects that are stronger on the average within the HF model than in Nature and that differ in details, depending on the Skyrme parametrization, from Nature.  Principally, dedicated shell corrections should be developed for each of the Skyrme parametrizations, but that is not feasible with that level of scrutiny as for data.

In the situation of an energy functional smoothed out over shell effects,
making the functional continuous in nucleon number and isospin, different relations would emerge between a variety of quantities used in describing a nuclear system, as was discussed in the preceding Section.  Because of such relations, specific results might be arrived at in different ways, e.g.~the symmetry coefficient for a~system might be obtained from energy differences \emph{or} from nucleonic densities.  Depending on the method, however, the results derived in one or another way might be less or more  susceptible to shell corrections.  Thus, if the derived results were smooth when obtained without an application of shell corrections, they should not change much after an application of these corrections.  Given that we cannot afford to derive the shell corrections for every Skyrme interactions, we should look for a way of deriving the symmetry coefficients for individual nuclei, such that the coefficients come out smooth as a function of charge and mass numbers and are equivalent to the coefficients obtained from excitation energies to IAS, in the bulk limit.  Given the abundance of information that emerges from SHF calculations, finding such a method is plausible.

A hint at what the method of arriving at the symmetry coefficients could be, successful from the perspective above, comes from the fact that the asymmetry skins calculated in SHF are rather smooth functions of nucleon numbers \cite{dobaczewski_neutron_1996,sarriguren_nuclear_2007}.  In this context, in the continuous limit of the energy functional, we have shown in Section \ref{sec:HohenbergKohn} that the asymmetry coefficient could be obtained in terms of an integral over asymmetric density, cf.~Eq.~\eqref{eq:Aaa}.  The asymmetric density on its own might be obtained in terms of the ratio of the density of transverse isospin to the transverse component of isospin chemical potential and, for switched off Coulomb interactions, in terms of the ratio of neutron-proton density difference to the third component of the chemical potential, cf.~Eq.~\eqref{eq:rho3} and~I.  The problem with the chemical potential is that it is likely highly sensitive to shell effects, when obtained from changes in energy.  Fortunately, the issue of chemical potential can be circumvented, with the chemical potential divided out, when recognizing that the net density changes slowly inside a nucleus, allowing to use local relations such as for uniform matter, particularly under an~integration that smoothes out effects of any potential density oscillations.  In~essence, using \eqref{eq:rhoax}, one can calculate nuclear capacitance for the interior at $r < r_c$, where $r_c$ is some cut-off radius, and one can obtain the full capacitance by exploiting the fact that isospin distributes itself in proportion to capacitance.  Thus, on one hand we have
\beq
\label{eq:arcS}
\frac{a_a}{a_a^V \, A} \int_0^{r_c} \text{d} {\pmb r} \, \rho_a ({\pmb r}) \simeq
\frac{a_a}{A} \int_0^{r_c} \text{d} {\pmb r} \, \frac{\rho}{S(\rho)} \big({\pmb r}\big) \, ,
\eeq
and, on the other,
\beq
\label{eq:arcT}
\frac{a_a}{a_a^V \, A} \int_0^{r_c} \text{d} {\pmb r} \, \rho_a ({\pmb r}) =
\frac{ \int_0^{r_c} \text{d} {\pmb r} \, \rho_a ({\pmb r})}{ \int \text{d} {\pmb r} \, \rho_a ({\pmb r})}
= \frac{ \int_0^{r_c} \text{d} {\pmb r} \, \frac{2 a_a^V}{\mu_i} \, \rho_i ({\pmb r})}{ \int \text{d} {\pmb r} \,  \frac{2 a_a^V}{\mu_i} \, \rho_i ({\pmb r})} = \frac{1}{2 T_i} \int_0^{r_c} \text{d} {\pmb r} \, \rho_i ({\pmb r}) \, .
\eeq
Upon equating the r.h.s.\ of \eqref{eq:arcS} and \eqref{eq:arcT}, the coefficient $a_a$ may be found.  To get insight, when concentrating on the third component of isospin, for switched off Coulomb interactions, the combination of equations above yields for the capacitance for asymmetry:
\beq
\label{eq:Aaac}
\frac{A}{a_a} \simeq \frac{N-Z}{\int_0^{r_c} \text{d} {\pmb r} \big[ \rho_n - \rho_p \big]({\pmb r})} \, \int_0^{r_c} \text{d} {\pmb r} \,  \frac{\rho}{S(\rho)} \big({\pmb r}\big) \, .
\eeq
Equation \eqref{eq:Aaac} expresses the net capacitance of the system as the capacitance of the interior, calculated using asymmetric density for nuclear matter, scaled with the ratio of net asymmetry to the asymmetry for the interior.

Obviously, the standard SHF calculations are done for vanishing transverse isospin.  Regarding the third isospin component, the Coulomb potential interferes in the relation between isospin density and chemical potential and switching off of the Coulomb interactions is undesirable, given the Coulomb effect on the symmetry coefficient, that one may want to capture.  Fortunately, given the slow variation, particularly in nuclear interior, of the Coulomb potential with position, the effect of the Coulomb potential may be reduced to a~relatively trivial one, of the renormalization of the local chemical potential in Eq.~\eqref{eq:rho3loc}.  In addition, in the continuous limit of the energy functional, the chemical potential may be obtained from density, in particular using the interior, rather than changes in the energy.  By integrating both sides of Eq.~\eqref{eq:rho3loc} over the interior at $r < r_c$, we find
\beq
\label{eq:mu3C}
\mu_3 \int_0^{r_c} \text{d}{\pmb r} \, \frac{\rho}{S(\rho)} ({\pmb r}) \simeq 2  \int_0^{r_c} \text{d}{\pmb r} \,
\big[ \rho_n - \rho_p    \big] ({\pmb r}) - \frac{1}{2} \int_0^{r_c} \text{d}{\pmb r} \, \Phi({\pmb r})
\, \frac{\rho}{S(\rho)} ({\pmb r}) \, .
\eeq
In terms of the chemical potential $\mu_3$, the capacitance for asymmetry next follows from~\eqref{eq:Aaa} and \eqref{eq:rho3loc} as
\beq
\label{eq:capC}
\frac{A}{a_a} \simeq  \int_0^{r_c} \text{d}{\pmb r} \, \frac{\rho}{S(\rho)} ({\pmb r}) + 2 \int_{r_c}^\infty \text{d}{\pmb r} \, \frac{\rho_n - \rho_p}{\mu_3 + \frac{1}{2} \Phi} \big( {\pmb r}  \big) \, .
\eeq

The following should be noted regarding Eqs.~\eqref{eq:mu3C} and \eqref{eq:capC}.  If the Coulomb potential were shifted by some constant in both equations, the capacitance would not change.  It is actually only the variation of the potential across the nuclear volume, interplaying with the variation in the neutron-proton density difference, that matters in the determination of the capacitance.  If the potential is put to zero, or to another constant, the capacitance from~\eqref{eq:mu3C} and \eqref{eq:capC} reduces to that from Eq.~\eqref{eq:Aaac}.  In practical calculations of the capacitance, we tend to take nuclei for which the chemical potential $\mu_3$ is large compared to the variation of~$\Phi$.

\begin{figure}
\centerline{\includegraphics[width=.75\linewidth]{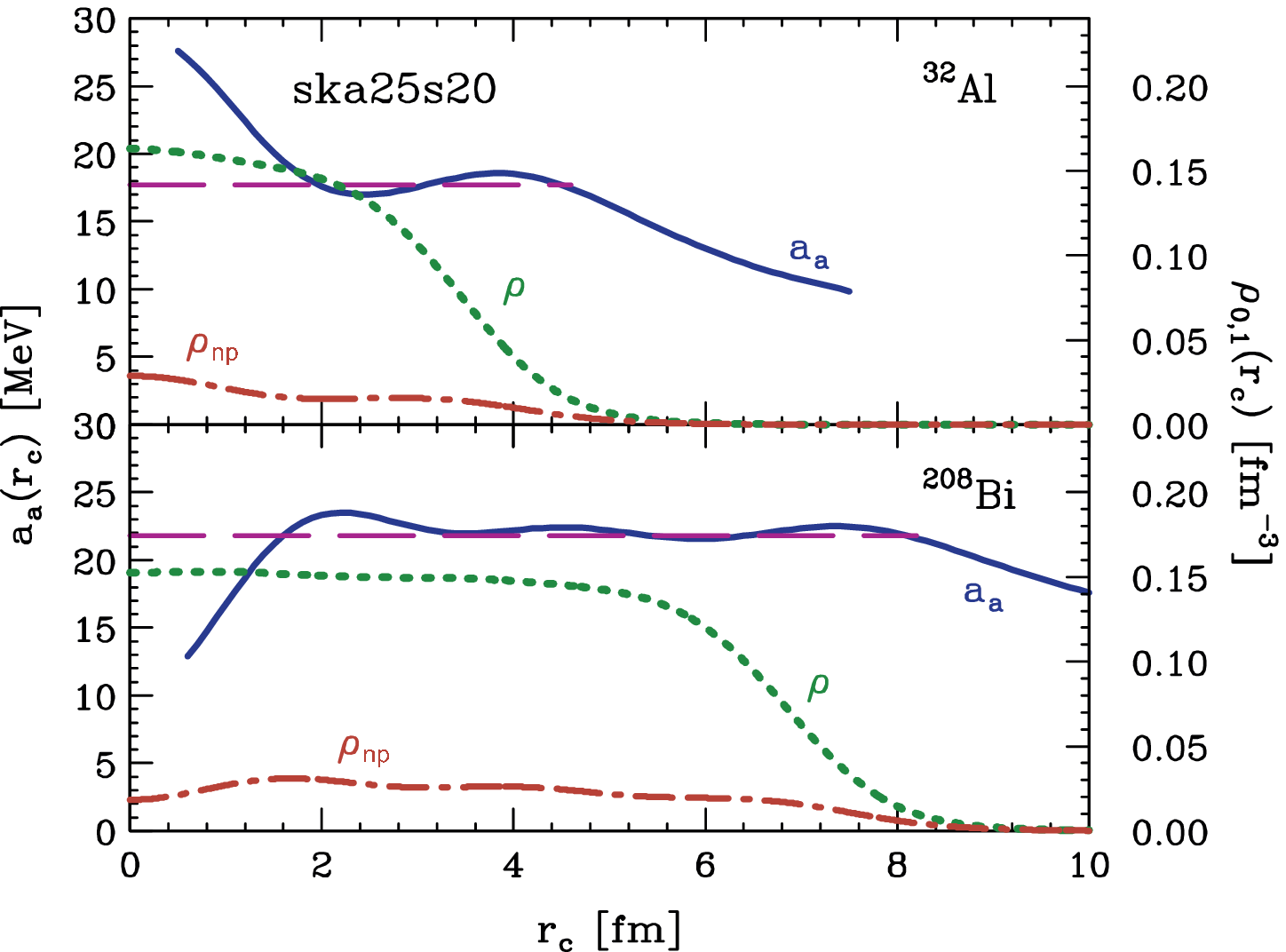}}
\caption{Densities $\rho$ and $\rho_{np}$ (right scales), represented by dotted and dash-dotted lines, respectively, and the asymmetry coefficient $a_a$ (left scales), represented by solid lines, calculated from Eqs.~\eqref{eq:mu3C} and \eqref{eq:capC}, as a function of the cut-off radius~$r_c$, for $^{32}$Al (top panel) and $^{208}$Bi (bottom panel) nuclei.  The results stem from SHF calculations with the ska25s20 \cite{brown08} parametrization.  Horizontal dashed lines represent coefficient values obtained with $\rho(r_c) = 0.10 \, \text{fm}^{-3}$.
}
\label{fig:asyrc}
\end{figure}

As practical measures of success of the above strategy, the sensitivity of the calculated coefficients to $r_c$ could be used as well as the size of shell effects in those coefficients.  The~sensitivity of the coefficient calculation to $r_c$ is illustrated in Fig.~\ref{fig:asyrc} from a SHF calculation with the exemplary ska25s20 \cite{brown08} interaction.  Nuclear densities, particularly the difference $\rho_{np}$, are affected by shell effects leading to some oscillations in the interior.  As more and more of the volume is taken into account, the effects of the oscillations get averaged out.  The~oscillations tend to be less pronounced and averaging is more effective in a heavy than in a medium-mass nucleus.  Beyond nuclear surface, specifically at net density $\rho \lesssim 0.04 \, \text{fm}^{-3}$, cf.~I, the local density approximation \eqref{eq:rho3loc} breaks down, and the coefficient from~\eqref{eq:mu3C} and~\eqref{eq:capC} generally begins to underestimate the true value of the coefficient.  With this, the optimal value of $r_c$ should be in the vicinity of the nuclear surface;
a too low $r_c$ might be signaled by excessive shell oscillations as a function of mass and charge numbers in the inferred $a_a$.
Upon carrying out tests with the quality of coefficient evaluation in half-infinite nuclear matter of~I, we decided on $r_c$ corresponding to the net density of $0.10 \, \text{fm}^{-3}$, i.e.
\beq
\label{eq:rc}
\rho(r_c) = 0.10 \, \text{fm}^{-3} \, .
\eeq

\begin{figure}
\centerline{\includegraphics[width=.57\linewidth]{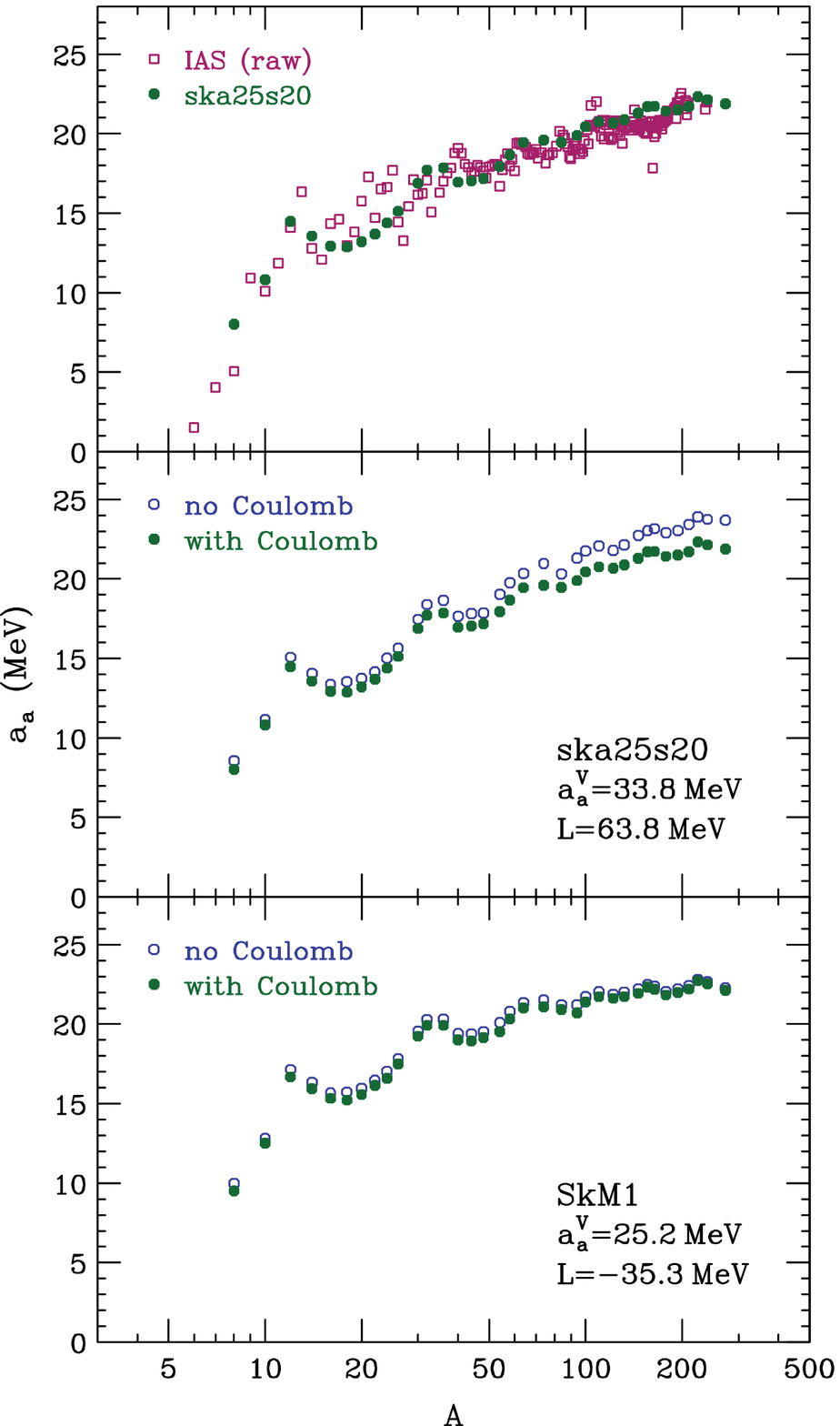}}
\caption{Symmetry coefficients $a_a$ obtained for individual~$A$.  The~top panel shows the coefficients extracted from the SHF calculations with the ska25s20 \cite{brown08} interaction (filled symbols), as well as the coefficients extracted from the measured excitation energies to IAS (open), when \emph{no} microscopic corrections are employed.  The middle and bottom panels, respectively, show coefficients from the SHF calculations with ska25s20 and SkM1 \cite{gomez:1995} interaction parametrizations, either incorporating (closed symbols) or not (open) Coulomb interactions, obtained from either Eq.~\eqref{eq:mu3C} or \eqref{eq:capC}, with $r_c$ from Eq.~\eqref{eq:rc}.
}
\label{fig:asyCou}
\end{figure}

Symmetry coefficients extracted following Eqs.~\eqref{eq:mu3C} and \eqref{eq:capC}, for nuclei with $\eta=(0.17-0.31)$, using $r_c$ from~\eqref{eq:rc}, are represented for two sample interaction parametrizations in Fig.~\ref{fig:asyCou}.  In the top panel we show the coefficients extracted from the calculations with the ska25s20 interaction~\cite{brown08}, together with the coefficients from IAS data, extracted when disregarding microscopic corrections (the same as in Fig.~\ref{fig:asyma_ns}).  The goal is to underscore the practical suppression of shell effects achieved for the SHF coefficients, when using densities.  The~effects are relatively weak for $A \gtrsim 30$.  Have we attempted to extract those coefficients using energy differences within the SHF calculations, the shell effects could have been so strong that the panel might seem uniformly filled with the results!  The use of $r_c$ from Eq.~\eqref{eq:rc} is additionally tested in the following subsection.

The remaining panels in Fig.~\ref{fig:asyCou} illustrate other essential characteristics of the symmetry coefficients resulting from the SHF calculations.  When the symmetry energy changes slowly at subnormal densities, such as for the SkM1 parametrization \cite{gomez:1995}, the coefficients for heavy nuclei come close to $a_a^V$ for normal matter.  On the other hand, when the energy changes quickly, as in the case of the ska25s20 \cite{brown08} parametrization with high~$L$, the coefficients are significantly reduced compared to $a_a^V$, even for heavy nuclei.  Moreover, a faster variation with density generally translates into a faster variation of $a_a$ with~$A$.  Besides standard results, Fig.~\ref{fig:asyCou} shows also results obtained when Coulomb interactions are switched off.  Obviously, Coulomb interactions may only have a significant impact on the coefficients in heavier nuclei.  However, the impact is further conditioned on the significant dependence of the symmetry energy on density - in the case of SkM1 in Fig.~\ref{fig:asyCou}, the impact is nearly none.  For typical parameterizations, though, such as ska25s20, the increase in the size of a nucleus on account of Coulomb interactions results in a drop of the coefficient values for heavy nuclei.  As can be seen in Fig.~\ref{fig:asyCou}, the differences in the mass dependencies for ska25s20 and SkM1 are stronger when the Coulomb interactions are switched off.  The surface tension, increasing on the average the density in lighter nuclei, is expected to reduce differences in mass dependencies for different~$L$, similarly to the Coulomb interactions.

\subsection{Towards Properties of Nuclear Matter}

One potential utility of the SHF calculations is in testing the degree to which a model-independent quest for the properties of nuclear matter, following the volume-surface separation~\cite{Danielewicz:2003dd,Danielewicz:2007pf}, can succeed.  Towards that end, we carry out the SHF calculations~\cite{reinhard:014309} with Coulomb interactions switched off and with nuclear masses far exceeding those in Nature, such as $A = 125 \, 000$.  The lack of competition between the Coulomb and nuclear interactions, facilitates developing a perspective on the large-$A$ limit.

\begin{figure}
\centerline{\includegraphics[width=\linewidth]{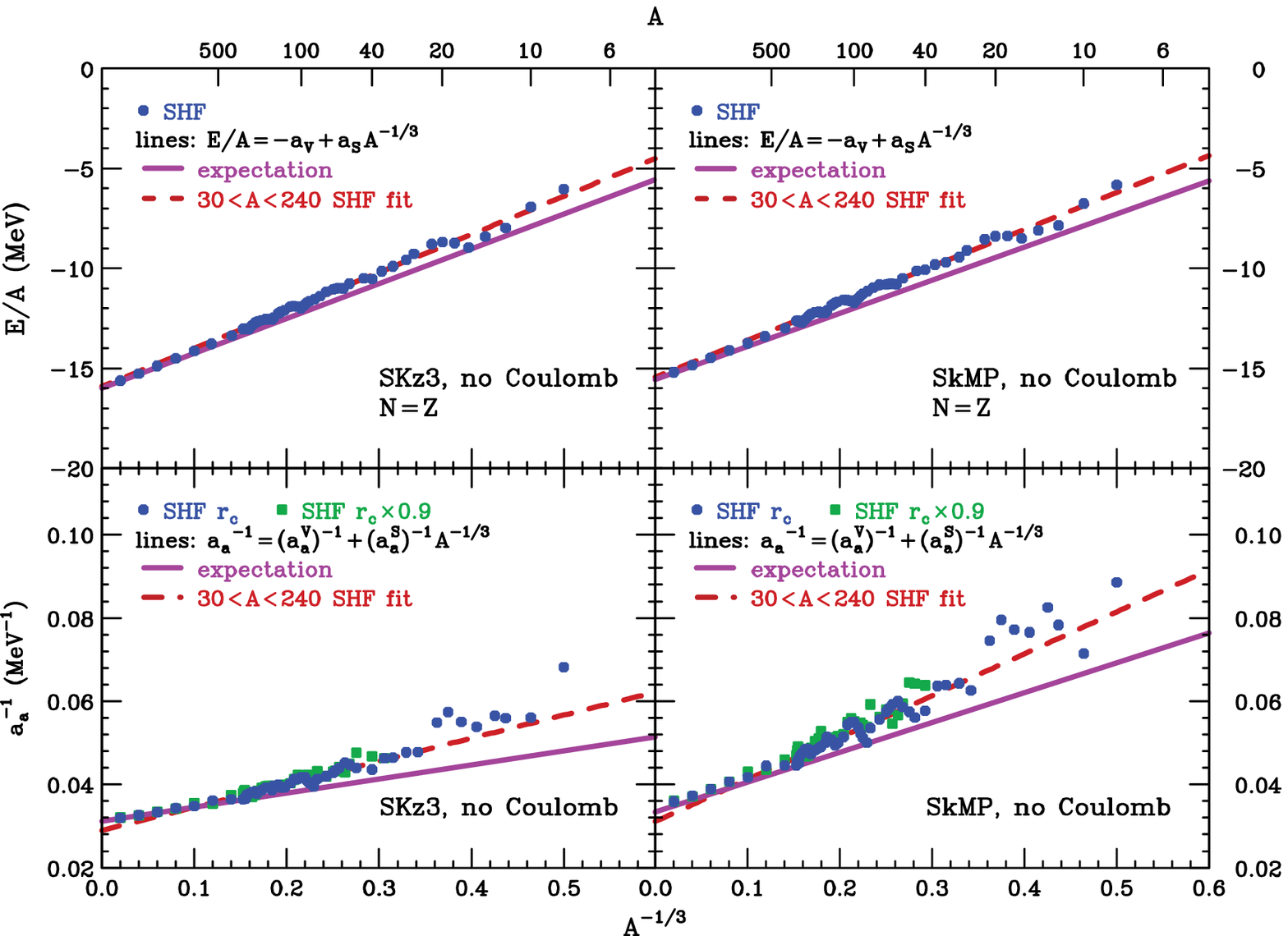}}
\caption{Tests of volume-surface decomposition for energy in the SHF calculations with Coulomb interactions switched off, for sample SKz3~\cite{PhysRevC.66.014303} (left panels) and SkMP \cite{PhysRevC.40.2834} (right panels) Skyrme parametrizations.  The top panels represent energy per nucleon for $N=Z$ nuclei, while the bottom panels represent inverse symmetry coefficient, $1/a_a$, for $\eta \sim 0.2 $ nuclei, both as a function of $A^{-1/3}$.  The~symbols represent results of spherical SHF calculations.  For the symmetry coefficient, results obtained for $r_c$ of Eq.~\eqref{eq:rc} reduced by the factor of 0.9 (squares) are shown for heavier masses in addition to the standard results (circles).
The solid lines represent expectations based on the half-infinite matter calculations of~I.  The compressional effect of Eq.~\eqref{eq:aaSp} is included in the surface symmetry coefficient for the lines.  The dashed lines show linear fits to the results in the mass region $30 < A <240$.
}
\label{fig:Bed_smple}
\end{figure}

The large-$A$ expansions, for both isoscalar and isovector sectors of the nuclear energy, are tested for sample Skyrme parametrizations, SKz3 and SkMP, in Fig.~\ref{fig:Bed_smple}.  The top panels in that figure show the energy per nucleon for symmetric $N=Z$ nuclei as a function of~$A^{-1/3}$.  The symbols represent results of the spherical SHF calculations.  With the absolute values of the energy per nucleon being shown in the panels, the shell effects turn out to be relatively minor.  The~solid lines show expectations based on half-infinite matter calculations of I, assuming the volume-surface decomposition for the energy~$E_0$.  While the energies of symmetric nuclei from the spherical SHF calculations do not quite follow these expectations, the deviations are not  very large.   The dependence of energies on $A^{-1/3}$ is roughly linear for the nuclei occurring in Nature.  For reference, dashed lines in the top panels show linear fits to the energies, for the two interactions, made in the mass region of $30 < A < 240$, mimicking the IAS analysis before.  In terms of variability with mass number~$A$, for the Skyrme interactions meeting our stability criteria (cf.~Appendix \ref{Appendix:SkyrmeStability}), we generally find the so-fitted $a_S$-values just $\sim (10$-$15)$\% larger that those established for half-infinite matter.  The fitted volume parameters $a_V$ typically turn out to be only $\sim 0.5$\% larger than those known for an infinite system.  Without much expected impact of the shell effects and with no practical need for the corresponding corrections potentially bringing in systematic errors, these SHF results render support for the opportunity of learning from the experimental energies, without much model-dependence, on the properties of an infinite {\em symmetric} system.

The bottom panels in Fig.~\ref{fig:Bed_smple} show next the inverse of the (a)symmetry coefficient, from the Hartree-Fock calculations with the same two Skyrme parametrizations as the top panels, again plotted as a function of $A^{-1/3}$.  The results, obtained from the spherical calculations for $\eta \sim 0.2$ nuclei, with~$r_c$ of Eq.~\eqref{eq:rc}, are represented by circles.  In addition, for heavier nuclei, the results obtained for $r_c$ reduced by the factor of $0.9$ are represented by squares in the panels.  For heavier nuclei, that reduction moves $r_c$ completely out of the surface region, while retaining significant volume for averaging out the shell effects within the interior integrations~\eqref{eq:Aaac}.  If our procedure for calculating the symmetry coefficient is correct, then the results for the symmetry coefficient should not change significantly for heavy nuclei, with such a change in~$r_c$.  We can see in the bottom panels of Fig.~\ref{fig:Bed_smple} that indeed the coefficients do not change much with the $r_c$-change.

The solid lines in the bottom panels represent the expectations, regarding inverse symmetry-coefficients, based on the volume-surface decomposition and the half-infinite matter results of~I.  At the largest $A$, the spherical SHF calculations follow those expectations, but as values of $A$ drop, some deviations develop.  The dashed lines in the panels represent linear fits to the spherical results in the mass region of $30 < A < 240$.  The fitted surface symmetry coefficients $a_a^S$ typically deviate more from the expectations than their isoscalar counterparts and can be lower from these expectations by as much as $35\%$.  The volume symmetry coefficients $a_a^V$ tend to be much better reproduced by the fits.  However, we have already observed in Sec.~\ref{sec:SymIAS} a fragility of the latter coefficients under the uncertainties in the microscopic corrections, making the inferred $a_a^V$-values  span essentially the whole range of practical possibilities.  The current and former findings bring into question the strategy of inferring the coefficients of an infinite system through a fit and, therefore, gaining knowledge about the latter system in a model-independent manner.  Under the circumstances, a strategy set up as model-dependent from the start might, in the end, deliver less biased conclusions.

Given the widespread use of SHF results in describing nuclei, a direct comparison between the SHF and IAS results is a natural candidate for such a model-dependent strategy.  Not all Skyrme parametrizations employed in the literature, though, can be
comparably useful for the purpose.  In our own calculations, for many of those parametrizations we found a disturbing lack of intrinsic consistency in that the expectations developed on the basis of half-infinite matter, cf.~I, did not satisfactorily agree with the spherical calculations even for the largest of the investigated~$A$.  Such finding were usually connected with large variations in the calculational results from a given parametrization for moderate and low~$A$.  This was far more common for the characteristics isolated within the isovector rather than the isoscalar sector.  For some quite mundane nuclei, the spherical calculations with problematic parametrizations did not converge at all.
The~majority of parametrizations producing such outcomes turned out to violate one or more of the Landau long-wavelength stability conditions, cf.~Appendix~\ref{Appendix:SkyrmeStability}.  However, some parametrizations exhibiting such symptoms did not actually violate the Landau conditions, but produced combinations of interaction parameter values suspected in the short wavelength instabilities \cite{Lesinski:2006cu}.  This led to the analysis of short-wavelength instabilities for Skyrme interactions contained in Appendix \ref{Appendix:SkyrmeStability}.

According to Appendix~\ref{Appendix:SkyrmeStability}, an insistence on absolute stability of the systems described in terms of Skyrme interactions would eliminate the vast majority of Skyrme parametrizations from considerations.  Correspondingly, we adopt a pragmatic approach, accepting all Skyrme parametrizations for which the conditions, under which the instabilities must develop, are sufficiently removed from the typical conditions in the calculations that we carry out for our systems.  By imposing requirements on the stability conditions, see Appendix \ref{Appendix:SkyrmeStability}, rather than on aspects of outcomes from calculations, of interest to us, we hopefully minimize the bias on the latter.  Imposing of variety of other constraints \cite{dutra_skyrme_2012} on the interactions could be tempting, but we fear an excessive dilution of the sample of the interactions and reduced flexibility in the symmetry energy, given the limited number of parameters in the interactions and their entanglement in the various physical characteristics of nuclear systems.

\section{Constraints on Symmetry Characteristics of Nuclear Matter}
\label{sec:constraints}

The fragility of the conclusions on the $a_a^V$-$a_a^S$ properties of half-infinite matter, drawn from IAS results, is paralleled by a similar fragility of conclusions on the symmetry energy at normal density, phrased in terms of parameters $a_a^V$ and $L$, when one tries to bypass the $a_a^V$-$a_a^S$ fits to the IAS results and confront the IAS results with the results of SHF calculations directly.  The parallel fragility should not be surprising, given the strong and intuitively understandable correlation between the $a_a^S$ and $L$ parameters for the Skyrme interactions, cf.~I: high values of $L$ tend to be tied to low values of $a_a^S$.  Under the circumstances, we start out by reexamining the strategy of learning about isovector properties of nuclear matter.

\subsection{Pearson Correlation Coefficient}

Nominally retreating in the scope, we ask whether the IAS results can produce {\em any} significant constraints on the symmetry energy of infinite matter, whether at normal (i.e.~$a_a^V$) or lower density, given the portion of the matter at subnormal density in the nuclei.  At~the qualitative level, such a question can be addressed in terms of the so-called Pearson correlation coefficient \cite{pearson_liii._1901} for a sample (in the current context see also~\cite{PhysRevC.81.051303}):
\beq
\text{r}_{XY} = \frac{\big\langle ( X - \langle X \rangle ) \, ( Y - \langle Y \rangle ) \big\rangle }
{\sqrt{\big\langle ( X - \langle X \rangle )^2\big\rangle \, \big\langle ( Y - \langle Y \rangle )^2\big\rangle  }} \, .
\label{eq:rXY}
\eeq
Here, $X$ and $Y$ are two variables suspected of being tied to each other.  The coefficient is seen as covariance of the two variables, normalized with dispersions.  The coefficient is particularly well suited for picking up a linear correlation between variables.  The definition imposes limits $-1 \le \text{r}_{XY} \le 1$.  The values of $|\text{r}_{XY}|$ close to~1 signify a very tight correlation between $X$ and $Y$ (indeed $\text{r}_{XX}=1$), while values close to~0 signify essential lack of a~correlation.

\begin{figure}
\centerline{\includegraphics[width=.8\linewidth]{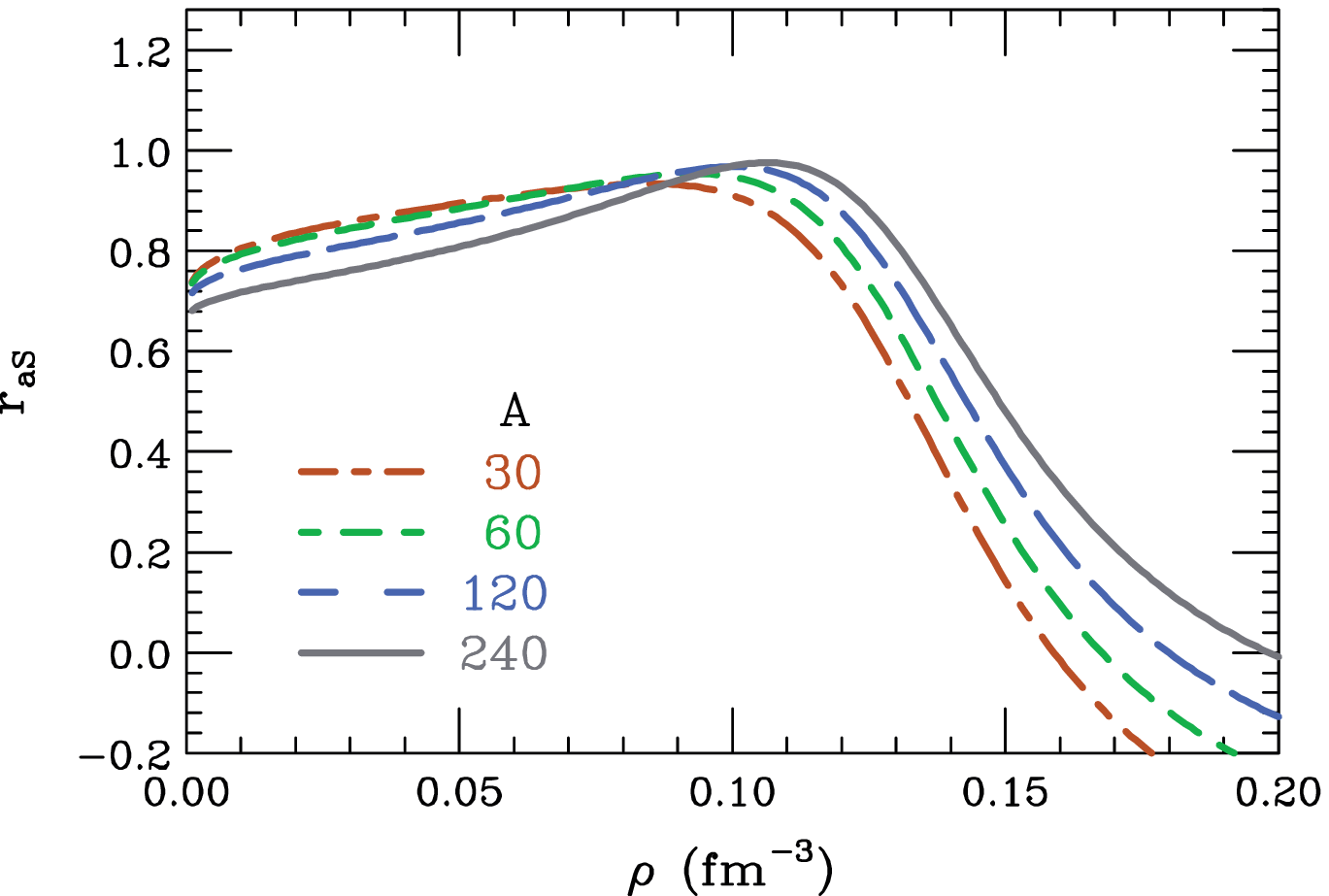}}
\caption{Pearson correlation coefficient \cite{pearson_liii._1901}, Eq.~\eqref{eq:rXY}, between the asymmetry coefficient $a_a(A)$ for a given mass $A$ and the value of symmetry energy in uniform matter $S(\rho)$, for Skyrme interactions, plotted, at different indicated $A$, against the density~$\rho$.
}
\label{fig:ACO}
\end{figure}

\begin{figure}
\centerline{\includegraphics[width=.60\linewidth]{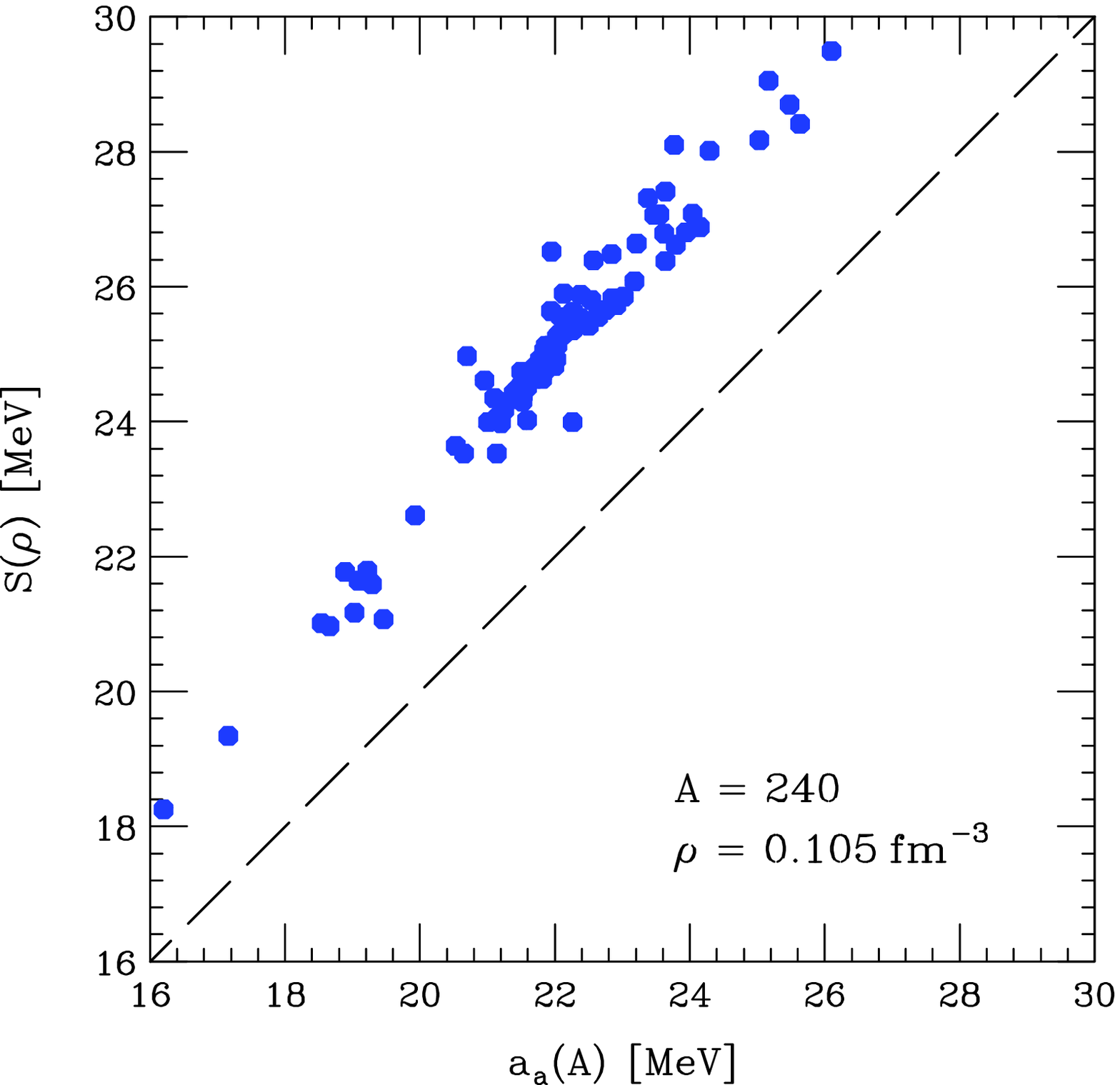}}
\caption{Values of the symmetry in uniform matter, $S(\rho)$, at $\rho = 0.105 \, \text{fm}^{-3}$, for the Skyrme interactions, plotted against the values of the symmetry coefficient, $a_a(A)$, for $A=240$.  Dashed line representing $S(\rho) = a_a(A)$ serves as a guide to the eye.  The $A$-$\rho$ combinations representing lower values of $r_{aS}$ in Fig.~\ref{fig:ACO} generally yield coarser $a_a$-$S$ correlations than in the current figure.
}
\label{fig:a240105}
\end{figure}

In testing the potential connection between IAS results and the values of the symmetry energy at different densities, we examine the correlation coefficient between the values of asymmetry coefficient at selected $A$ and the symmetry energy in uniform matter at different densities for the Skyrme interactions, i.e.\ $X \equiv a_a(A)$ and $Y \equiv S(\rho)$ in Eq.~\eqref{eq:rXY}.  The~correlation coefficient is displayed at different~$A$ against~$\rho$ in Fig.~\ref{fig:ACO}.  It is apparent in the figure that the coefficient strongly depends on~$A$ in the vicinity of $\rho_0 \simeq 0.16 \, \text{fm}^{-3}$.  The coefficient is rather low there and achieves a significant magnitude only for the largest~$A$.  On the other hand, surprisingly, the correlation becomes very strong at subnormal densities.  In fact, for $A=240$ the coefficient reaches a~maximum as high as 0.98 at $\rho \simeq 0.106 \, \text{fm}^{-3}$, demonstrating that the value of $a_a(240)$ is an excellent predictor of $S(0.106 \, \text{fm}^{-3})$.  The correlation for the region of the latter maximum is illustrated in Fig.~\ref{fig:a240105} and it is seen that the correlation is indeed very tight and linear in nature.  Past the maximum for $A=240$ in Fig.~\ref{fig:ACO}, the Pearson coefficient drops, but not by much, towards low densities.  As the mass number $A$ for the $a_a$ coefficient is reduced, the~maximum for~$\text{r}_{aS}$ shifts to lower densities and in particular to $0.084 \, \text{fm}^{-3}$ for $A=30$.  The value of $\text{r}_{aS}$ at the maximum drops somewhat, but remains high, with the value of 0.93 for $A=30$.    The~origin of the strong $a_a(A)$-$S(\rho)$ correlation, evident from the low-density side in Fig.~\ref{fig:ACO}, will be discussed once the consequences of that correlation will be explored.

In the past, a supposition was put forward \cite{centelles_nuclear_2009} and exploited in \cite{liu_nuclear_2010}, see also~\cite{PhysRevC.83.044308}, that the values of~$a_a(A)$ at specific $A$ represent values of $S(\rho)$ at specific $\rho$.  However, that supposition was based on the $a_a^V$-$a_a^S$ representation from half-infinite matter, rather than an actual calculation of~$a_a(A)$ within SHF.  Figure \ref{fig:Bed_smple} illustrates significant disagreements with the $a_a^V$-$a_a^S$ representation.  Moreover, justification for the specific use of the supposition above would require a narrow peak in~$\text{r}_{aS}$ curves in Fig.~\ref{fig:ACO} at each~$A$.  However, the maxima are steep only from their high-density sides and are at best mildly pronounced from the low density side~- the~$A=30$ correlation is, in fact, nearly flat from the low-$\rho$ side.  Finally, according to the use of the supposition, one should have an approximate equality of the the symmetry coefficient~$a_a(A)$ and the symmetry energy in uniform matter $S(\rho)$ at the maximum.  Taking the case of the most pronounced maximum in Fig.~\ref{fig:ACO}, it~can be seen in Fig.~\ref{fig:a240105} that the strong correlation does not imply, though, an~equality between the coefficient and the specific symmetry-energy value.

\subsection{Constraints on the Density-Dependence of Symmetry Energy}

Given the high values of the Pearson correlation coefficients at subnormal densities, between the values of the symmetry coefficient and the symmetry energy, we next attempt to map out the values of the symmetry energy by confronting the coefficients from SHF calculations and from the IAS data.  To minimize the role of any uncompensated microscopic effects, necessarily different for the coefficients extracted in different manner and not judged in the comparison, we concentrate on the mass region of $A \ge 30$ and we compare there the results from $a_a^V$-$a_a^S$ interpolations, rather than directly the coefficient values for any~$A$.  The~microscopic effects and their subtraction end up dominating, though, the uncertainties in the $a_a^V$-$a_a^S$ representations, both on the IAS and SHF sides.

\begin{figure}
\centerline{\includegraphics[width=.72\linewidth]{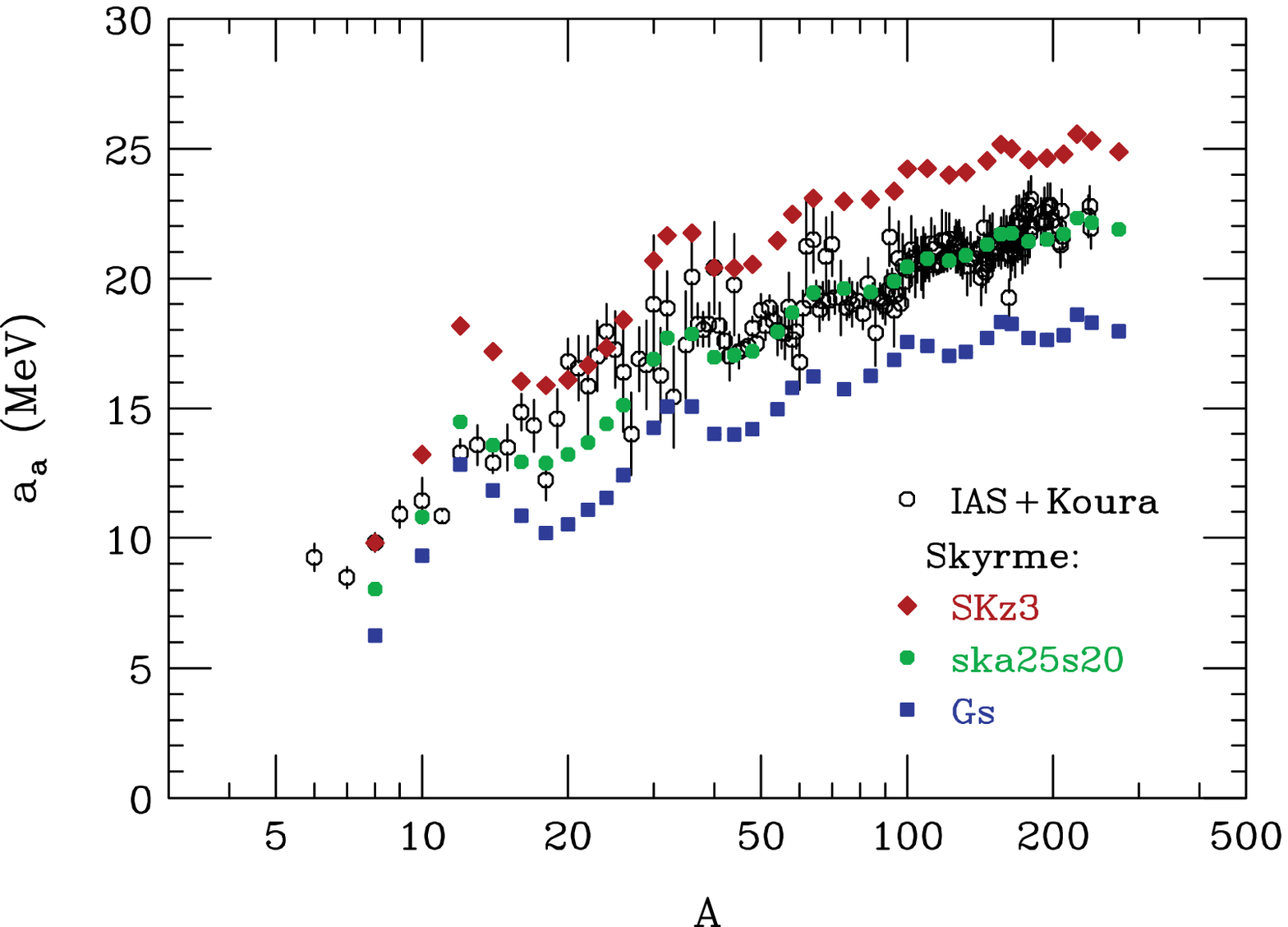}}
\caption{Symmetry coefficients $a_a$ obtained for individual~$A$ from the measured excitation energies to IAS (open symbols) and from the SHF calculations with different indicated Skyrme interactions (closed symbols).  The ska25s20 parametrization \cite{brown08} yields coefficients that agree with IAS within uncertainties, while the two other parametrizations \cite{PhysRevC.33.335,PhysRevC.66.014303} do not.
}
\label{fig:asyma_smpl}
\end{figure}

We estimate the uncertainties in the smooth parametrization of the IAS results by examining the results from applying the three different sets of shell corrections \cite{Moller:1993ed,Koura:2005,Groote1976418}.  We~find $(\delta a_a^\text{IAS})^\text{mic} \sim 0.41 \, \text{MeV}$ both at $A \sim 30$ and $A \sim 240$.  In extracting the coefficients using densities from SHF calculations, as discussed in~Sec.~\ref{subsec:aaSHF}, we find variations in the coefficients at the level of $(a_a^\text{SHF})^\text{mic} \sim 0.47 \, \text{MeV}$ at $A \sim 30$ and $(a_a^\text{SHF})^\text{mic} \sim 0.27 \, \text{MeV}$ at $A \sim 240$, on account of the interplay of oscillations in the densities and arbitrariness in the choice of~$r_c$.  Altogether, the uncertainties in the determination of coefficients from the IAS and SHF sides could account for discrepancies of $\delta a_a \sim 0.61 \, \text{MeV}$ at $A \sim 30$ and $\delta a_a \sim 0.49 \, \text{MeV}$ at $A \sim 240$, between the smooth $a_a^V$-$a_a^S$ parametrizations of IAS and SHF results.  Figure \ref{fig:asyma_smpl} shows examples of coefficients from Skyrme interactions, as a function of mass, with one set agreeing with coefficients from IAS, within the discussed uncertainties, and two other sets disagreeing.  One more SHF set agreeing with IAS is illustrated in the top panel of Fig.~\ref{fig:asyCou}.

\begin{figure}
\centerline{\includegraphics[width=.66\linewidth]{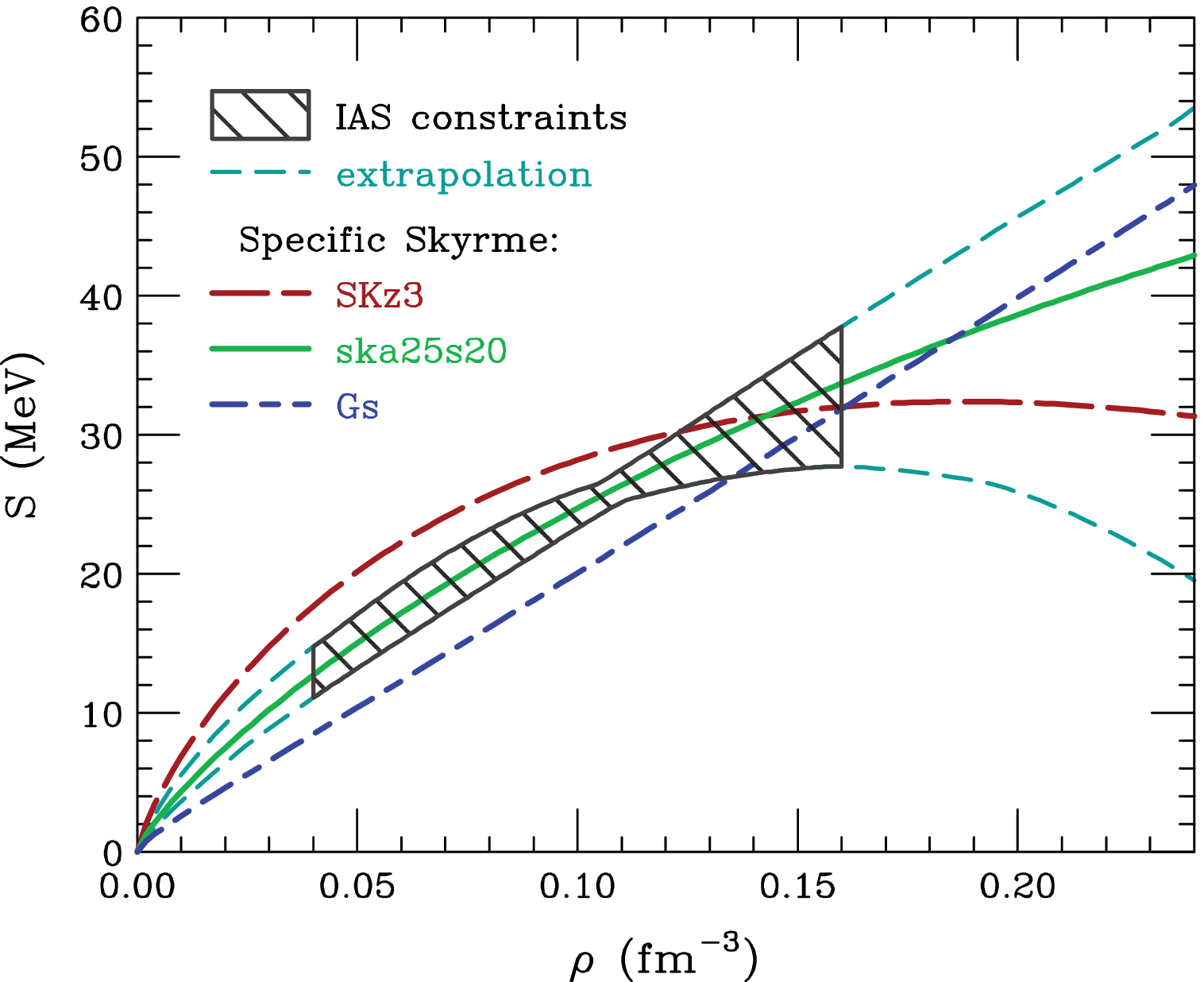}}
\caption{Symmetry energy in uniform matter as a function of density.  The hatched region represents IAS constraints.  The short-dashed lines represent extrapolations of that region to supranormal, $\rho > \rho_0$, and low, $\rho < \rho_0/4$, densities.  The solid, long-dashed and short-long-dashed lines represent the symmetry energies for the three Skyrme parametrizations represented in Fig.~\ref{fig:asyma_smpl} (with symmetry coefficients).
}
\label{fig:symec}
\end{figure}

\begin{figure}
\centerline{\includegraphics[width=.60\linewidth]{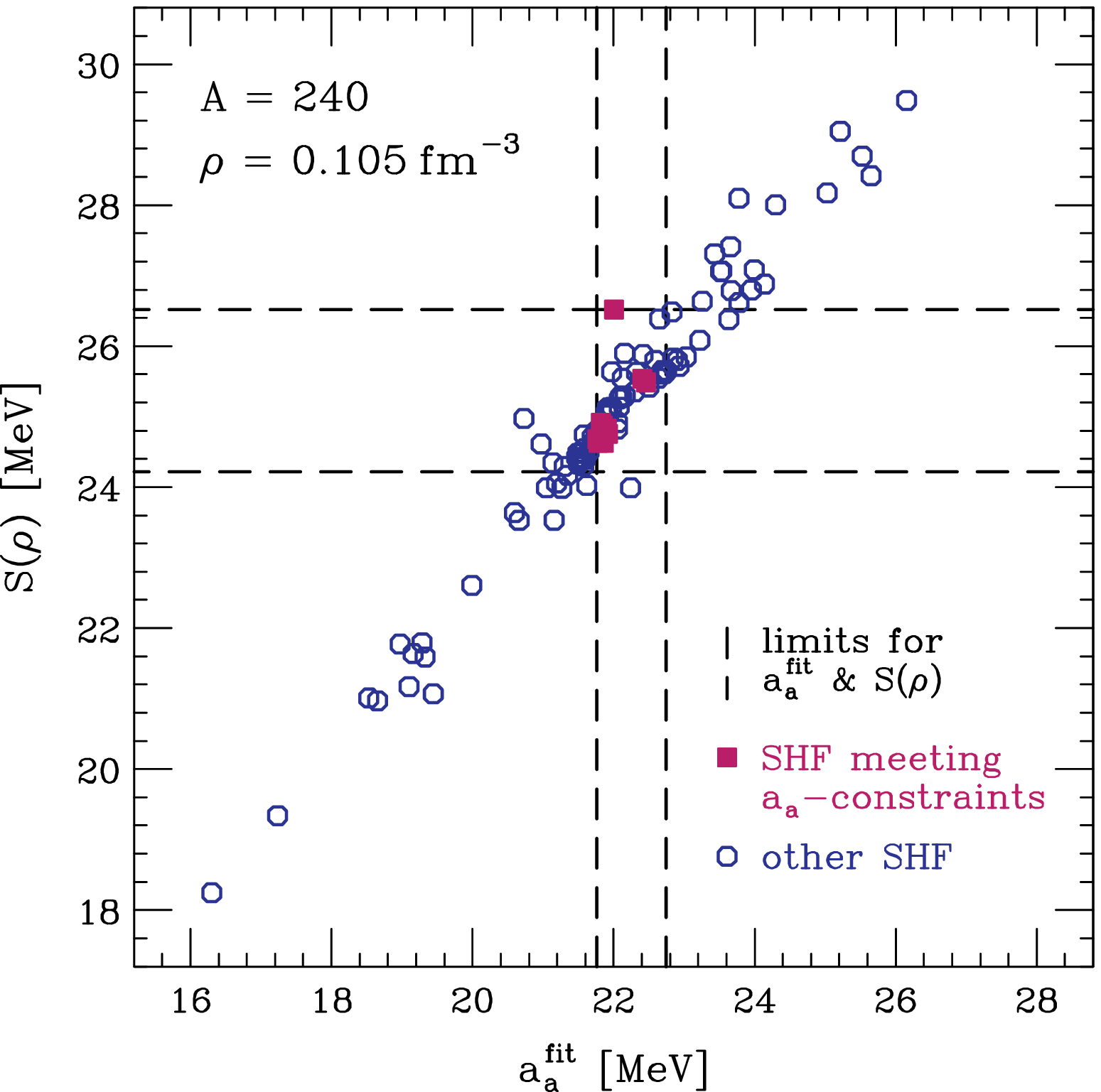}}
\caption{Analog of Fig.~\ref{fig:a240105}, with values of the symmetry in uniform matter, $S(\rho)$, at $\rho = 0.105 \, \text{fm}^{-3}$, plotted against the symmetry coefficient, $a_a^\text{fit}(A)$, at $A=240$, with the last obtained from the surface-volume fit to the results of SHF calculations.  For heavy nuclei the systematic of direct and fitted coefficients is not much different.  Filled squares represent those SHF calculations which meet the $a_a$-constraints across the $A=(30\,$--$\,240)$ mass region.  Open circles represent the remaining SHF results.  The dashed vertical and horizontal lines represent, respectively, the uncertainty range in $a_a^\text{fit}(240)$ and deduced limits on $S(0.105 \, \text{fm}^{-3}$).
}
\label{fig:af240105}
\end{figure}

\begin{figure}
\centerline{\includegraphics[width=.64\linewidth]{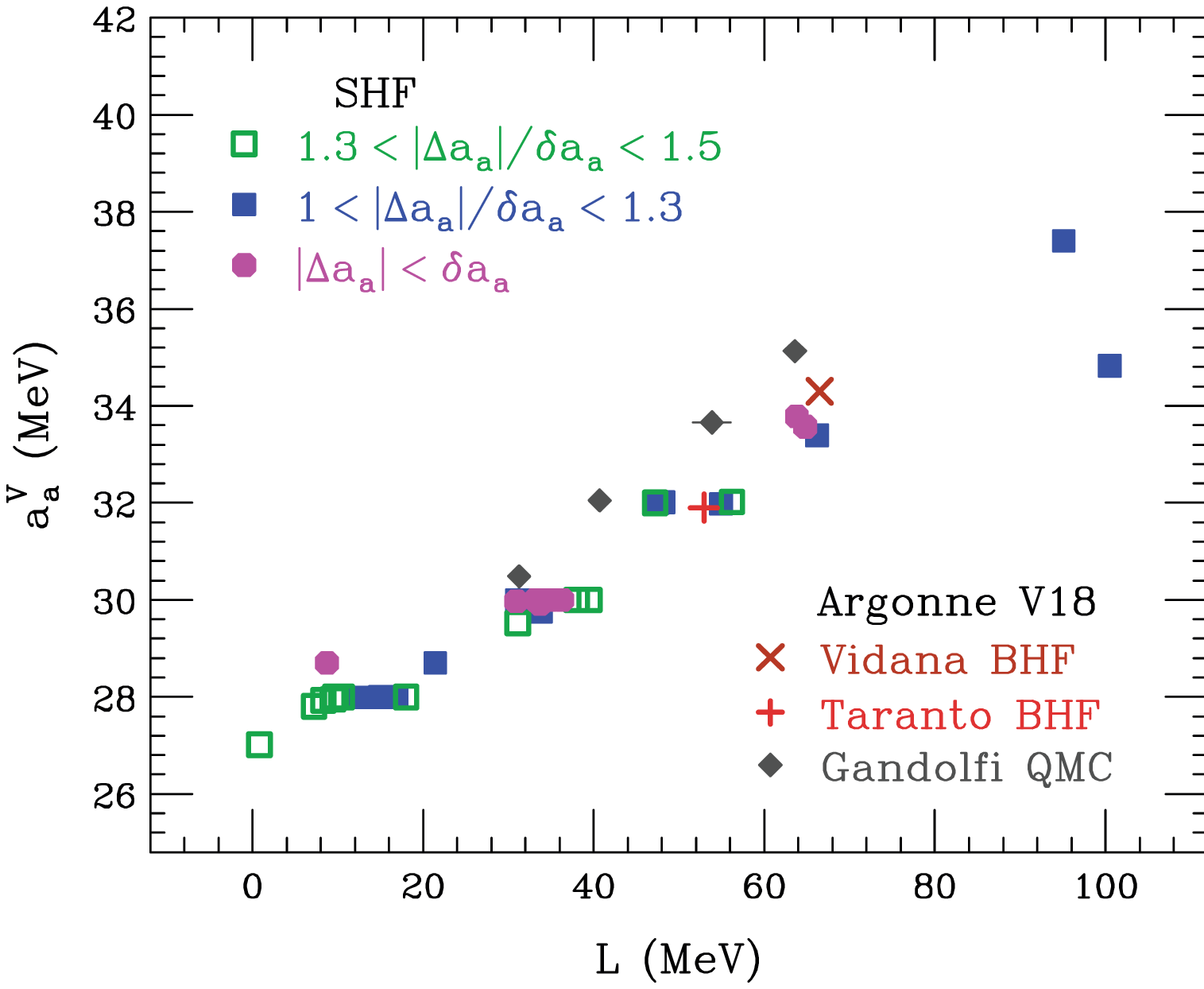}}
\caption{Correlation between the volume symmetry coefficient, $a_a^V = S(\rho_0)$, and the slope parameter $L$ of the symmetry energy at~$\rho_0$.  Filled circles represent those Skyrme interactions that yield symmetry coefficients which agree with the IAS results within errors.  Filled and open squares represent those additional Skyrme interactions for which the symmetry coefficients begin to agree with the IAS results, after the coefficient errors get inflated by 30\% and by 50\%, respectively.  The~crosses and filled diamonds represent microscopic results based on the Argonne V18 interaction \cite{PhysRevC.51.38}, obtained, respectively, within Brueckner-Hartree-Fock (BHF) calculations and within Quantum Monte-Carlo (QMC) calculations.  The two displayed BHF calculations, respectively, by Vida\~{n}a~\etal~\cite{PhysRevC.80.045806} (diagonal cross) and by Taranto~\etal~\cite{taranto_selecting_2013} (vertical crosss), differ in the details of the adopted 3N interactions.  The range of results within QMC, by Gandolfi~\etal~\cite{gandolfi_maximum_2011}, is similarly due to the exploration of different assumptions about the 3N force.  In the latter calculations, moreover, the 2N interactions were modified in the short-range portion of $\ell \ge 2$ partial waves, compared to the original V18.
}
\label{fig:aavl}
\end{figure}

The next logical step might be to map out those regions of the symmetry energy as a~function of density, $S(\rho)$, that represent the Skyrme interactions which agree with IAS.  The symmetry energy values for the exemplary Skyrme parametrizations of Fig.~\ref{fig:asyma_smpl} are shown in Fig.~\ref{fig:symec}.  The~problem that we face, in mapping out the constraint region for~$S(\rho)$, is that, by imposing the constraint of agreement with IAS, we thin out the sample of Skyrme parametrizations under consideration down to just 10.  To make things worse, the latter parametrizations bunch up in the space of characteristics of symmetry-energy, cf.~Figs.~\ref{fig:af240105} and~\ref{fig:aavl}, into just 3 groups.  Such a limited sampling set can imperil the plan to map out~$S(\rho)$, in that the allowed region of values may come out artificially narrow.  In Fig.~\ref{fig:aavl}, showing the $a_a^V$-$L$ parameters, we can observe a significant spreading out of the symmetry energy parameters for quite modest relaxation of the constraints of agreement with~IAS.  The broad range of the values in Fig.~\ref{fig:aavl} illustrates, beyond just Fig.~\ref{fig:ACO}, the general difficulty in assessing the symmetry energy around~$\rho_0$, using just IAS.

To cope with the limited sample of parametrizations that meet the $a_a$-constraints, we adopt the following reasoning. Have we had far more Skyrme parametrizations for our analysis, meeting or right outside of the IAS constraints, the span of the $a_a^V$, $L$ and $S(\rho)$ values, such as in Figs.~\ref{fig:af240105} and~\ref{fig:aavl}, would have been densely filled. In the future, it may be possible to generate the parametrizations in the symmetry-energy regions of interest to accomplish this~\cite{brown08}.  Still, for small ranges of parameters characterizing the symmetry energy, such as representing the uncertainties in $a_a^\text{fit}$, it should be possible to find trajectories within the Skyrme parameter space, joining parametrizations in the range defined by $\delta_{a_a}$, along which trajectories the symmetry-energy results would vary slowly, allowing for a linear interpolations in the symmetry energy results.   That slow variation might not have to pertain to the Skyrme parameter values.  Based on that idea, together with the Skyrme parametrizations that directly meet the $\delta_{a_a}$ constraints, we consider parametrizations that moderately miss the constraints, such as by 30\% or 50\%, see Fig.~\ref{fig:aavl}, and we follow linear interpolations between different pairs of the parametrizations.  To stay consistent with the linearity of the interpolation, we demand that the symmetry energy values, for the two parametrizations on which the interpolation is based, do not deviate by more than 15\% within the density range on which we focus, that will be elaborated upon.  The last condition turns out to have no bearing, however, on the final results.  The discussed linear interpolations provide us with correlations between the $a_a^\text{fit}$ and $S(\rho)$ values, beyond those for the existing Skyrme parametrizations, populating voids in the space of the characteristics of symmetry energy.  E.g.~in the case of the characteristics displayed in Fig.~\ref{fig:aavl}, those interpolations complete the correlation valley between the $a_a^V$ and $L$ values there.  In the context of IAS results alone, however, we are primarily interested in the region of subnormal densities, where $S(\rho)$ values exhibit a tighter correlation with $a_a$, as discussed in the context of Fig.~\ref{fig:ACO}, and which vary over a narrower range than at $\rho_0$.

Upon combining the spread of $S(\rho)$ values for the Skyrme parametrizations that meet the~$\delta_{a_a}$ constraints directly, with the spread from the interpolations that meet the constraints, we arrive at the constraints on $S(\rho)$ indicated by the hatched region in Fig.~\ref{fig:symec}.  The extent, to which the interpolations broaden the $S(\rho)$-range, may be inferred from Fig.~\ref{fig:af240105} where the values of $S(\rho)$ at a specific density are shown for the parametrizations that meet the constraints directly, together with the deduced range for $S(\rho)$.  The~$S(\rho)$-boundaries saturate when we incorporate incorporate interpolations with the Skyrme parametrizations that miss the constraints by about 30\% - there is no change in those boundaries, if we include all parametrizations that miss the constraints by 100\%.  Using structure data, we obviously cannot claim to constrain the symmetry energy at supranormal densities.  Also in I we found that the symmetry energy values at a local density impact the system properties only down to the density of about 1/4 of the normal.  Constraining the investigated system, to remain uniform at low densities, is a theoretical problem of its own.  The boundaries for $S(\rho)$ that we arrive at, either at lower densities or higher than normal, are extrapolations based on the Skyrme parametrizations.

Consistently with the Pearson-coefficient results in Fig.~\ref{fig:ACO}, using IAS we arrive at excellent constraints ($\pm 1.2 \, \text{MeV}$) on $S(\rho)$ in the vicinity of $\rho \sim 0.105 \, \text{fm}^{-3}$.  Towards lower densities, the constraints remain very good (with uncertainty less than $\pm 2.4 \, \text{MeV}$).  We should mention  that while we terminate the constraint claim at $\rho \sim 0.04 \, \text{fm}^{-3}$, with $S \sim 12.9 \, \text{MeV}$, the symmetry coefficient values can reach even lower values for the lightest nuclei, $A < 20$.  In the other density direction, above $\rho \simeq 0.13 \, \text{fm}^{-3}$, the constraints rapidly deteriorate and at $\rho_0 \simeq 0.16 \, \text{fm}^{-3}$, the symmetry energy could be anywhere in the range $(27.7 - 37.8) \, \text{MeV}$ and still be consistent with IAS.  In the extrapolation of IAS results to supranormal densities, following the Skyrme parametrization, the symmetry energy could be in the range from -3.6 to 73.6~MeV, i.e.\ the energy resolution gets completely erased.  Thus, while the IAS systematic provides quite narrow constraints at moderately subnormal densities, the constraints are very weak around normal density and, on its own, cannot serve as a basis for extrapolations to the high densities encountered in the interiors of neutron stars.

Simultaneous analysis of Figs.~\ref{fig:asyma_smpl} and \ref{fig:symec}, as well as Figs.~\ref{fig:ACO} and \ref{fig:aavl}, allows, further, for a~qualitative understanding of the systematic of the constraints on symmetry energy as a~function of density.  Thus, the symmetry coefficient value of $a_a \sim 22.2 \, \text{MeV}$ at $A \sim 240$ can be arrived at with a relatively low value of $a_a^V$, such as $a_a^V \sim 28 \, \text{MeV}$, combined with a~relatively weakly changing symmetry energy around $\rho_0$, characterized by $L \sim 10\, \text{MeV}$, or with a~rather high value of $a_a^V$, such as $a_a^V \sim 38 \, \text{MeV}$, combined with a~rapidly changing $S$ around $\rho_0$, characterized by $L \sim 95 \, \text{MeV}$, or with other combinations in-between.  A~drop in the symmetry energy with density, induces pushing of the neutron-proton asymmetry into the surface region of a nucleus, enhancing the importance of that region in determining the asymmetry coefficient and compensating for the large energy at $\rho_0$.  Whichever the case of $S(\rho)$, the possible continuous functions have to cross within the region of moderately subnormal densities, giving rise to a bottleneck type of region around $\rho \sim 0.105 \, \text{fm}^{-3}$, where the constraints become really tight.  The dropping of the symmetry coefficients with mass and the vanishing of the symmetry energy in the limit of $\rho \rightarrow 0$, evidenced in the drop of~$a_a$ with drop in $A$ below the fit region of $A>30$, help to keep the constraints tight down to low densities.  With regard to the bottleneck region, see further Refs.~\cite{alex_brown_neutron_2000,trippa_giant_2008,RocaMaza:2012mh}.

\begin{figure}
\centerline{\includegraphics[width=.64\linewidth]{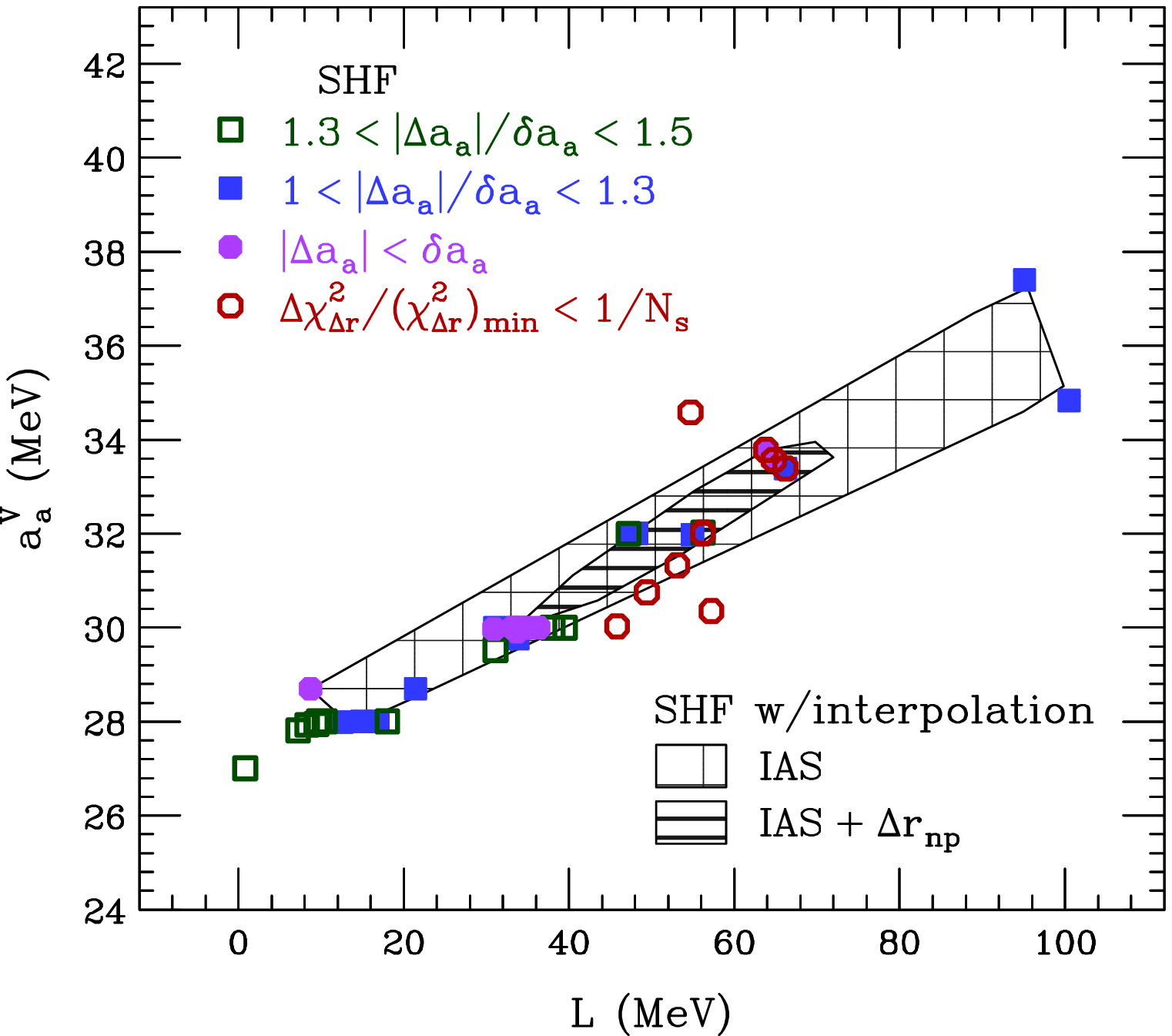}}
\caption{Skyrme interactions represented in the $a_a^V$-$L$ plane.  Filled circles and filled and open squares represent those interactions that yield symmetry coefficients which agree with the coefficients from IAS within, respectively, the estimated errors and the errors inflated by 30\% and~50\%.  Open circles represent those Skyrme interactions that yield $\chi^2$ for asymmetry skins of nuclei in Table~\ref{tab:skin}, within $(\chi^2)_\text{min}/N_s$ away from the $\chi^2$-minimum.  Here, $N_s=9$ is the number of nuclei with determined skin values in Table~\ref{tab:skin}.  The hatched areas represent hulls for interpolations between the results from different Skyrme interactions that are within or barely miss the constraints.  The outer and inner hulls are obtained, respectively, when either only IAS or both IAS and skin constraints are applied.
}
\label{fig:aavlr}

\end{figure}

\begin{figure}
\centerline{\includegraphics[width=.64\linewidth]{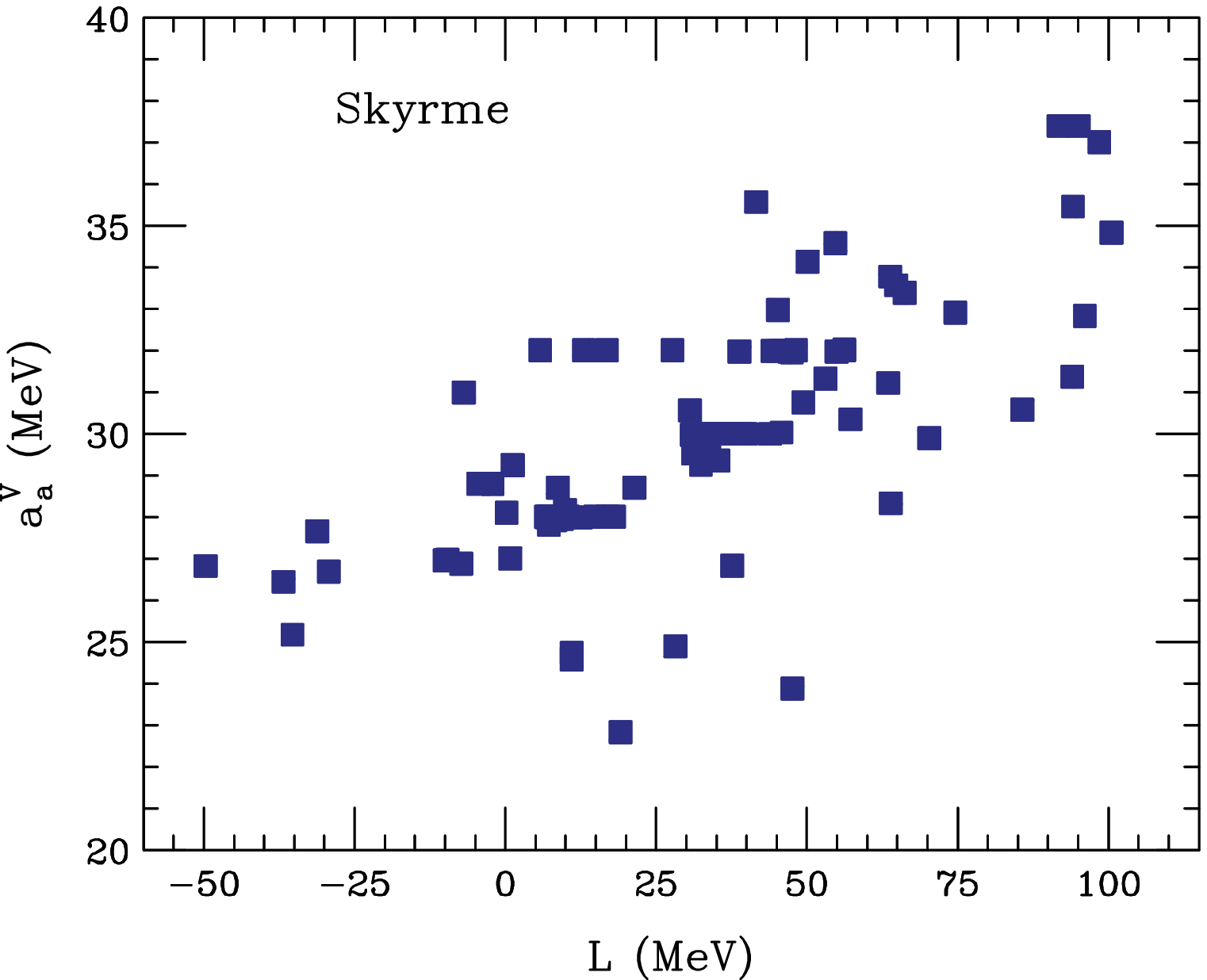}}
\caption{Skyrme interactions passing our stability criteria, represented in the plane of symmetry-energy parameters $a_a^V$ and $L$ at $\rho_0$, cf.~Eq.~\eqref{eq:SaL}.  Difference in the scales for the axes, compared to Figs.~\ref{fig:aavl} and~\ref{fig:aavlr}, should be noted.
}
\label{fig:aavlall}
\end{figure}

Even with the spreading resulting from relaxing the errors, the $a_a^V$ and $L$ parameters remain narrowly correlated along a line within the $a_a^V$-$L$ plane~\cite{PhysRevC.86.015803} in Fig.~\ref{fig:aavl}.  The latter line parallels a~narrow correlation line between the $a_a^V$ and $a_a^S$ values arrived at in \cite{Danielewicz:2003dd}, following a fit to the nuclear masses with a mass formula.  The first correlation is tied to the other, because $L$ is tied to $a_a^S$, cf.~I.  In the context of the correlation line, see further Refs.~\cite{lattimer_constraining_2012,lattimer_nuclear_2012}.
In Fig.~\ref{fig:aavlr}, we complement Fig.~\ref{fig:aavl} by showing the hull in $a_a^V$-$L$ plane for the interpolations between the results of Skyrme interactions that either satisfy or barely miss the IAS constraints.  Moreover,
in Fig.~\ref{fig:aavlall}, we show
for reference $a_a^V$-$L$ parameters for all Skyrme interactions in the literature, that pass our stability criteria and any interrelation between the parameters is far less pronounced than either in Fig.~\ref{fig:aavl} or in Fig.~\ref{fig:aavlr}.
  Besides the results following from our IAS analysis, Fig.~\ref{fig:aavl} displays the correlation between $S(\rho_0)$ and~$L$ values arrived at in microscopic calculations based on the Argonne V18 interaction~\cite{PhysRevC.51.38}, respectively by Vida\~na \etal\ \cite{PhysRevC.80.045806}, Taranto \etal~\cite{taranto_selecting_2013} and by Gandolfi~{\em et al.}~\cite{gandolfi_maximum_2011}.  While those microscopic results, testing the consequences of uncertainties in 3-body interactions, line up with our, it should be mentioned that Gandolfi~{\em et al.}, as common for microscopic calculations, define the symmetry energy as a difference between the energies for neutron and symmetric matter, rather than the coefficient in expansion with respect to squared asymmetry.  That difference in definition can give rise to a difference of $(1$--$2) \, \text{MeV}$ in $S(\rho_0)$, see~\cite{PhysRevC.80.045806} and I, and an~additional difference in~$L$.

While the constraints on the symmetry energy of uniform matter in Fig.~\ref{fig:symec}, following from~IAS, are quite narrow at moderately subnormal densities, their widening around normal density is quite disconcerting, though understandable.  Even the basic volume-surface formula \eqref{eq:aaVS} indicates that symmetry coefficients $a_a(A)$ stem from an interplay of the symmetry energy in the nuclear interior and in the surface where density changes.  In~\cite{Danielewicz:2003dd},  we found that the uncertainties in the phenomenological $a_a^V$-$a_a^S$ determination could be narrowed by combining the mass with the asymmetry-skin constraints.  Given the connection between $a_a^S$ and $L$, a similar narrowing of the constraints may be expected for the $a_a^V$ and $L$ parameters, see in particular~\cite{PhysRevC.83.044308}.

\include{table_skin}

\subsection{Auxiliary Constraints from Asymmetry Skins}

Asymmetry skins, quantified in terms of difference of rms radii between neutrons and protons,
\beq
\label{eq:Drnp}
\Delta r_\text{np} = r_\text{n}^\text{rms} - r_\text{p}^\text{rms} \, ,
\eeq
may be understood in terms of variation of the symmetry energy with density.  For the energy dropping with density, the nucleus can lower the net energy by expelling the neutron-proton asymmetry into the surface region.  In the simple macroscopic consideration~\cite{Danielewicz:2003dd}, the size of the skin directly reflects the ratio of the surface to net symmetry coefficient
\beq
\label{eq:DrnpEst}
\Delta r_\text{np} \simeq \frac{2}{3} \, r^\text{rms} \, \frac{a_a}{a_a^S \, A^{1/3}} \, \bigg( \eta - \frac{a_C}{12 a_a^V} \, \frac{Z}{A^{1/3}}       \bigg) \, ,
\eeq
where $r^\text{rms}$ is the nucleonic rms radius, $a_a$ is given by \eqref{eq:aaVS}, the subtraction from the relative asymmetry~$\eta$ represents a Coulomb correction to the symmetry-energy effects and
\beq
\frac{a_a}{a_a^S \, A^{1/3}} \equiv \frac{\frac{A^{2/3}}{a_a^S}}{\frac{A}{a_a}}
\eeq
represents the ratio of surface to net capacitances for asymmetry~\cite{Danielewicz:2003dd}.  The faster the drop of symmetry energy with density, or higher $L/a_a^V$, the lower $a_a^S/a_a^V$, cf.~I, and the larger~$\delta r_\text{np}$ according to \eqref{eq:DrnpEst}.  Notably, the macroscopic formula \eqref{eq:DrnpEst} predicts that the asymmetry skins test the absolute magnitude of the symmetry energy only through the Coulomb effects and otherwise reflect the {\em relative} magnitude of the characterization of the energy in the interior and in the surface. Given, however, the situation in Fig.~\ref{fig:symec}, with the bottleneck region of symmetry energy well constrained and poorly constrained normal density region, constraining additionally the pace of drop of the energy may significantly narrow the constraints within the normal region.

\begin{figure}
\centerline{\includegraphics[width=.8\linewidth]{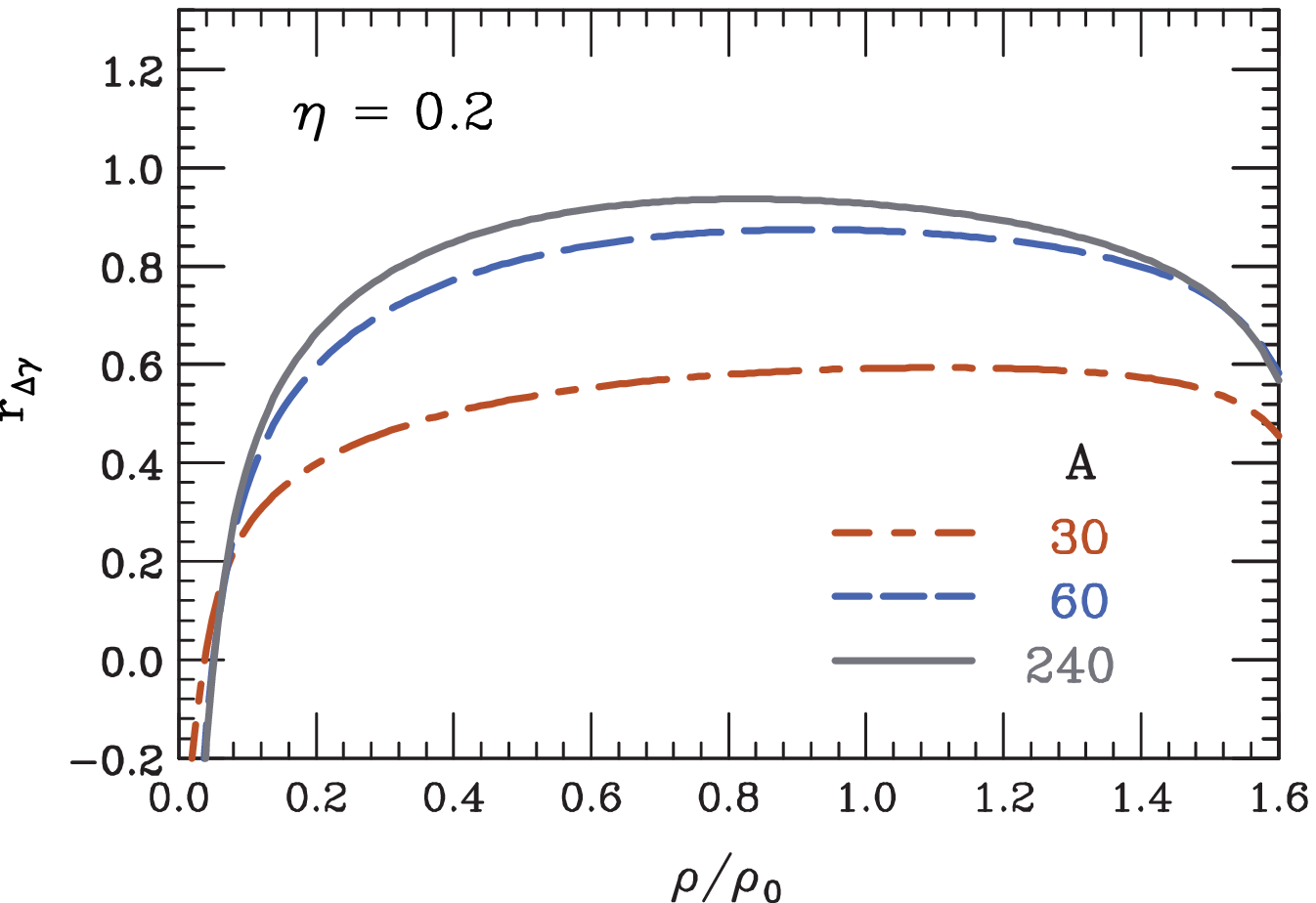}}
\caption{Pearson correlation coefficient, Eq.~\eqref{eq:rXY}, between the asymmetry skin, Eq.~\eqref{eq:Drnp}, for a~nucleus characterized by mass $A$ and asymmetry $\eta=0.2$, and the stiffness of symmetry energy in uniform matter, $\gamma(\rho)$ of Eq.~\eqref{eq:gammarho}, for Skyrme interactions, plotted at different indicated $A$ against the density~$\rho$.
}
\label{fig:rco}
\end{figure}

\begin{figure}
\centerline{\includegraphics[width=.8\linewidth]{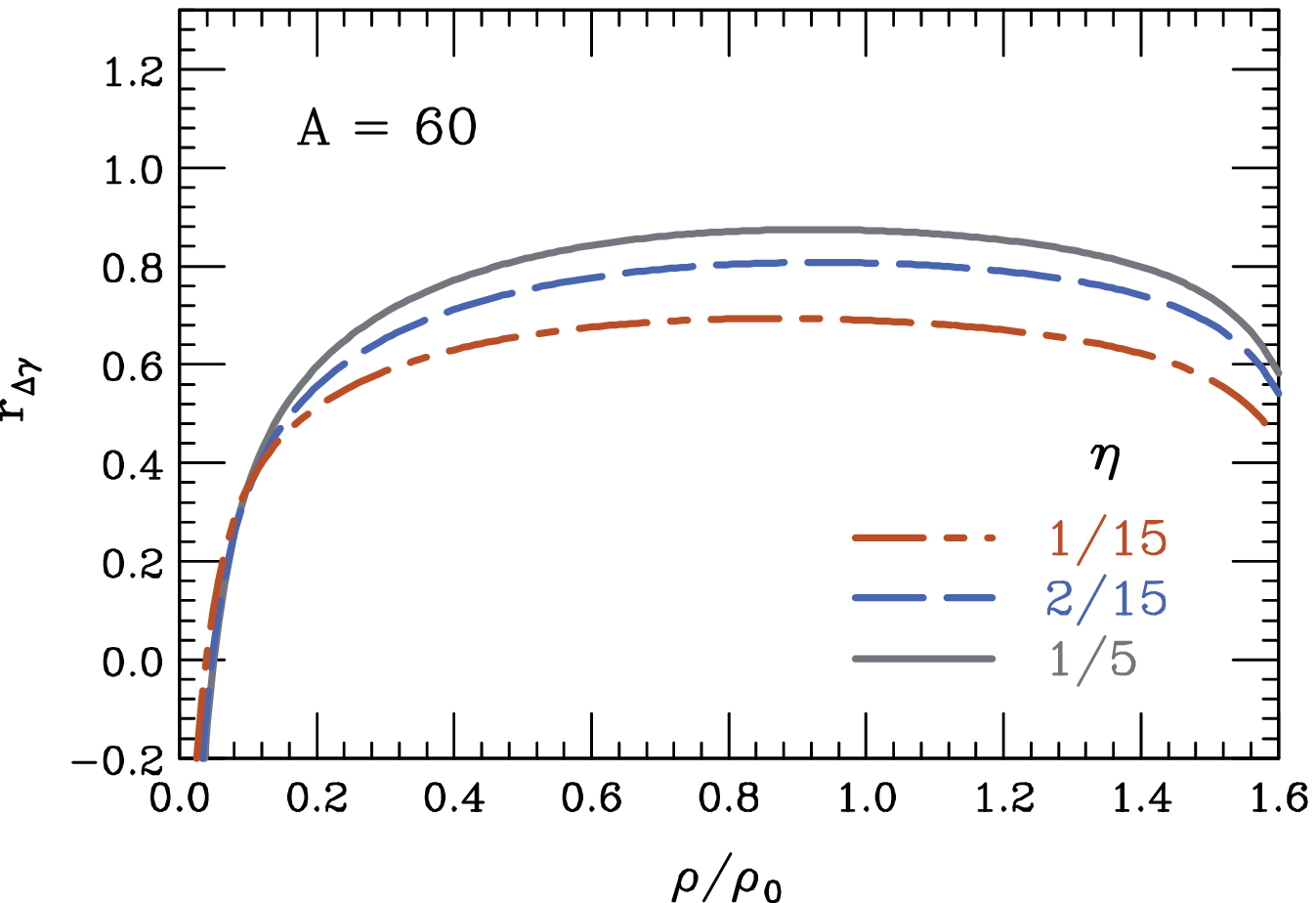}}
\caption{Pearson correlation coefficient, Eq.~\eqref{eq:rXY}, between the asymmetry skin, Eq.~\eqref{eq:Drnp}, for nuclei characterized by mass $A=60$ and different indicated asymmetry values, and the stiffness of symmetry energy in uniform matter, $\gamma(\rho)$ of Eq.~\eqref{eq:gammarho}, for Skyrme interactions, plotted against the density~$\rho$.
}
\label{fig:rcot}
\end{figure}

Following up on the qualitative consideration above, Figs.~\ref{fig:rco} and~\ref{fig:rcot} present the values of the Pearson coefficient r$_{\Delta \gamma}$ for the Skyrme interactions, between the skin size \eqref{eq:Drnp} and the dimensionless stiffness\footnote{After elasticity used for similar derivatives in the research literature.} of the symmetry energy defined as
\beq
\label{eq:gammarho}
\gamma (\rho) = \frac{\text{d} \, \log{S}}{\text{d} \, \log{\rho}} = \frac{\rho}{S} \, \frac{\text{d} \, S}{\text{d} \, \rho} \, ,
\eeq
playing the role of effective power in the power parametrization of the symmetry energy; at normal density
\beq
 L= 3 \gamma \, a_a^V \, .
\eeq
Figures~\ref{fig:rco} and~\ref{fig:rcot} both show the correlation coefficient against density $\rho$ for the stiffness and it is observed that the coefficient is relatively constant in the region of $(0.4$--$1.4) \, \rho_0$. The extension of that region to the supranormal densities obviously demonstrates the likely excessively simple form of the symmetry energy within the Skyrme interactions, cf.~Eq.~(101) of~I. Even at the quite typical asymmetry of $\eta=0.2$, the coefficient remains fairly high, r$_{\Delta \gamma} \sim 0.9$, within the coarsely constant-value region, and it depends weakly on $A$, dropping off only at $A < 60$.

The results in Figs.~\ref{fig:rco} and~\ref{fig:rcot} indicate that, provided the nuclear mass and asymmetry are both significant, the skin sizes can indeed constraint the stiffness of symmetry energy.  In~what follows, we exploit a variety of results on skins, from analysis of experimental data, to additionally limit the variation of symmetry energy with density.  The underlying measurements need to tie directly to the nucleonic radii and not to some other nuclear characteristics that get only correlated to the skins through structure theory.  We take $A \ge 48$ and we exclude results with excessive uncertainties compared to the likely magnitude of the skins.  In judging the quality of theoretical descriptions, such results of data analysis would act to inflate the quality of $\chi^2$ - in essence any symmetry energy in the theoretic calculations will yield skins close to zero for weakly asymmetric nuclei.  Based on \eqref{eq:DrnpEst}, we specifically demand that the errors on skins, for the results to be considered, meet the coarse condition
\beq
\delta_\Delta < 0.13 \, \eta \, r^\text{rms} \, ,
\eeq
which, in the end, turns out to be relatively forgiving with regard to the available data in the higher $A$-region.  When there are $N_v$ different skin values $\Delta_i$, $i = 1, \ldots, N_v$, claimed in the literature for a given nucleus, with uncertainties $\delta_i$, we examine the $\chi^2$ consistency of the set against the best estimate of the skin:
\beq
\overline{\Delta}_{np} = \frac{\sum_i \frac{\Delta_i}{\delta_i^2} }{\sum_i \frac{1}{\delta_i^2}} \, .
\eeq
If $\chi^2$ for the set,
\beq
\chi^2 = \sum_i  \frac{( \Delta_i - \overline {\Delta}_{np})^2}{\delta_i^2}  \, ,
\eeq
exceeds $N_v-1$, we augment the uncertainty for the skin, compared to the naive result from the claimed errors alone, so that the uncertainty of the skin from the combination becomes:
\beq
\label{eq:deltaDelta}
\delta_\Delta^2 = \text{max} \bigg[ \frac{1}{\sum_i \frac{1}{\delta_i^2}}, \frac{\overline{(\Delta- \overline{\Delta})^2}}{N_v-1}   \bigg] \, .
\eeq
Such renormalization is common in data analysis and can be derived from the assumption that claimed errors are underestimated by a common factor.  When the presence of an underestimation factor is relaxed at $\chi^2 > N_v-1$, the best estimate of that factor leads to the result \eqref{eq:deltaDelta}.  Other assumptions about error underestimation could be adopted, leading to bit different results for the best estimate of the skin and its uncertainty, but, given that the quoted errors do not vary much from one considered data analysis to another, further refinements are presumably not warranted.  The skins inferred from data, meeting the conditions listed earlier in this paragraph, as well as combined results, are provided in Table~\ref{tab:skin}.

\begin{figure}
\centerline{\includegraphics[width=.66\linewidth]{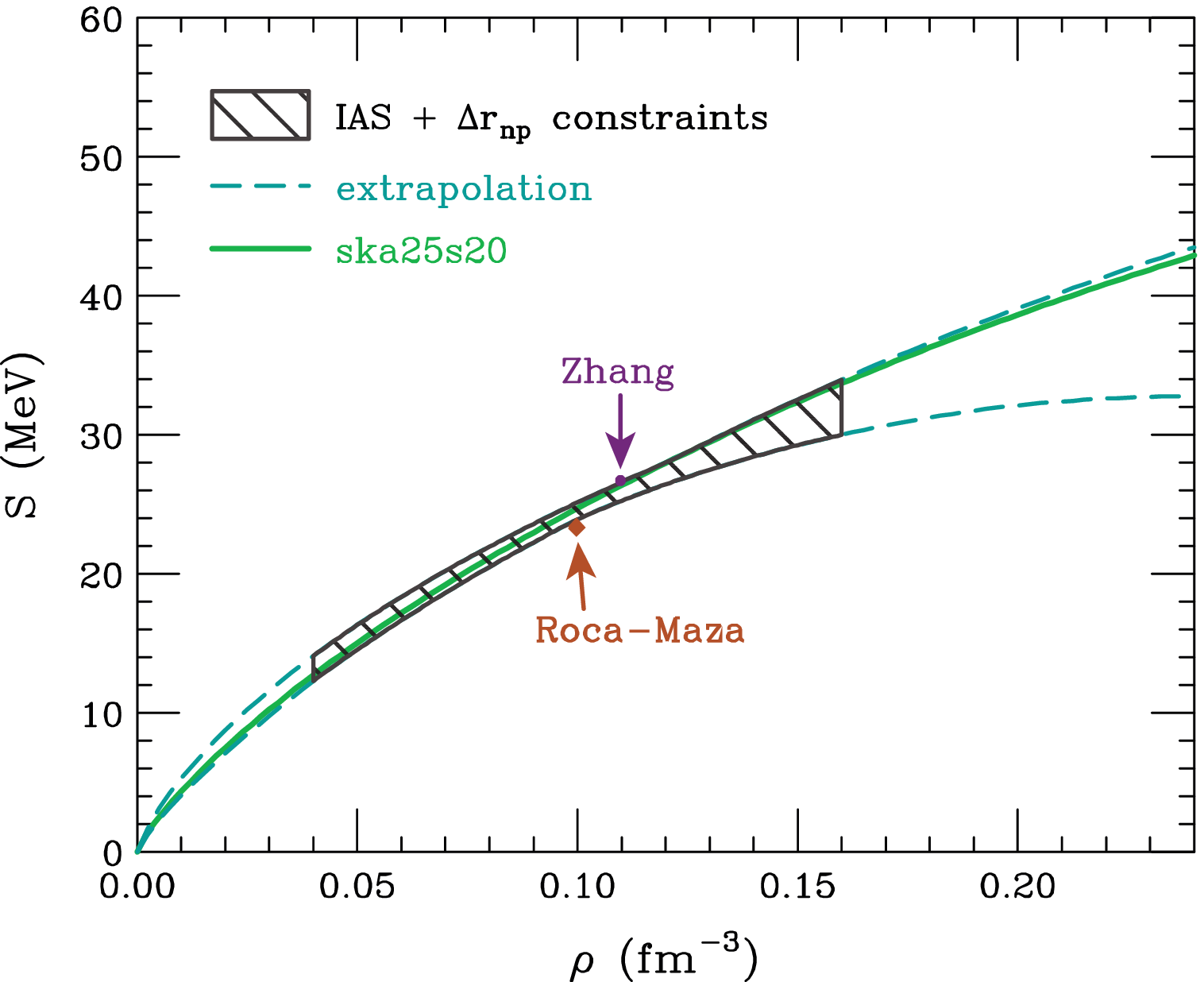}}
\caption{Symmetry energy in uniform matter as a function of density.  The hatched region represents our combination of IAS and skin constraints.  The short-dashed lines represent extrapolations of that region to supranormal, $\rho > \rho_0$, and low, $\rho < \rho_0/4$, densities.  The solid line represents the symmetry energy for the ska25s20 parametrization~\cite{brown08}.  The two symbols, circle and diamond, represent the values of symmetry energy at moderately subnormal densities deduced, respectively, by Roca-Maza \etal~\cite{RocaMaza:2012mh} and by Zhang~\etal~\cite{zhang_constraining_2013}.  The vertical sizes of those symbols represent the claimed errors on the deduced values of the symmetry energy.
}
\label{fig:symecd}
\end{figure}

\begin{figure}
\centerline{\includegraphics[width=.66\linewidth]{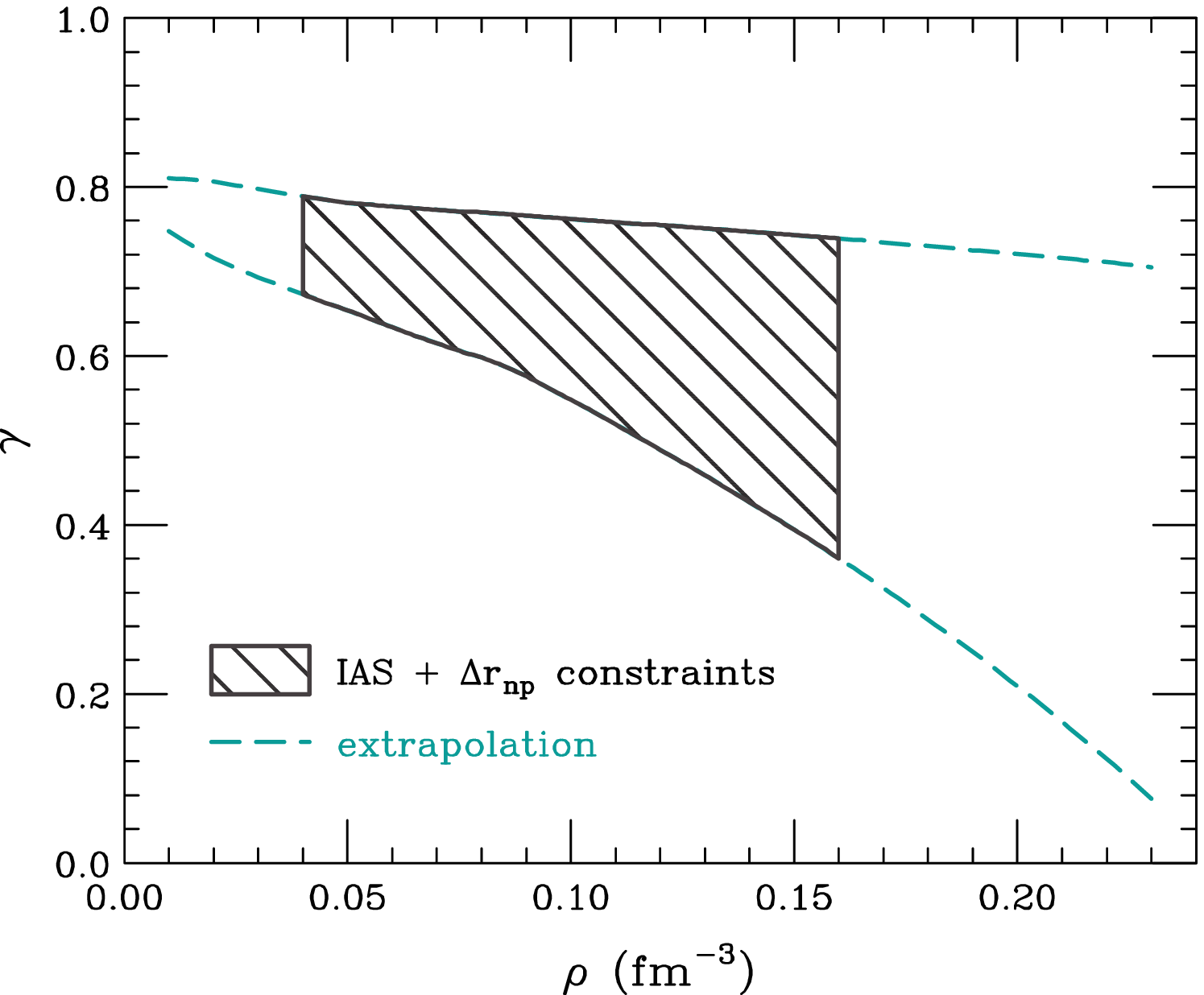}}
\caption{Stiffness of symmetry energy, Eq.~\eqref{eq:gammarho}, in uniform matter as a function of density.  The hatched region represents a combination of IAS and skin constraints.  The short-dashed lines represent extrapolations of that region to supranormal, $\rho > \rho_0$, and low, $\rho < \rho_0/4$, densities.
}
\label{fig:gamma}
\end{figure}

When we consider the ensemble of Skyrme interactions, we find that the minimal values of $\chi^2$ deviations, between the calculated asymmetry skins and those deduced from measurements in Table~\ref{tab:skin}, concentrate in the vicinity of the $a_a^V$-$L$ correlation line deduced from~IAS, see Fig.~\ref{fig:aavlr}.  Notably, the asymmetry skins are generally not used in adjusting the Skyrme parameters in the literature.  The minimal values of average deviations square are $\chi^2/N_s \sim 1.5$, where $N_s=9$ is the number of nuclei  in Table~\ref{tab:skin}.  While the minimal average square exceeds~1, the error renormalizations, carried out for two cases in Table~\ref{tab:skin}, underscore the difficulties in assessing errors of deduced asymmetry skins - it is plausible that the errors remain also underestimated for other nuclei where additional skin values are not available for a cross-comparison.  Continuing with the idea of $\chi^2$-renormalization, we indicate in the $a_a^V$-$L$ plane in Fig.~\ref{fig:aavlr} those Skyrme interactions for which  $\big(\chi^2 - (\chi^2)_\text{min} \big) < (\chi^2)_\text{min}/N_s$.
Two of the interactions, ska25s20 and ska35s20 by Brown~\cite{brown08}, simultaneously meet the IAS and skin constraints.  Otherwise, as before, we interpolate results between the Skyrme interactions that either meet or barely miss the two constraints and we arrive at the narrower of the hulls in Fig.~\ref{fig:aavlr}.   The skin values representing that hull are provided in the last column of Table~\ref{tab:skin}.  It can be observed that the arrived at errors are fairly tight.  Finally, Figs.~\ref{fig:symecd} and~\ref{fig:gamma}  display results on the symmetry energy as a function of density, representing the Skyrme interactions and their interpolations, which meet both the IAS and skin constraints.  Figure~\ref{fig:symecd} displays the values of the energy and Fig.~\ref{fig:gamma} - the values of the symmetry-energy stiffness.

As is apparent in comparing Figs.~\ref{fig:symec} and~\ref{fig:symecd}, the very narrow constraints on the symmetry-energy values at densities below $0.13 \, \text{fm}^{-3}$ become even more narrow after applying the skin constraints, never exceeding $\pm 1.1 \, \text{MeV}$ in that subnormal density region.  In~the vicinity of the normal density, the constraints narrow too, relative to IAS alone, shrinking down to the potentially useful range of $\pm 2.2 \, \text{MeV}$ at $0.16 \, \text{fm}^{-3}$: $S(\rho_0) = (30.0\,$--$\,34.3) \, \text{MeV}$.  The comparison of Figs.~\ref{fig:gamma} and~\ref{fig:symecd} demonstrates how much more difficult it is to establish constraints on the slope relative to the constraints on the values.  Still for the slope, as for the values, the situation is better at deeper subnormal densities than in the vicinity of normal.  Irrespectively of the difficulties, Fig.~\ref{fig:gamma} indicates that an upper limit of about 0.8 can be placed on the stiffness of the symmetry energy at subnormal and moderately supranormal densities.  Interestingly, the stiffness of 2/3, such as a for the nonrelativistic Fermi-energy contribution to the symmetry energy, expected to dominate at $\rho \rightarrow 0$, is consistent with the constraints at most densities.  Given, however, the limits on $S$ around $\rho_0$ and elsewhere, the stiffness needs to be higher than 2/3 within some density range.  Figure~\ref{fig:gamma} suggests that the stiffness is more likely to exceed 2/3 at deeply subnormal density region than around normal density.  The drop in stiffness with density would be consistent with a contribution of 3N forces to the symmetry energy.  Thus, for nuclear saturation the predominant effect of~3N forces must be repulsive.  However, in combination of the Pauli principle and the short-range of nuclear interactions, the contribution of three-nucleon forces to net energy should be stronger in symmetric than asymmetric matter.  Hence, the 3N forces should come in with a negative sign into the symmetry energy.  The rise in the importance of the 3N forces with density should then contribute to a~decrease in the slope of symmetry energy with increase in density.\footnote{Notably, in Ref.~\cite{gandolfi_maximum_2011} the opposite seems to be true, as the symmetry energy and its slope increase as the strength of the 3N forces is increased, but there the energy of symmetric matter is kept fixed, and only the energy of neutron matter is changed, when the strength of the 3N forces is changed.}  Weakening of 2N interactions with an increase in relative momenta may have a similar effect on both the symmetry energy and energy of symmetric matter as the 3N forces.

\subsection{Comparison to Other Results in the Literature}

\begin{figure}
\centerline{\includegraphics[width=.66\linewidth]{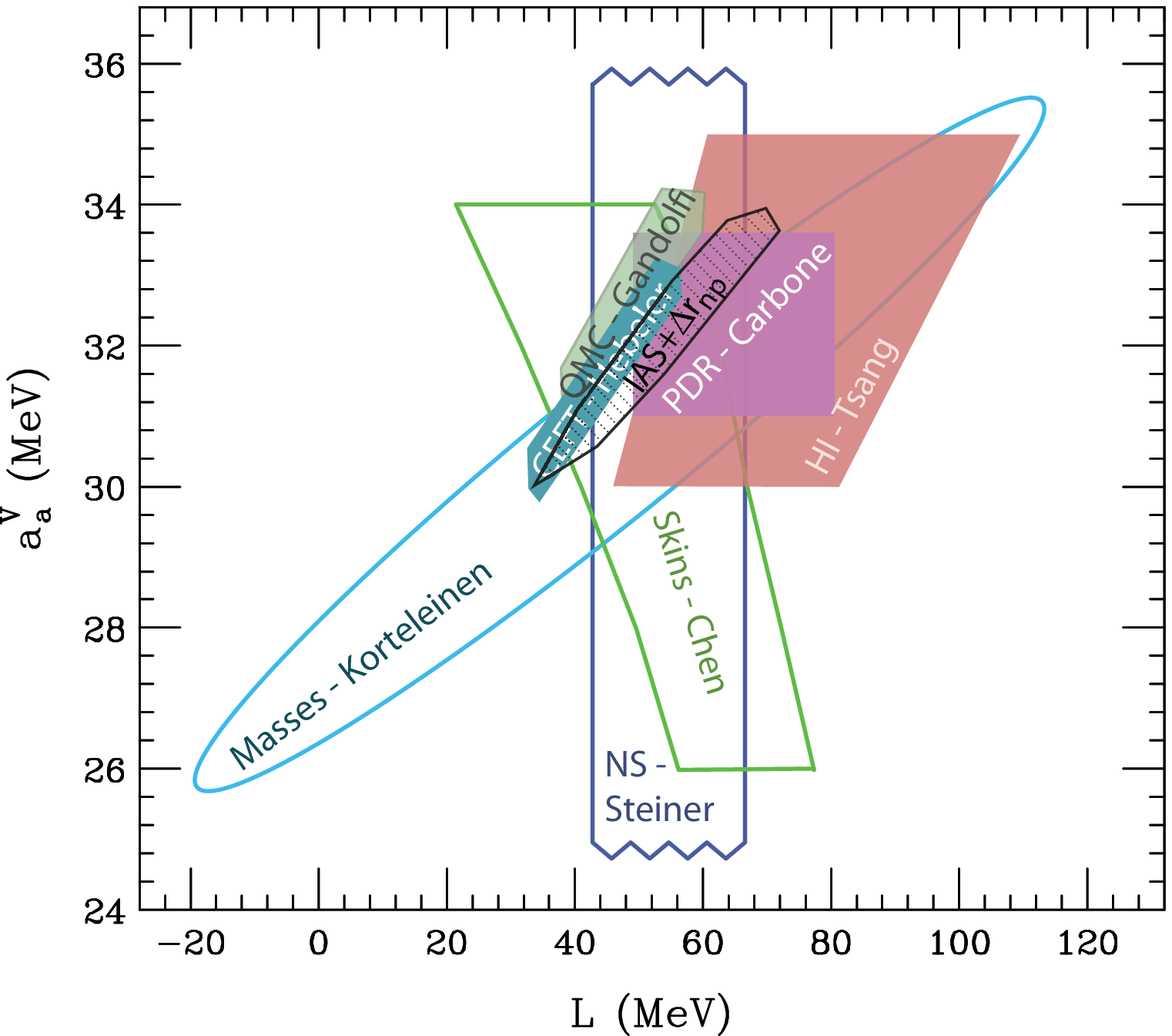}}
\caption{Constraints, from different sources, on the symmetry-energy parameters at $\rho_0$, $a_a^V \equiv S(\rho_0)$ and $L$, after \cite{PhysRevLett.102.122701} and \cite{lattimer_constraining_2012}.  Included are predictions from neutron-matter calculations by Gandolfi~\etal~\cite{gandolfi_maximum_2011}, within QMC, shown already in Fig.~\ref{fig:aavl}, and by Heberle \etal~\cite{hebeler_constraints_2010}, within chiral effective field theory (CEFT).  From Fig.~\ref{fig:aavl}, we further reproduce here our own IAS + skins constraints.  Other illustrated constraints, deduced from observables, include those from skins by Chen \etal~\cite{chen_density_2010}, from nuclear masses by Korteleinen \etal~\cite{kortelainen_nuclear_2010}, pygmy dipole resonance (PDR) by Carbone~\etal~\cite{carbone_constraints_2010}, heavy-ion collisions (HI) by Tsang~\etal~\cite{PhysRevLett.102.122701}, and from neutron-star (NS) observations by Steiner \etal~\cite{0004-637X-722-1-33}.
}
\label{fig:aavlcomp}
\end{figure}

In Fig.~\ref{fig:aavlcomp}, we compare our constraints on the symmetry-energy parameters at $\rho_0$, $a_a^V$~and~$L$, to selected constraints recently arrived in literature~\cite{gandolfi_maximum_2011,hebeler_constraints_2010,chen_density_2010,kortelainen_nuclear_2010,carbone_constraints_2010,PhysRevLett.102.122701,0004-637X-722-1-33}.
It is apparent that a good deal of convergence emerges between the constraints arrived by various means.  Rather astounding is the parallel nature and overlap of our constraints and those stemming from nuclear-matter calculations~\cite{gandolfi_maximum_2011,hebeler_constraints_2010} displayed in Fig.~\ref{fig:aavlcomp}, and also \cite{taranto_selecting_2013,PhysRevC.80.045806} in~Fig.~\ref{fig:symec}.

On the first sight, the similarity between the $a_a^V$-$L$ correlations produced by IAS constraints and by microscopic calculations might seem just a coincidence, but on second look, the similarity can be understood.  Thus, $a_a$-coefficients can be reproduced with symmetry energies where strong variations in the normal-density region are compensated by opposite but much more moderate variations in the farther subnormal region, over a wider range of densities than close to normal.  In the microscopic calculations, the 2N low-momentum interactions are generally settled and variations are explored for the 3N interactions (or higher momenta).  With this, the lower-density part of the symmetry energy is subjected to far lesser uncertainty than higher density.  In both cases then the $a_a^V$-$L$ correlations emerge in a situation, where the lower-density part of the symmetry energy varies only moderately and the stronger variations are reserved for the normal and supranormal region.

Concerning subnormal densities, narrow constraints, indicated in Fig.~\ref{fig:symecd}, have been put on the symmetry energy $S(\rho)$, at $\rho \approx 0.11 \, \text{fm}^{-3}$, by Zhang and Chen~\cite{zhang_constraining_2013} using skins and ground-state binding energies, and at $\rho = 0.1 \, \text{fm}^{-3}$, by Roca-Maza~\etal~\cite{RocaMaza:2012mh} using properties of isovector giant
quadrupole resonance (also earlier by Trippa~\etal~\cite{trippa_giant_2008} using properties of giant dipole resonance).  As is apparent in the figure, those constraints are mutually contradictory and our own constraints are lodged in-between.  Both Refs.~\cite{RocaMaza:2012mh} and~\cite{zhang_constraining_2013} assume an equality between the strongly correlated $S(\rho)$ and $a_a(A)$, disavowed in the context of Fig.~\ref{fig:a240105}.

\begin{figure}
\centerline{\includegraphics[width=.66\linewidth]{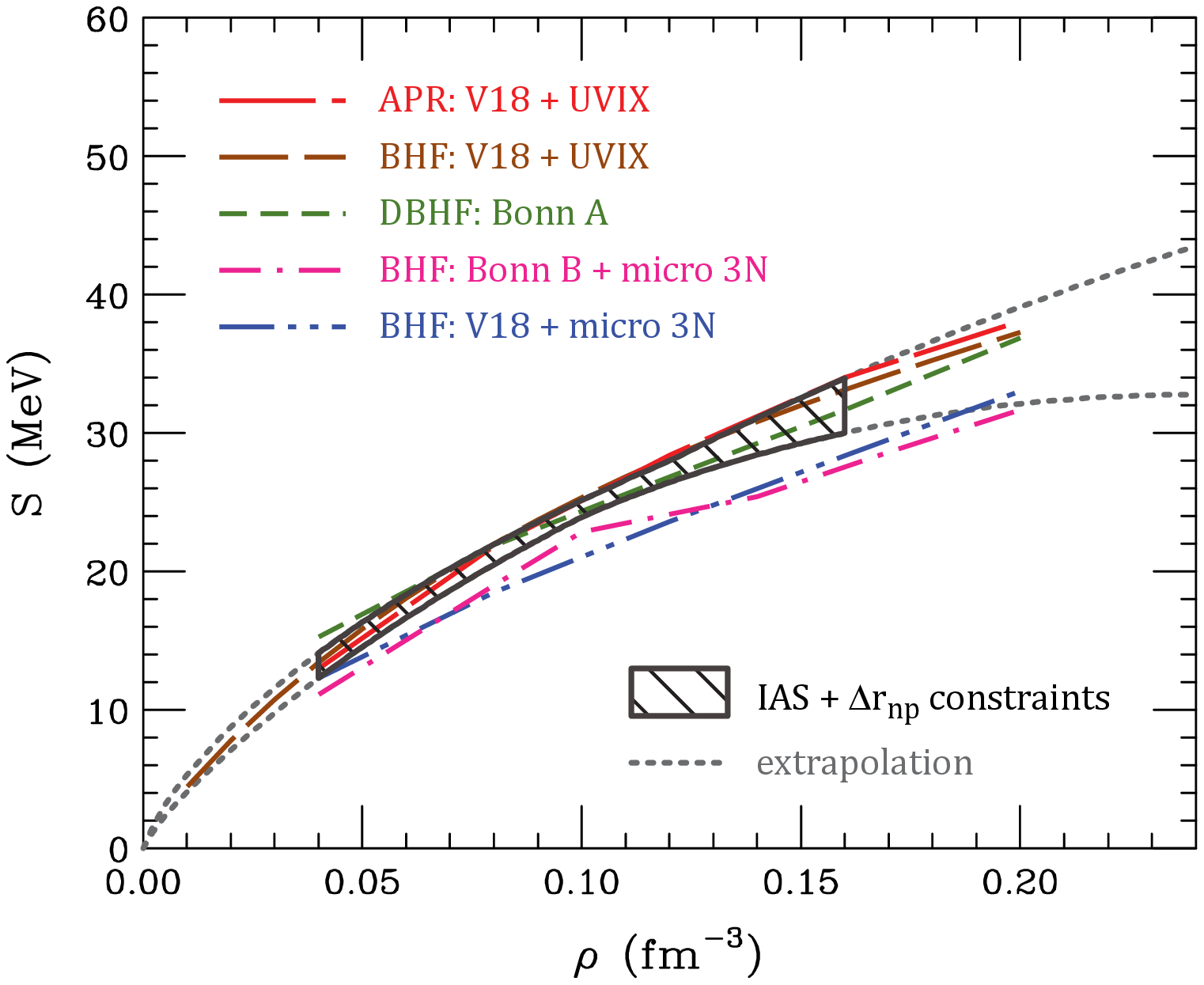}}
\caption{Symmetry energy in uniform matter as a function of density.  The hatched region represents a combination of IAS and skin constraints.  The dotted lines represent extrapolations of that region to supranormal, $\rho > \rho_0$, and low, $\rho < \rho_0/4$, densities.  The remaining lines represent symmetry energy within traditional microscopic calculations \cite{taranto_selecting_2013}.  In the microscopic calculations, it is easiest to obtain the symmetry energy as a difference between the energies of neutron and symmetric matter.  The long- and medium-dashed lines represent calculations with Argonne V18 NN and 3N UVIX interaction \cite{PhysRevC.51.38}, within variational \cite{akmal_equation_1998} and BHF calculations~\cite{taranto_selecting_2013}, respectively.  In the variational case, a~phenomenological adjustment of the results was made \cite{akmal_equation_1998}, but not in BHF.  The short-dashed lines represents results from relativistic Dirac-Brucker-Hartree-Fock (DBHF) method with Bonn A NN interaction \cite{gross-boelting_covariant_1999}.  The dash-dotted and dash-double-dotted lines represent the BHF calculations with a microscopic 3N interaction and either Bonn B or V18 interaction \cite{taranto_selecting_2013}.
}
\label{fig:bruco}
\end{figure}

Regarding the whole subnormal region, there had been, obviously, different microscopic calculations carried out in the past.  In Fig.~\ref{fig:bruco} we compare our IAS + skins constraints on $S(\rho)$ to the microscopic results from \cite{taranto_selecting_2013}, which include those authors' own results and other from literature.  The situation in the comparison seems quite good, since 3 out of~5 microscopic prediction are in an essential agreement with our constraints.   There can be questions, however, as to what to conclude further from the comparison.   On account of significant momentum-dependence of interactions in relativistic DBHF calculations~\cite{gross-boelting_covariant_1999}, consistent with heavy-ion (HI) flow analyses (e.g.~\cite{danielewicz_determination_2000}), the saturation of nuclear matter and description of symmetry energy do not require 3N interactions.  The variational calculations with V18 interaction, denoted APR in the figure, incorporate relativistic corrections, while the analogous BHF calculations do not.  Those corrections, however, largely cancel~\cite{akmal_equation_1998} in arriving at the symmetry energy as difference of energies for neutron and symmetric matter.

As discussed, skins test the strength of symmetry energy in the subnormal relative to normal density region.  Our result for the skin of $^{208}$Pb, of $0.179 \pm 0.023 \, \text{fm}$, from combining IAS constraints and all skin results, is not particularly controversial, consistent with many other conclusions in the literature.  However, an interesting situation develops in comparing the low to high mass end, in that the nonrelativistic SHF theory can predict significantly higher skin size for $^{48}$Ca, such as our $0.218 \pm 0.015 \, \text{fm}$, while the relativistic mean-field theories yield~\cite{piekarewicz_electric_2012} smaller or similar skin size for $^{48}$Ca compared to $^{208}$Pb,  in the interesting range of predictions.  It could be that the stronger shell effects, at the low-mass end, affect the systematic ties between the symmetry energy and skin size even more than evidenced in the Pearson coefficient in Fig.~\ref{fig:rco}.  I.e.~with the inclusion of different types of models, the drop-off of the Pearson coefficient at the low-mass end could be even more dramatic.

\begin{figure}
\centerline{\includegraphics[width=.66\linewidth]{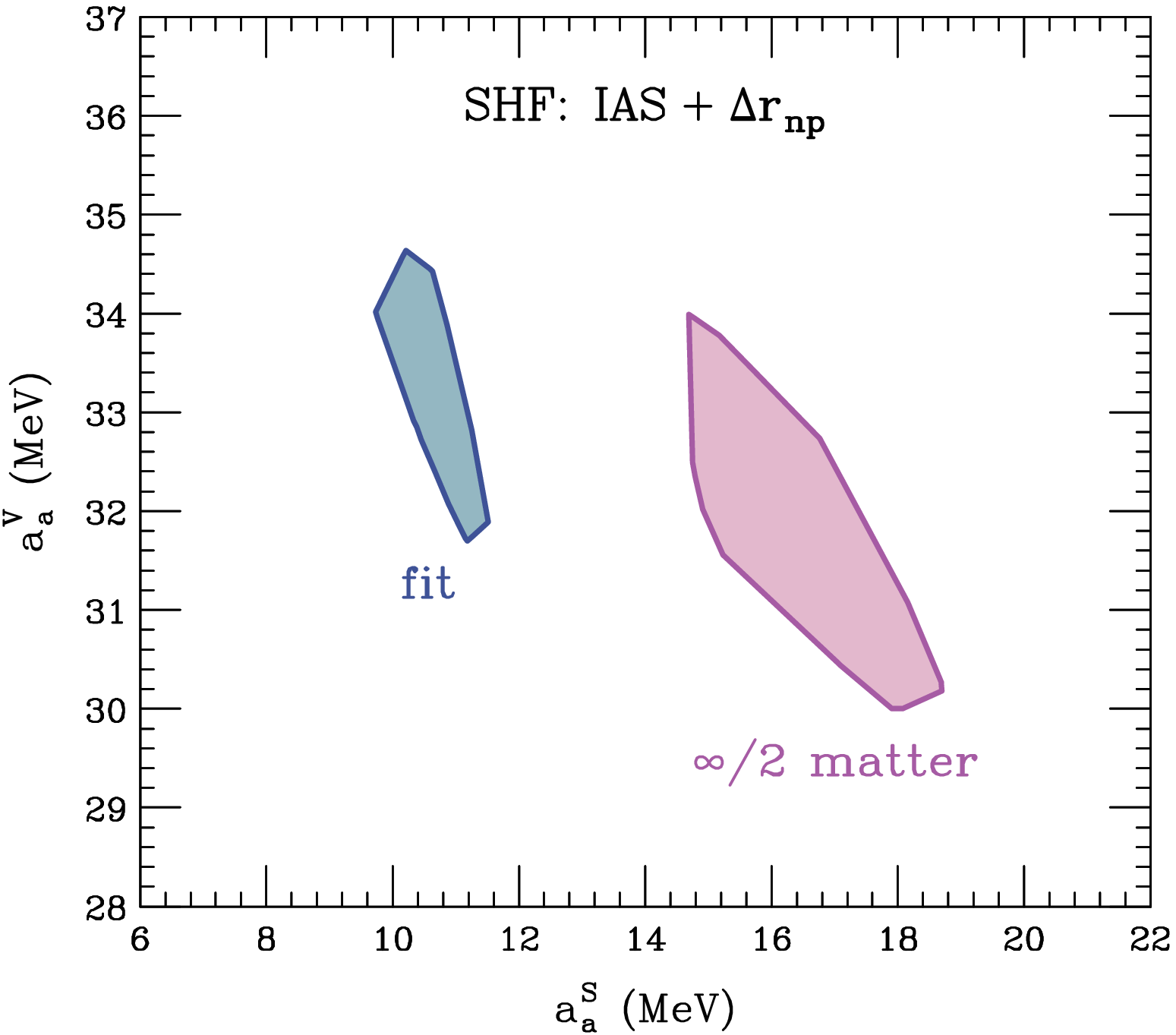}}
\caption{Within the plane of volume symmetry-energy coefficient $a_a^V$ vs surface $a_a^S$, the two hatched areas represent the Skyrme interactions and their interpolations, which conform with the IAS and skin constraints.  The left area shows the parameter values resulting from fitting the mass-dependence of the symmetry coefficients with a volume-surface formula, for the specific interactions and their interpolations.  The right area shows the corresponding coefficient values for the half-infinite matter.
}
\label{fig:AVAS}
\end{figure}

The final issue, that we need to address here, is of the comparison to our own preliminary IAS results from I, that have been quoted by others.  In~I, we examined the properties of half-infinite matter, arriving at values of $a_a^S$ for different Skyrme interactions, aside from the trivial~$a_a^V$.  We then used our preliminary fits to IAS results with the macroscopic volume-surface formula, and the $a_a^S$-$L$ correlation from half-infinite matter, to arrive at the nuclear-matter characteristics of $a_a^V = (31.5 \,$--$\, 33.5) \, \text{MeV}$, $a_a^S = (9.5 \,$--$\, 12) \, \text{MeV}$ and $L = (78 \,$--$\, 111) \, \text{MeV}$.  In doing so, we did not appreciate (a)~the impact that the shell corrections can have on the extracted values of $a_a(A)$ and (b)~the discrepancy between the values of~$a_a^S$ fitted to reproduce the~range of finite systems and the values of $a_a^S$ for half-infinite matter.  Uncertainties in the shell corrections give rise to a spread in the fitted values of $a_a^V$ and $a_a^S$ which remain, in spite of the spread, highly correlated.  In describing $a_a(A)$ in terms of SHF and employing the asymmetry skins as auxiliary constraints, we effectively narrowed the spread of coefficient values back towards the original estimate of the spread, prior to accounting for the impact of shell corrections.  Still, the discrepancy between the dependence $a_a(A)$, calculated for finite systems, and that expected on the basis of the half-infinite matter calculations combined with the macroscopic volume-surface formula, tends to be quite substantial, as evidenced in Fig.~\ref{fig:Bed_smple}.  Analogous discrepancy, between the expected and calculated dependence $\frac{E_0}{A}(A)$ for symmetric matter, tends to be far more limited, as evidenced again in Fig.~\ref{fig:Bed_smple}.  Faster  dependence of $a_a$ on mass, than expected in the volume-surface formula with parameters from half-infinite matter, leads to $a_a^S$ underestimated on the basis of the fit, see Fig.~\ref{fig:AVAS}, and to overestimated~$L$ from $a_a^S$-$L$ correlation.  Surprisingly, the volume parameter $a_a^V$ turns out be reasonably well estimated within the volume-surface fit.  In terms of specific value ranges, our current estimates for half-infinite matter are of $a_a^V = (30.2 \,$--$\, 33.7)\, \text{MeV}$ and $a_a^S = (14.8 \, $--$\, 18.5) \, \text{MeV}$, with the coefficients anticorrelated along the diagonal of their combined region.  We estimate the slope to be in the range $L= (35\,$--$\,70)\, \text{MeV}$, with $L$ strongly correlated with $a_a^V$ along the diagonal of their combined region.  More detailed information, though, can be found in terms of hulls in Figs.~\ref{fig:aavl}, \ref{fig:aavlcomp} and~\ref{fig:AVAS}.

\section{Conclusions}

To sum up, in this paper we extracted symmetry coefficients on a nucleus-by-nucleus and on isobaric chain-by-chain basis, by following charge invariance of nuclear interactions and by fitting excitation spectra (Sec.~\ref{sec:SymIAS} and Appendix~\ref{apx:Stat}) to states being either isobaric analogs of ground states or of low-lying states of other nuclei within the isobaric chain.  We~assumed that the specific excitation energies reflect a combination of symmetry energy and microscopic effects.  The symmetry coefficients were extracted when either correcting or not for the microscopic effects, with the microscopic corrections taken from literature.  No~matter which set of corrections from the literature we used, and even without the corrections, we~found the general tendency for the symmetry coefficients to decrease with nuclear mass, from $a_a \sim 22 \, \text{MeV}$ at $A \sim 240$ down to $a_a \sim 15 \, \text{MeV}$ at $A \sim 20$.  From the three employed sets of microscopic corrections, the median and most systematically changing results for the symmetry coefficients were produced by the corrections developed by Koura \etal~\cite{Koura:2005}.  We~found no evidence for a~dependence of the coefficients on charge number within an isobaric chain, consistently with charge invariance. (For heavy nuclei, the coefficients should be impacted by Coulomb effects, but the the stretches of the isobaric chain we had at hand for investigation were relatively short.)  Except for the heaviest nuclei, where detailed assessment was difficult, we found support, when employing the microscopic corrections~\cite{Koura:2005}, for the presumption that the symmetry-energy contribution to the energies is proportional to the net isospin squared.  For masses $A \gtrsim 30$, we found that the mass-dependence of the symmetry coefficient could be well described in terms of a macroscopic volume-surface formula, but details in the combination of volume and surface symmetry coefficients depended on specific microscopic corrections employed.  After reaching for SHF calculations (Sec.~\ref{sec:SHF}), and including auxiliary information from asymmetry skins (Sec.~\ref{sec:constraints}), one acceptable combination of the coefficients, that represents the left hull in Fig.~\ref{fig:AVAS}, becomes $a_a^V = 33.2 \, \text{MeV}$ and $a_a^S = 10.7 \, \text{MeV}$.

Within the Hohenberg-Kohn functional approach, we extended (Sec.~\ref{sec:HohenbergKohn}) our results from~I, by examining an interplay of the symmetry and Coulomb energies within the energy of a~system.  We demonstrated that, while the symmetry energy and Coulomb terms couple, for a~changing third isospin component, in a manner that potentially impedes the determination of the symmetry coefficient from energy differences between associated states, no such coupling, to within the interesting order, is present for changing transverse isospin components.  That lack of the coupling boosts the theoretic case for the determination of symmetry coefficients from excitation energies to IAS of ground states of other nuclei in the isobaric chain, as the excitation leaves the third component of isospin unaltered, while changing the square of transverse isospin components.  The Coulomb interactions can principally affect, though, the magnitude of the symmetry coefficient for a~given nucleus.  Importantly, the asymmetric density (neutron-proton difference) will adjust itself, in the lowest-energy state, not just to symmetry-energy but also to Coulomb effects.  The impact of the Coulomb interactions on the asymmetric density simplifies due to the long-range nature of those interactions.  The latter allowed us to arrive at formulas which made it possible to read off the symmetry coefficient from the shape of the asymmetric density when available.

We employed our Hohenberg-Kohn results in a~combination with spherically-symmetric SHF calculations~\cite{Reinhard:1991,reinhard:014309}.  For the Skyrme interactions, we identified a variety of absolute short-wavelength instabilities (Appendix~\ref{Appendix:SkyrmeStability}).  For individual interactions, those instabilities could be even more pronounced at moderate wavelengths~\cite{Lesinski:2006cu}.  Our search for the instabilities was forced by practicalities of the SHF calculations.  In our further considerations, we retained only those interaction parametrizations that were long-wavelength stable and that were either short-wavelength stable or for which the absolute short-wavelength instabilities occurred for sufficiently far-away densities to be triggered.  We exploited the SHF calculations in two different way.  On one hand, we tested (Sec.~\ref{sec:SHF}) the
general expectation that the mass-dependence of symmetry coefficients reflects the density-dependence of symmetry energy in uniform matter, combined with density variation across nuclear surface, and that the fits to mass-dependence can produce features of half-infinite nuclear matter in a~model-independent way.  Surprisingly, we found that the symmetry energies, which were widely varying in their density dependence in nuclear matter, were associated in the SHF calculations with the relatively similar mass dependencies for symmetry coefficients.  This finding necessarily limits the discriminating power of the mass dependence of the symmetry coefficients found from IAS.  Another finding was of significant discrepancies between the actual mass dependencies of coefficients from SHF calculations and the expectations based on the macroscopic volume-surface formula and half-infinite matter SHF calculations of~I.  That finding limits the principal ability to learn about nuclear matter from the symmetry coefficients, in a model-independent manner.

Following the first-stage findings within SHF, we proceeded to detailed comparisons between the coefficients extracted from excitations to IAS and from the SHF calculations (Sec.~\ref{sec:constraints}).  Assessment with Pearson coefficients indicated that we should be able to probe the symmetry-energy values in the vicinity and below the bottleneck density value of $\sim 0.11 \, \text{fm}^{-3}$.  Indeed, the comparison of results from IAS data and calculations produced
tight ($\pm 2.4 \, \text{MeV}$) constraints on the energy values at densities below $0.13 \, \text{fm}^{-3}$.  However, towards the normal density, the constraints widened dramatically.  Essentially, the typical values of symmetry coefficients for nuclei could be arrived at with relatively low values of symmetry energy at normal density, combined with slow changes in the subnormal region, and with rather high values of symmetry energy, but dropping rapidly in the subnormal region.  Crossing of different continuous behaviors, on its own, gives rise to the bottleneck region.  Given the thinning in the sample of Skyrme interactions meeting IAS (and stability) constraints, we had to resort to interpolations between results of the Skyrme interactions for which the results differed little.  Eventually, tailored interactions need to be developed, filling densely the region of immediate vicinity of the constraints.

With the above, even when the constraints widen around normal density, the value $a_a^V$ and slope parameter $L$ of the symmetry energy remain strongly correlated.  To limit the uncertainty around normal density, we reached for the empirical values of asymmetry skins, determined using hadronic probes.  The skins reflect the pace of drop in the symmetry energy across the nuclear surface, since that drop makes it energetic advantageous to push out more of the net asymmetry into the surface.  Upon incorporating the skin measurements, we managed to limit the uncertainty range for the symmetry energy at normal density to $a_a^V = (30.2 \,$--$\, 33.7)\, \text{MeV}$ and for the slope parameter to $L= (35\,$--$\,70)\, \text{MeV}$, with those values strongly correlated along the diagonal of their combined region.  Our~correlation ridge turns out to be close to the correlation ridges arrived at within the theoretic QMC and CEFT calculations.  Inclusion of the skin constraints narrows also uncertainties in the symmetry energy values at subnormal densities, down to $\pm 1.1 \, \text{MeV}$ below $0.13 \, \text{fm}^{-3}$.  Several microscopic calculations of the symmetry energy in the literature, including variational, BHF and DBHF, fall well within our constraint region for the density dependence of the energy.

 A~byproduct of the incorporating of the ensemble skin constraints are the narrowed SHF predictions for the skins of individual nuclei, including $0.179 \pm 0.023 \, \text{fm}$ for $^{208}$Pb.  It should be mentioned that the skin values from hadronic probes are not without controversies.  Indeed we ourselves needed to employ error renormalizations when dealing with the empirical skins and with their consequences and there might be model SHF biases involved there too.  In~this context, an independent assessment of the skins, using electroweak observables \cite{prex_collaboration_measurement_2012}, is highly desired.

\acknowledgements
We benefited from different forms of communication with a number of colleagues and from direct assistance.  In particular, we are grateful to P.-G.~Reinhard for letting us use his code, for spherical SHF calculations of nuclei of arbitrary size, and for explanations on using that code.  We thank Jirina Rikovska Stone for the collaboration in maintaining the database of Skyrme parametrizations.  We thank Brent Barker and Jun Hong for critical remarks helping us to improve this work.  We further thank the HiRA Group and Jacek Dobaczewski for stimulating discussions.   We particularly benefited moreover from communications with Augusto Macchiavelli, Sanjay Reddy and Fiorella Burgio.  This work was supported by the U.S.\ National Science Foundation under Grants PHY-0800026 and PHY-1068571, and by JUSTIPEN under the U.S.\ Department of Energy Grant DEFG02-06ER41407.

\appendix
\section{Statistical Analysis of Excitation Energies}
\label{apx:Stat}

We discuss here the details of our statistical strategy in analyzing measured excitation energies to ground-state IAS.  Let $E_i'$, $i=1, 2, \ldots, n$, represent the energy, corrected for microscopic effects, of an $i$'th considered excited IAS state, within the spectrum of a given nucleus, $E_i' \equiv (E_\text{IAS}')_i$, and let $E_0'$ represent the corrected energy of the $T= |T_z|$ state, normally the ground state.  Each of the energies contains some unaccounted microscopic contribution $\delta_i$, of unknown sign, that fluctuates from a state to a state with a typical root-mean-square value $\delta_\text{rms}$ that may depend on nuclear mass and charge numbers and on magnitude of excitation energy.  We are concerned here with the impact of those fluctuations onto the conclusions on symmetry energy, rather than with uncertainties in the measurement of energies of the states that are usually significantly smaller.

With the above, the covariance matrix from averaging over likely values of fully corrected excitation energies, $E_i^{\prime\prime} = E_i^\prime - \delta_i$, is
\beq
\left< \big(E_i'-E_i^{\prime\prime} \big)\left(E_j' - E_j^{\prime\prime}\right) \right> = \delta_\text{rms}^2 \, \delta_{ij} \, ,
\eeq
and the covariance matrix for excitation energies is
\beq
\left<\big[ (E_\text{IAS}^{*\prime})_i - (E_\text{IAS}^{*\prime\prime})_i\big] \big[(E_\text{IAS}^{*\prime})_j - (E_\text{IAS}^{*\prime\prime})_j\big]   \right>
= \delta_\text{rms}^2 \, \left( \delta_{ij} + 1 \right) \equiv \Sigma_{ij} \, ,
\eeq
where we assume a weak dependence of the fluctuations on excitation energy.  Different excitation energies above get correlated through the fluctuation in the ground-state energy subtracted from the energies of IAS states.

When fitting correlated data, the $\chi^2$ function, that needs to be minimized, is constructed out of the covariance matrix:
\beq
\label{eq:chi2def}
\chi^2 = \sum_{i,j=1}^n \big[ (E_\text{IAS}^{*\prime})_i - \Delta E_1(A,T_i) \big] \, \Sigma_{ij}^{-1} \,
\big[ (E_\text{IAS}^{*\prime})_j - \Delta E_1(A,T_j) \big] \, .
\eeq
For magnitude of fluctuations independent of excitation energy, the covariance matrix is easily inverted, producing
\beq
\label{eq:Siginv}
\Sigma_{ij}^{-1} = \frac{1}{\delta_\text{rms}^2} \, \Big( \delta_{ij} - \frac{1}{n+1} \Big) \, .
\eeq
With this, we find from \eqref{eq:chi2def}:
\beq
\label{eq:chi2rms}
\delta_\text{rms}^2 \, \chi^2 = \sum_{i=1}^n \big[ (E_\text{IAS}^{*\prime})_i - \Delta E_1(A,T_i) \big]^2
- \frac{1}{n+1} \Big(   \sum_{i=1}^n  \big[ (E_\text{IAS}^{*\prime})_i - \Delta E_1(A,T_i) \big]    \Big)^2 \, ,
\eeq
which is behind Eq.~\eqref{eq:chimod}.

Within individual isobaric chains, we find the effects of isospin symmetry to be so strong that, when the $Z$-dependence of our results is not the focus, to the extent possible, we combine information from individual isobars, to construct a {\em single spectrum} representing an isobar with the lowest $|T_z|$ within those nuclei in the chain for which there are IAS data.  In~this, we exploit information only on relative energies within the spectra of individual original isobars -- the need to resort to assumptions on displacement energies in the chain does not arise.  In few cases, a full combination of the states within the chain is not possible and we work with disconnected spectral links in the chain, that are unrelated by isospin symmetry.

When combining information from different isobaric chains, or from disconnected spectral links in one isobaric chain, the covariance matrix is block-diagonal and the inversion \eqref{eq:Siginv} applies within one subspace, with $n$ representing the number of the excited states in that subspace.  The $\chi^2$ function, to be minimized, then results from the sum of the terms such as on the r.h.s.\ of \eqref{eq:chi2rms}.  When the sought bulk symmetry-energy contribution to net energy, $E_1$, is linear in its parameters, as in Eqs.\ \eqref{eq:Eaaz} and \eqref{eq:IASb}, then the optimal parameters from $\chi^2$-minimization and their errors are easily arrived at in analytic form such as in Eqs.\ \eqref{eq:deltaaz} and \eqref{eq:azerror}.

\section{Stability of the Skyrme Interactions}
\label{Appendix:SkyrmeStability}

Effective nuclear interactions of the Skyrme zero-range form \cite{THR19581959615,PhysRevC.5.626} are commonly employed in nuclear practice.  They give rise to mean fields that depend only on local densities, which simplifies practical calculations.  Motivated in part by practical reasons, zero-range expansion is pursued within the microscopically-based effective theories of nuclear forces~\cite{1999PrPNP..43..337V}, with the zero-range terms acting alone or supplementing pion exchange.  Some of the Skyrme interactions put forward in the literature are known to be unstable in the long-wavelength limit, with instabilities evidenced in the violation of Landau stability criteria~\cite{PinesNoziers,VanGiai1981379,PhysRevC.66.014303,1984NuPhA.420..297T,vuong07}
\beq
\label{eq:LandauStab}
\chi_\ell > -(2 \ell+1) \, .
\eeq
Here, $\chi_\ell$ represents a Landau parameter of multipolarity $\ell$.  Still claims have been put forward that particular violations can obstruct the use of the nominally unstable interactions in dynamic, but not in static calculations \cite{PL123B.139}.

In carrying out calculations within the SHF model \cite{THR19581959615,PhysRevC.5.626}, using numerical codes by Reinhard {\em et al.} \cite{Reinhard:1991,reinhard:014309}, with the goal of extracting bulk nuclear characteristics, we found that we could not get a convergence for some of the interaction parameterizations in the literature, e.g.\ v070, SkP and SKz0, with the codes crashing for many of the nuclei of interest, both for nuclei occurring in Nature and those with unrealistically large masses.  In~the latter class of our calculations, the Coulomb interactions were switched off.  The~specific parameterizations were stable in the long wavelength limit, but the employed zero-range expansion for the interactions suggested a possibility of short-wavelength instabilities, which we pursue here.  In Ref.~\cite{Lesinski:2006cu} short-wavelength instabilities have been indeed observed for some of the studied zero-range interactions and characteristics of the effective interactions that enhance likelihood of short-wavelength instabilities were identified.  Here, we derive formal necessary conditions that must be met by the Skyrme interactions to prevent short-wavelength instabilities.  We further carry out an assessment of the interactions proposed in the literature with regard to the stability.  About {\em 40\% of those interactions} give rise to instabilities, either in the long- or short-wave limit or in both, even when only a minimal set of terms is included in the Skyrme energy energy functional.

Within the SHF model, the energy of a nuclear system is~\cite{Reinhard:1991}
\beq
E = E_\text{kin} + E_\text{Skyrme} + E_C + E_\text{pair} \, ,
\label{eq:ESkyrme}
\eeq
where the two last terms $E_C$ and $E_\text{pair}$ are, respectively, the Coulomb and pairing energies.
Both the kinetic and the model nuclear interaction term may be expressed in terms of local densities constructed from single-particle wavefunctions.  Thus, we have
\beq
E_\text{kin} =  \int \text{d} {\pmb r}  \,  e_\text{kin}({\pmb r}) =
\int \text{d} {\pmb r}  \,  \frac{\hbar^2}{2 m} \, \tau \, ,
\eeq
and
\beq
\tau =  \tau_n + \tau_p \, ,
\eeq
with
\beq
\label{eq:tauq=}
\tau_q ({\pmb r})  = \sum_\alpha n_\alpha^q \, {\pmb \nabla} \phi_\alpha^\dagger ({\pmb r}) \, {\pmb \nabla} \phi_\alpha ({\pmb r}) \, .
\eeq
In the above, the summations are carried out separately over neutron and proton states, $\phi_\alpha$ are the single-particle orbitals and $n_\alpha$ are occupations of the single-particle states $\alpha$.  The~Skyrme interaction energy-density may be decomposed into the isoscalar $e_0$ and isovector $e_1$ contributions:
\beq
\label{eq:ESk=}
E_\text{Skyrme}  \equiv   \int \text{d} {\pmb r}  \,  e_\text{Skyrme}({\pmb r}) =
\sum_T e_T ({\pmb r}) \, .
\eeq
For the simplest form of the functional, the above contributions are~\cite{Reinhard:1991,bender_gamow-teller_2002}
\beq
\label{eq:eT=}
e_T = C_T^\tau \, \rho_{\underline{T}} \, \tau_{\underline{T}} + C_T^\rho \, \rho_{\underline{T}}^2 + C_T^{\nabla \rho} \, \big( {\pmb \nabla} \rho_{\underline{T}}    \big)^2 + C_T^{\nabla J} \, \rho_{\underline{T}} \, \sum_{\mu \nu \xi} \epsilon_{\mu \nu \xi} {\nabla}^\mu \, {J}_{\underline{T}}^{\nu \xi} \, .
\eeq
The densities $\rho_{\underline{T}}$, $\tau_{\underline{T}}$ and $J_{\underline{T}}$ are the isoscalar and isovector combinations of nucleonic densities displayed in Eq.~\eqref{eq:tauq=}, and of those below
\begin{align}
\label{eq:rhoq=}
\rho_q ({\pmb r}) & = \sum_\alpha  n_\alpha^q \, \phi_\alpha^\dagger ({\pmb r}) \, \phi_\alpha ({\pmb r}) \\
\label{eq:Jq=}
{J}_q^{\nu \mu} ({\pmb r}) & = \frac{1}{2i} \sum_\alpha n_\alpha^q \, \phi_\alpha^\dagger ({\pmb r}) \,
\big( \overrightarrow{\pmb \nabla} - \overleftarrow{\pmb \nabla} \big)^\nu \, {\sigma}^\mu  \,
\phi_\alpha ({\pmb r}) \,  .
\end{align}
Specific combinations are
\beq
\rho_{\underline{0}}  = \rho_n + \rho_p \equiv \rho \, ,  \hspace*{2em}
\tau_{\underline{0}}   \equiv \tau \, ,  \hspace*{2em}
{J}_{\underline{0}}  = {J}_n + {J}_p \equiv {J} \, ,  \\
\eeq
and
\beq
\rho_{\underline{1}}  = \rho_n - \rho_p \equiv \rho_{np} \, ,   \hspace*{2em}
\tau_{\underline{1}}  = \tau_n - \tau_p \equiv \tau_{np} \, ,   \hspace*{2em}
{J}_{\underline{1}}  = {J}_n - {J}_p \equiv {J}_{np} \, .  \\
\eeq
The coefficients $C_T$ could, principally, be all density dependent~\cite{bender_gamow-teller_2002}, but in the Skyrme parametrization that dependence is limited - either a constant coefficient is used or a specific power of net density is added.  In terms of the standard Skyrme parameters \cite{THR19581959615,PhysRevC.5.626,Reinhard:1991}, the coefficients in the functional \eqref{eq:ESk=} and \eqref{eq:eT=} are
\begin{align}
& C_0^\tau  = \frac{3}{16} \, t_1 + \frac{1}{16} \, t_2 \, (5 + 4 x_2) \, ,\\
& C_1^\tau  = - \frac{1}{16} \, t_1 \, (1 + 2 x_1) + \frac{1}{16} \, t_2 \, (1 + 2 x_2) \, ,\\
& C_0^\rho = \frac{3}{8} \, t_0 + \frac{3}{48} \, t_3 \, \rho^\alpha \, , \\
& C_1^\rho = - \frac{1}{8} \, t_0 \, (1 + 2 x_0) - \frac{1}{48} \, t_3 \, (1 + 2 x_3 ) \, \rho^\alpha \, , \\
& C_0^{\nabla \rho} = \frac{9}{64} \, t_1 - \frac{1}{64} \, t_2 \, (5 + 4 x_2) \, , \\
& C_1^{\nabla \rho} = - \frac{3}{64} \, t_1 \, (1 + 2 x_1) - \frac{1}{64} \, t_2 \,
(1 + 2 x_2) \, , \\
& C_0^{\nabla J} = - \frac{3}{4} \, W_0 \, , \\
& C_1^{\nabla J} = - \frac{1}{4} \, W_0 \, .
\end{align}

The densities $\rho_{\underline{T}}$, $\tau_{\underline{T}}$ and $J_{\underline{T}}$, employed in the basic form \eqref{eq:eT=} of the functional, are all invariant under time reversal.
Beyond its basic form \eqref{eq:eT=}, the Skyrme functional may be supplemented with interaction terms dependent on densities that are odd under time reversal~\cite{bender_gamow-teller_2002,Stone2007587}.  In practical applications, the latter densities, and the associated contributions to the energy, tend to be of the order of $1/A$ relative to the densities and currents even in time, that are incorporated in~\eqref{eq:eT=}.  However, terms that are odd in time also need to be added to~\eqref{eq:eT=} to ensure Galilean invariance~\cite{Engel1975215}, even without invoking new interaction contributions.  Moreover, tensorial spin-orbit terms, proportional to $J^2$, may be accounted for in~\eqref{eq:eT=}.  In more elaborate Skyrme parameterizations, the isoscalar and isovector spin-orbit coefficients~$C_T$ are independent from each other, with
\begin{align}
& C_0^{\nabla J} = - b_4 - \frac{1}{2} \, b_4' \, , \\
& C_1^{\nabla J} = - \frac{1}{2} \, b_4' \, .
\end{align}
When we employ such a more elaborate parametrization in the combination with the basic SHF code of~\cite{Reinhard:1991} set up for interdependent coefficients, we take
\beq
W_0 = \frac{2}{3} \, \big( 2 \, b_4 +  b_4' \big) \, .
\eeq

The Skyrme parameters and/or the coefficients in the energy functional \eqref{eq:eT=} can be analytically related, see e.g.~\cite{bender_gamow-teller_2002,vuong07}, to different physical quantities characterizing nuclear systems, including the symmetry energy for uniform matter.  Nucleonic effective masses $m_q^*$ follow, in particular, from combining the isoscalar and isovector mass parameters $B_T$:
\beq
\label{eq:Bq}
\frac{\hbar^2}{2m_q^*} = B_q = B_0 \pm B_1 = \frac{\hbar^2}{2m} + C_0^\tau \, \rho \pm C_1^\tau \, \rho_{np}
\, ,
\eeq
where the upper sign pertains to neutrons and lower to protons.

For the Skyrme interactions, the Landau coefficients, governing long-wavelength stability of described systems, vanish for multipolarites $\ell \ge 2$.  From the remaining coefficients, only the $\ell = 0$ coefficients are large enough to make the violation of conditions \eqref{eq:LandauStab} a real possibility.  In~addition, for the basic form of the functional~\eqref{eq:eT=}, the coefficients related to spin $G_0$ and~$G_0'$ vanish, limiting the number of Landau coefficients of interest for the stability to two: isoscalar $F_0$ and isovector~$F_0'$.  In symmetric matter, these coefficients are \cite{bender_gamow-teller_2002}:
\beq
F_0 = \frac{2 m^*}{\hbar^2} \, \bigg[  \frac{1 }{\pi^2} \, k_F \, \bigg(2 C_0^\rho + 4 \rho \, \frac{\text{d} C_0^\rho}{\text{d} \rho} + \rho^2 \, \frac{\text{d}^2 C_0^\rho}{\text{d}\rho^2}      \bigg)                    + 3 \rho \,  C_0^\tau  \bigg] \, ,
\eeq
and
\beq
F_0' = \frac{2 m^*}{\hbar^2} \, \Big( \frac{2 }{\pi^2} \, k_F \, C_1^\rho + 3 \rho \,  C_1^\tau \Big) \, .
\eeq

In Table \ref{tab:skystab} we provide a variety of parameters that decide on the stability of predictions following from different Skyrme interactions proposed in the literature, for systems close to $n$-$p$ symmetry.  The Skyrme interactions have been compiled by Jirina Stone \cite{Stone2007587} and supplemented by us, cf.~I.   The stability parameters are grouped into two groups, those associated with the basic functional~\eqref{eq:eT=} and those pertaining to an expanded functional.  The parameter values that imply an instability are emphasized by displaying them in {\em cursive} font.  The names for the Skyrme parameterizations that fail any of the stability criteria, even for the basic functional, are also emphasized using the cursive font.  The names for the Skyrme parameterizations that pass all the stability criteria, including those for the extended functional, are emphasized using a bold font.  The first two parameters listed in Table \ref{tab:skystab} are $F_0$ and $F_0'$.  We can see in the Table that for all the listed Skyrme parameterizations
$F_0 > -1$, cf.~\eqref{eq:LandauStab}, which implies that the described symmetric normal matter is stable with respect to long wavelength isoscalar perturbations of the occupied Fermi spheres~\cite{PinesNoziers}.  We can further see that for nearly all parameterizations $F_0' > -1$, which implies that the described matter is stable with respect to analogous isovector perturbations~\cite{PinesNoziers}.  Further in the Table we list the effective neutron and proton masses, in the limit of neutron matter.  The latter are of importance for the short-wavelength stability of systems described by the Skyrme interactions, to be discussed next.

In analyzing the short-wavelength stability, we will presume that a self-consistent solution of the SHF equations \cite{Reinhard:1991} for a system was found.  For that solution, some occupied low-energy orbitals, generally involving low-momentum components, $k \lesssim k_F $, represent the nuclear system.  We shall verify whether populating orbitals with high-momentum components, $k \gg k_F$, or changing momentum content of the occupied orbitals towards high values, might lower the net system energy for the Skyrme interactions, possibly indiscriminately.  For the sake of our analysis, within the energy functional~\eqref{eq:ESkyrme} we emphasize the terms that are of second-order in gradients, most sensitive to high-momentum content:
\beq
\label{eq:Egrad}
E = \int \text{d} {\pmb r} \, \sum_T \bigg\lbrace B_T \, \tau_{\underline{T}}
+ C_T^{\nabla \rho} \, \big( {\pmb \nabla} \rho_{\underline{T}}    \big)^2
+ C_T^{\nabla J} \, \rho_{\underline{T}} \, \sum_{\mu \nu \xi} \epsilon_{\mu \nu \xi} {\nabla}^\mu \, {J}_{\underline{T}}^{\nu \xi} + \ldots \bigg\rbrace \, .
\eeq
General expectation is that negative coefficients for the squares of density gradients within the energy functional might lead to short-wavelength instabilities.  However, such terms need to interplay with the terms involving kinetic-energy density that may compensate for the negative terms and stabilize the system.
In Ref.~\cite{Lesinski:2006cu}, though, it was observed that large
negative values of $C_1^{\nabla \rho}$ for Skyrme interactions are indeed conducive to instabilities.  Another observation there was that a~large positive neutron-proton mass splitting, $m_n^* - m_p^*$, in the limit of neutron matter, tends to be associated with instabilities.

In what follows, we shall consider the response of net energy to different short-wavelength changes within a hypothetical system.  Following observations in \cite{Lesinski:2006cu}, we first consider a~situation where a~short wavelength oscillation develops, with neutron and proton standing wavefunctions being out of phase by $\pi/2$:
\beq
\begin{split}
\label{eq:phi}
\phi_{{\pmb k} \lambda n} ({\pmb r}) & \simeq {\mathcal A}  \cos{( {\pmb k} {\pmb r} )} \, , \\
\phi_{{\pmb k} \lambda p} ({\pmb r}) & \simeq {\mathcal A}   \sin{( {\pmb k} {\pmb r} )} \, ,
\end{split}
\eeq
within some region $\Delta V$ that is originally nearly free from particles.
Here, $\lambda$ represents spin projection along some chosen direction and at first we consider oscillations that are spin-independent.
We assume that $k$ may take on arbitrarily large values, while ${\mathcal A}$ can take values that are large enough for the associated contributions to dominate the local particle densities and for those contributions to additionally dominate the inertial parameters \eqref{eq:Bq}.  We will come back to the last issue later on.  With~\eqref{eq:phi}, the local nucleon densities are
\beq
\rho_n ({\pmb r}) \simeq 2 {\mathcal A}^2  \cos^2 {( {\pmb k} {\pmb r} )} \, , \hspace*{2em}
\rho_p ({\pmb r}) \simeq 2 {\mathcal A}^2  \sin^2 {( {\pmb k} {\pmb r} )} \, ,
\eeq
while the net density and relative density difference are
\beq
\label{eq:rho_A}
\rho ({\pmb r}) \equiv \rho_{\underline{0}} ({\pmb r}) \simeq 2 {\mathcal A}^2 \, , \hspace*{2em}
\rho_{np} ({\pmb r}) \equiv \rho_{\underline{1}} ({\pmb r}) \simeq 2 {\mathcal A}^2 \cos{(2 {\pmb k} {\pmb r})} \, .
\eeq
With \eqref{eq:phi} and \eqref{eq:rho_A} inserted into the energy functional \eqref{eq:Egrad}, the dominant contribution, from the region where \eqref{eq:phi} applies, behaves as $\propto {\mathcal A}^4 \, k^2$, for large ${\mathcal A}$ and $k$.  To establish whether the system is stable against the short wavelength oscillations in \eqref{eq:phi}, we need to establish the sign of the net multiplicative factor for the discussed leading term in the energy.

On account of using the same orbitals for the two spin directions, the net contribution to the energy from spin-dependent terms in \eqref{eq:Egrad} is zero.  Those terms that do not involve gradients in \eqref{eq:Egrad} do not contribute to the leading behavior at all.  With \eqref{eq:rho_A}, the {$C_{0}^{\nabla \rho}$-term} in~\eqref{eq:Egrad} does not contribute to the leading behavior either.  The nucleon numbers contributing to some region $\Delta V$, where the modification \eqref{eq:phi} arises, are generally pulled from the remainder of the system.  If that remainder is largely undisturbed, the energy cost due to removing the nucleon numbers $\Delta N$ and $\Delta Z$ from that remainder is
\beq
(\Delta E)_F \simeq -\mu_n \, \Delta N - \mu_p \, \Delta Z \, ,
\eeq
and that cost does not contribute to the leading behavior within the energy change, either.  On the other hand, with \eqref{eq:rho_A}, the $C_{1}^{\nabla \rho}$-term within the energy \eqref{eq:Egrad} contributes
\beq
(\Delta E)_{1}^{\nabla \rho} = \int_{\Delta V} \text{d} {\pmb r}  \, C_{1}^{\nabla \rho}  \, \big({\pmb \nabla} \rho_{np} \big)^2
\simeq \int_{\Delta V} \text{d} {\pmb r}  \, C_{1}^{\nabla \rho} \, 16 k^2 \, {\mathcal A}^4 \sin^2{ (  2 {\pmb k} {\pmb r} )}
\simeq \Delta V  \, C_{1}^{\nabla \rho} \, 8 k^2 \, {\mathcal A}^4 \, ,
\eeq
where the integration is over the region $\Delta V$ where \eqref{eq:phi} holds. Remaining contributions to the leading behavior come from the terms involving kinetic-energy density.

With \eqref{eq:phi}, the kinetic-energy density-factors are:
\beq
\tau_n ({\pmb r}) \simeq 2 {\mathcal A}^2 \, k^2 \sin^2{({\pmb k} {\pmb r})} \, ,
\hspace*{2em}
\tau_p ({\pmb r}) \simeq 2 {\mathcal A}^2 \, k^2 \cos^2{({\pmb k} {\pmb r})} \, ,
\eeq
yielding
\beq
\tau ({\pmb r}) \simeq 2 {\mathcal A}^2 \, k^2  \, ,
\hspace*{2em}
\tau_{np} ({\pmb r}) \simeq - 2 {\mathcal A}^2 \, k^2 \cos{(2 {\pmb k} {\pmb r})} \, .
\eeq
The terms involving kinetic energy density then contribute
\beq
(\Delta E)_{0}^{\tau} \simeq  \Delta V  \, C_0^\tau \,   4 {\mathcal A}^4 \, k^2 \, ,
\eeq
and
\beq
(\Delta E)_{1}^{\tau} \simeq \int_{\Delta V} \text{d} {\pmb r}  \, C_1^\tau \,
(-4 {\mathcal A}^4 \, k^2) \cos^2{(2 {\pmb k} {\pmb r})} \simeq - \Delta V \, C_1^\tau \, 2
{\mathcal A}^4 \, k^2 \, .
\eeq

Combining the leading terms in energy for the oscillations \eqref{eq:phi}, we get for the net leading contribution to the energy
\beq
\Delta E \simeq \Delta V \, \Psi_1^{\nabla \rho} \, 8 {\mathcal A}^4 \, k^2 \, ,
\eeq
where
\beq
\label{eq:Psi1Dr}
\Psi_1^{\nabla \rho} = C_1^{\nabla \rho} + \frac{1}{2} \, C_0^\tau - \frac{1}{4} \, C_1^\tau \, ,
\eeq
and where we normalized the multiplicative factor according to $C_1^{\nabla \rho}$.  Obviously, for the stability of the system with respect to the oscillations of the type \eqref{eq:phi}, we must have
\beq
\Psi_1^{\nabla \rho} \ge 0 \, .
\eeq
We display the $\Psi_1^{\nabla \rho}$ parameters for different Skyrme interactions in Table \ref{tab:skystab}.  As is apparent, many of the Skyrme parameterizations are unstable with respect to a development of oscillations of the type \eqref{eq:phi}.  Interactions unstable in that particular manner include the LNS and SkP interactions assessed in Ref.~\cite{Lesinski:2006cu} and the SKz0 and SKzm1 interactions for which we could find no convergence in~I, at any finite asymmetry.  Besides oscillating out of phase, the neutron and proton densities shot up in magnitude in the regions of instability within the calculations of Ref.~\cite{Lesinski:2006cu}.  Clearly, as presumed in \cite{Lesinski:2006cu}, a large negative value of $C_1^{\nabla \rho}$ is conducive to the investigated type of instability but, in the end, decisive is an interplay between the gradient term and the effective mass in \eqref{eq:Psi1Dr}.

We complement the analysis of isovector short-wavelength instability with an analysis of similar isoscalar instability.  We now consider orbitals that acquire, in some region~$\Delta V$ originally largely free from particles, the approximate form
\beq
\label{eq:phi+}
\phi_{{\pmb k} \lambda q} ({\pmb r})  \simeq {\mathcal A}  \cos{( {\pmb k} {\pmb r} )} \, ,
\eeq
independent of the species,
where, as before, $k$ is large and ${\mathcal A}$ is sufficiently large to make the contribution of those orbitals locally dominant.  Within analogous considerations to those before, the isoscalar gradient term now contributes to the leading contribution of the region~$\Delta V$ to the energy, but the isovector gradient term does not.  On combining the gradient and effective mass terms, we obtain
\beq
\Delta E \simeq \Delta V \, \Psi_0^{\nabla \rho} \, 8 {\mathcal A}^4 \, k^2 \, ,
\eeq
where
\beq
\label{eq:Psi0Dr}
\Psi_0^{\nabla \rho} = C_0^{\nabla \rho} + \frac{1}{4} \, C_0^\tau  \, .
\eeq
For stability against an indiscriminate rise in short-wavelength isoscalar density oscillations, we obviously need
\beq
\label{eq:Psi0Drge0}
\Psi_0^{\nabla \rho} \ge 0 \, .
\eeq
The values of the short-wavelength stability parameter $\Psi_0^{\nabla \rho}$ are again provided for different Skyrme interactions in Table \ref{tab:skystab} and it is seen that all proposed interactions meet the stability criterion \eqref{eq:Psi0Drge0}.

Choices of the potentially unstable orbitals so far emphasized the role of density gradients, relative to effective-mass, in generating the short-wavelength instabilities.  However, short-wavelength instabilities may also arise in association with the effective mass alone.  Thus, high-$k$ oscillations may develop for different orbitals of one species, while out of phase:
\beq
\label{eq:outofphase}
\phi_{{\pmb k} \lambda q}^{(1)} ({\pmb r})  \simeq {\mathcal A} \,  \sin{( {\pmb k} {\pmb r} )} \, , \hspace*{2em}
\text{and} \hspace*{2em} \phi_{{\pmb k} \lambda q}^{(2)} ({\pmb r})  \simeq {\mathcal A} \,  \cos{( {\pmb k} {\pmb r} )} \, .
\eeq
Such development gives rise to a significant kinetic energy density $\tau_q$ for nucleons $q$ in combination with an approximately constant density for those nucleons.  We arrive at the most stringent conditions on the parameters of the Skyrme interactions, in the context of~\eqref{eq:outofphase}, when assuming that the large density of kinetic energy for $q$ is correlated with enhanced density either for the same or the other nucleons.  With this, we explore the limits of
\beq
\tau_{np} \approx \pm \tau \hspace*{2em} \text{and} \hspace*{2em} \rho_{np} \approx \pm \rho \, ,
\eeq
where the signs in the two relations are independent and both $\rho$ and $\tau$ are large.  The requirement of the positive leading term in energy produces then the condition
\beq
\label{eq:C01t}
C_0^\tau \pm C_1^\tau \ge 0 \hspace*{2em} \text{or} \hspace*{2em} C_0^\tau \ge \big| C_1^\tau \big| \, .
\eeq

The condition \eqref{eq:C01t} is equivalent to a demand that inverse nucleon effective masses are equal to or greater than those in free space,
\beq
\label{eq:mq*}
\frac{1}{m_q^*} \ge \frac{1}{m} \, .
\eeq
Clearly, when the mass parameter is linear in density and falls below the parameter in free space at one density, then, at sufficiently high density, it will turn to negative values allowing for an indiscriminate growth in high-$k$ modes.
In Table \ref{tab:skystab} we list the ratios of $m_q^*/m$ for neutrons and protons in neutron matter at normal density.  The ratios that exceed unity signal an effective-mass instability in the energy.  As~to the neutron-proton mass difference as a possible signal for instability \cite{Lesinski:2006cu}, we have for the difference of inertial parameters
\beq
B_n - B_p = \frac{1}{2m_n^*} - \frac{1}{2m_p^*} = \frac{m_p^* - m_n^* }{2m_n^* \, m_p^* } =
2 C_1^\tau \, \rho_{np}  \, .
\eeq
While we find no direct tie between the magnitude and sign of $C_1^\tau$ and a short-wavelength instability, clearly an unusually high $|C_1^\tau|$, in the context of a relatively large $|m_n^* - m_p^*|$, may lead to a violation of the inequality~\eqref{eq:C01t}.  Interactions that violate the stability conditions~\eqref{eq:C01t}, equivalent to \eqref{eq:mq*}, include most of the interactions developed in Ref.~\cite{arXiv:0804.3385} and utilized in \cite{PhysRevC.81.051303} to analyze correlations between observables tied to symmetry energy.

We find no contributions to the leading short-wavelength from the spin-orbit $\nabla J$ term in~\eqref{eq:eT=} and no associated short-wavelength stability condition.  Overall, with all the stability conditions for the basic functional taken together, we can observe in Table~\ref{tab:skystab} that close to 40\% of the Skyrme parameterizations proposed in the literature are unstable right at the level of the basic functional \eqref{eq:ESk=} with \eqref{eq:eT=}.

While the basic functional \eqref{eq:eT=} is of our primary interest, for completeness we discuss the stability of systems described employing a more complete energy functional with addition to $e_T$ of  \cite{bender_gamow-teller_2002,Stone2007587}:
\beq
\begin{split}
\label{eq:eTDelta}
e_T^\Delta = & - C_T^\tau \, \rho_{\underline{T}} \, {\pmb j}_{\underline{T}}^2
+ C_T^{\nabla J} \, {\pmb \sigma}_{\underline{T}} \, \big({\pmb \nabla} \times {\pmb j}_{\underline{T}}   \big) \\
&  + C_T^J \, \Big[ \sum_{\mu \nu}  \big( {J}_{\underline{T}}^{\mu \nu}  \big)^2  - {\pmb \sigma}_{\underline{T}} \, {\pmb \tau}_{\underline{T}} \Big]
  + C_T^\sigma \, {\pmb \sigma}_{\underline{T}}^2
+ C_T^{\nabla \sigma} \sum_{\mu \nu} \big( {\nabla}^\mu \, {\sigma}_{\underline{T}}^\nu \big)^2 \, .
\end{split}
\eeq
Here, the additionally appearing densities are constructed, as before, from nucleonic densities:
\begin{align}
\label{eq:j=}
{\pmb j}_q ({\pmb r})  & = \frac{1}{2i} \sum_\alpha  n_\alpha^q \, \phi_\alpha^\dagger ({\pmb r})\, \big( \overrightarrow{\pmb \nabla} - \overleftarrow{\pmb \nabla} \big) \, \phi_\alpha ({\pmb r}) \, , \\
\label{eq:sigmaq=}
{\pmb \sigma}_q ({\pmb r}) & = \sum_\alpha  n_\alpha^q \, \phi_\alpha^\dagger ({\pmb r})\, {\pmb \sigma} \, \phi_\alpha ({\pmb r}) \, , \\
\label{eq:tauvq=}
{\pmb \tau}_q ({\pmb r}) &  = \sum_{\alpha \, \mu} n_\alpha^q \, {\nabla}^\mu \phi_\alpha^\dagger ({\pmb r})\, {\pmb \sigma} \, {\nabla}^\mu \phi_\alpha ({\pmb r}) \,  .
\end{align}
All of the densities above change sign under time reversal.
The first two terms in \eqref{eq:eTDelta} ensure Galilean covariance for the terms already appearing in \eqref{eq:eT=}. The third term is tensor spin-orbit term.  The two remaining terms are spin-spin interaction and spin-gradient terms.  The new interaction coefficients in \eqref{eq:eTDelta} may be sought without imposing any relation \cite{hellemans:287,PhysRevC.76.014312,chamel_spin_2010} to the even-time coefficients in \eqref{eq:eT=} or these new coefficients can be derived from the original Skyrme interaction~\cite{THR19581959615,PhysRevC.5.626} that yields \cite{bender_gamow-teller_2002}:
\begin{align}
& C_0^J  =  \frac{1}{16} \, t_1 \, (1 - 2 x_1) - \frac{1}{16} \, t_2 \, (1 + 2 x_2) \, ,\\
& C_1^J  =  \frac{1}{16} \, t_1 - \frac{1}{16} \, t_2 \, ,\\
& C_0^\sigma = - \frac{1}{8} \, t_0 \, (1 - 2 x_0) - \frac{1}{48} \, t_3 \, (1 - 2 x_3 ) \, \rho^\alpha \, ,\\
& C_1^\sigma = - \frac{1}{8} \, t_0  - \frac{1}{48} \, t_3  \, \rho^\alpha \, , \\
& C_0^{\nabla \sigma} = - \frac{3}{64} \, t_1 \, (1 - 2 x_1) - \frac{1}{64} \, t_2 \, (1 + 2 x_2) \, , \\
& C_1^{\nabla \sigma} = - \frac{3}{64} \, t_1 - \frac{1}{64} \, t_2 \, .
\end{align}
We use these latter naive Skyrme-coefficient substitutions, when assessing the stability of systems described by Skyrme interactions with the energy density \eqref{eq:eTDelta} included.
Venturing outside of the original Skyrme interaction, an additional tensor term to that in \eqref{eq:eTDelta} may be considered~\cite{PhysRevC.76.014312,hellemans_tensor_2011}.
An addition of \eqref{eq:eTDelta} to the energy density gives rise to effective masses dependent on spin orientation.  Thus, in matter locally polarized along the $z$-axis the effective masses follow from
\beq
\label{eq:Bql}
\frac{\hbar^2}{2m_{q \lambda}^*} = B_{q \lambda} = \frac{\hbar^2}{2m} + C_0^\tau \, \rho + q \, C_1^\tau \, \rho_1 - \lambda \big( C_0^J \, \sigma_0^z + q \, C_1^J \, \sigma_1^z          \big)
\, ,
\eeq
where on the r.h.s.\ $q=1$ and $-1$, for a neutron and proton, respectively, and $\lambda =1$ and $-1$ for spin up and spin down.  Potential for ferromagnetic instabilities in neutron matter, tied to effective masses, has been pursued by several authors \cite{kutschera_polarized_1994,isayev_spin-ordered_2006,chamel_spin_2010}.

The addition of \eqref{eq:eTDelta} to the energy density of a system gives rise to finite Landau parameters~$G_0$ and $G_0'$ that are tied to spin oscillations, respectively isoscalar and isovector in nature:
\begin{align}
& G_0 = \frac{2 m^*}{\hbar^2} \, \Big( \frac{2 }{\pi^2} \, k_F \, C_0^\sigma - 3 \rho \,  C_0^J \Big) \, , \\
& G_0' = \frac{2 m^*}{\hbar^2} \, \Big( \frac{2 }{\pi^2} \, k_F \, C_1^\sigma - 3 \rho \,  C_1^J \Big) \, .
\end{align}
These two Landau parameters, provided here for symmetric matter, govern the long wavelength stability of described systems, along with the parameters $F_0$ and $F_0'$.  Values of these stability parameters are provided Table \ref{tab:skystab} when using the Skyrme values for the coefficients~$C^\sigma$.  Again some violations of the lower bound of $-1$ are observed.  The offending interactions include the T Skyrme parametrization and most of the SkI parameterizations.

As to the short-wavelength instabilities, the addition of the Galilean correction terms in~\eqref{eq:eTDelta} does not alter the so-far established stability conditions.  However, two other added gradient terms can give rise to new types of short-wavelength spin instabilities.  We start with isoscalar oscillations that emphasize the role of the spin-gradient term and we consider the growth of orbitals with spins locally oriented in the positive and negative direction of the $z$-axis, respectively,
\beq
\label{eq:phisig}
\phi_{{\pmb k} \, \uparrow \, q} ({\pmb r}) \simeq {\mathcal A}  \cos{( {\pmb k} {\pmb r} )} \, , \hspace*{6em}
\phi_{{\pmb k} \, \downarrow \, q} ({\pmb r}) \simeq {\mathcal A}   \sin{( {\pmb k} {\pmb r} )} \, .
\eeq
These orbitals yield
\begin{align}
\label{eq:rhos0}
\rho & \simeq 2 {\mathcal A}^2 \,  ,&   \tau &  \simeq 2 {\mathcal A}^2 \, k^2 \, ,  \\
\sigma_0^z  & \simeq 2 {\mathcal A}^2 \cos{(2 {\pmb k} {\pmb r})} \,  , &  \tau_0^z & \simeq - 2 {\mathcal A}^2 \, k^2 \cos{(2 {\pmb k} {\pmb r})}  \, ,
\end{align}
and $J^{\mu \nu} \simeq 0$.
Upon inserting these results into the energy functional and generally following analogous strategy to that before, we arrive at the following stability condition with respect to the growth of short-wavelength isoscalar spin oscillations:
\beq
\label{eq:psi0ns}
\psi_0^{\nabla \sigma} = C_0^{\nabla \sigma} + \frac{1}{2} \, C_0^\tau + \frac{1}{4} \, C_0^J \ge 0 \, .
\eeq

Exploring the stability with respect to the formation of short-wavelength spin isovector oscillations, we consider growth of orbitals with spins locally oriented in the positive and negative direction of the $z$-axis, respectively, out of phase for neutrons and protons:
\beq
\begin{split}
\label{eq:phisiv}
\phi_{{\pmb k} \, \uparrow \, n} ({\pmb r}) & \simeq \phi_{{k} \, \downarrow \, p} ({\pmb r})  \simeq {\mathcal A}  \cos{( {\pmb k} {\pmb r} )} \, , \\
\phi_{{\pmb k} \, \downarrow \, \, n} ({\pmb r}) & \simeq \phi_{{k} \, \uparrow \, p} ({\pmb r})  \simeq {\mathcal A}   \sin{( {\pmb k} {\pmb r} )} \, .
\end{split}
\eeq
To the leading order, within the considerations we pursue, these orbitals once again produce the densities of
\eqref{eq:rhos0} as well as
\begin{align}
\sigma_{1}^z  & \simeq 2 {\mathcal A}^2 \cos{(2 {\pmb k} {\pmb r})} \,  , &  \tau_{1}^z & \simeq - 2 {\mathcal A}^2 \, k^2 \cos{(2 {\pmb k} {\pmb r})}  \, ,
\end{align}
with $\sigma_0^z, \, \tau_0^z , \, J_q^{\mu \nu} \simeq 0$.  Inserting these densities into the energy functional produces a stability condition against the growth of short-wavelength isovector spin oscillations, mirroring that in Eq.~\eqref{eq:psi0ns}, in the form
\beq
\label{eq:psi1ns}
\psi_1^{\nabla \sigma} = C_1^{\nabla \sigma} + \frac{1}{2} \, C_0^\tau + \frac{1}{4} \, C_1^J \ge 0 \, .
\eeq

To emphasize the role of effective mass in \eqref{eq:Bql}, we consider a situation where two orbitals of one species undergo traveling-wave modification with spins oriented in the $+z$ direction:
\beq
\label{eq:phiq0}
\phi_{{k} \, \uparrow \, \, q} ({\pmb r}) \simeq \frac{\mathcal A}{\sqrt{2}} \,  \text{e}^{ i {\pmb k} {\pmb r} } \, , \hspace*{6em}
\phi_{{-k} \, \downarrow \, q} ({\pmb r})  \simeq \frac{\mathcal A}{\sqrt{2}} \,  \text{e}^{ - i {\pmb k} {\pmb r} } \, .
\eeq
We arrive at most stringent effective-mass constraints when these modifications take place against a
high-density background of the same or opposite species, fully polarized along the same or opposite direction.  The requirement of stability against an indiscriminate growth, ${\mathcal A} \rightarrow \infty$, of modifications of the type~\eqref{eq:phiq0} under such conditions, turns out to be mathematically equivalent to the requirement that the mass parameter for some density $\rho$,
\beq
\label{eq:Bql*}
\frac{\hbar^2}{2m_{q,\lambda}^*} = \frac{\hbar^2}{2m} +
\rho \Big[  C_0^\tau + q \, C_1^\tau - \lambda \big( C_0^J + q \, C_1^J \big) \Big]
\, ,
\eeq
never falls below the value in free space, i.e.
\beq
\label{eq:1m*}
\bigg( \frac{1}{m_{q,\lambda}^*} \bigg)_\text{min} \ge \frac{1}{m} \, ,
\eeq
see also \cite{kutschera_polarized_1994,isayev_spin-ordered_2006,chamel_spin_2010}, where the effective-mass stability conditions are less stringent.  Also, it is apparently not realized in these references that the violation of an effective-mass condition at high density in neutron matter impacts the low density neutron matter and symmetric matter as well, rendering the latter systems metastable.  Activating the $J^2$ term in \eqref{eq:eTDelta} with different orbital modifications, in place of ${\pmb \sigma \tau}$ with \eqref{eq:phiq0}, produces conclusions that are consistent with \eqref{eq:1m*}.

Parameter values pertaining to the high-$k$ stability, when the tensorial terms and those with time-odd densities are included in the energy functional, are again provided in Table~\ref{tab:skystab}.  Numerous parameterizations violate spin stability conditions.  In particular, all SkI parameterizations, as well as all SLy parameterizations (not just SLy5 considered in \cite{Lesinski:2006cu}) violate the short-wavelength vector-isoscalar condition \eqref{eq:psi0ns}.  In the Table we also provide the values of $\big(m/m^*\big)_\text{min}$ at normal density.  Naively one would have expected the latter to represent inverses of maximal values of effective masses, over spin and isospin, at that density.  While this true for most of the Skyrme parameterizations, for a number of them the effective masses can reach infinity and cross to negative values already at subnormal density, e.g.\ for the T parametrization, hence presentation of the inverse values of effective-mass reduction.

Overall, application of the high-$k$ stability criteria related to the odd-time terms has a~devastating effect on the collection of the parameterizations, with few parameterizations surviving the test of meeting all, both low- and high-$k$ stability criteria for even- and odd-time terms.  The ZR parameterizations passing the stability tests have other under undesirable properties, including pathological symmetry energies.  Besides the issues of long- and short-wavelength instabilities addressed in Table~\ref{tab:skystab}, one should also mention the possibility of instabilities for intermediate wavelengths, such as investigated within the random-phase-approximation (RPA) by Lesinski \etal\ \cite{Lesinski:2006cu}.  One should note, though, that any instabilities identified within RPA might principally get stabilized by some system reorganization, such as within the region of nuclear liquid-gas phase transition.  As studied here, though, the high-$k$ instabilities are absolute.  While instabilities leading to a reorganization cannot be physically excluded, the absolute instabilities can.

The Skyrme interaction, in its various versions, is meant to be a low-momentum theory.  In practical calculations with that interaction, though, the low-momentum aspect is not enforced in any systematic way.  Whether a specific solution may venture into a region of high-momentum instability can depend on symmetries imposed on the solution, as obviously on the demand of vanishing of all or of certain time-odd densities, and on numerical details, such as on the densities and angular momenta that the system is allowed to explore, and on the numerical mesh employed in the solution, that limits the supported wavevectors.  Even the physical size of the system can matter and in our own practice we found a higher likelihood of problems with large systems, exceeding the nuclei in Nature, than with small.  Finally, finer details of the instability are likely to matter in practice, particularly when supranormal densities are involved, as the region of instability is likely to be separated from a metastable region at lower densities by an energy barrier within the space of system parameters.

The latter issue is particularly straightforward to assess for effective-mass instabilities, as the mass parameters are linear in density.  Thus, the critical density for some effective mass turning negative is
\beq
\label{eq:rhoct}
\rho_c^\tau = \frac{\rho_0}{1 - \big( \frac{m}{m^*(\rho_0)}\big)_\text{min}} \, .
\eeq
If we take e.g.\ the basic time-even version of Ska35s15 interaction, with $m^*/m = 1.017$, just slightly above unity, the critical density for instability is at the safe $\rho_c^\tau = 60 \, \rho_0$!  On the other hand, if we activate the time-odd portion of the energy functional, one of the effective masses goes up to $m^*/m = 2.46$ and the critical density goes down to the precarious $\rho_c^\tau = 1.69 \, \rho_0$.  Low critical densities, if not even subnormal, for effective masses turning to negative when spin densities are activated, is a problem for most of the Skyrme interactions, as readily seen from Table \ref{tab:skystab}.

As to the instabilities with respect to density oscillations, critical average densities for participating orbitals may be assessed.  Thus, with the orbital density given by \eqref{eq:rho_A} in the scalar-isovector case, the full energy to the order of $k^2$ from the region where the orbital modification occurs is
\beq
\Delta E \simeq \Delta V \, \rho \, k^2 \, \Big\lbrace  \frac{\hbar^2}{2 m} + 2 \rho \, \Psi_1^{\nabla \rho}       \Big\rbrace \, .
\eeq
The $k^2$ coefficient remains positive up to the critical density of
\beq
\label{eq:rhocDr}
\rho_c^{\nabla \rho}  \simeq - \frac{\hbar^2}{4 m \, \Psi_1^{\nabla \rho}} \, .
\eeq
E.g.\ for the SkP interaction with $\Psi_1^{\nabla \rho} = -23.9 \, \text{MeV} \, \text{fm}^5$, the critical density is $\rho_c^{\nabla \rho}  \simeq 2.7 \, \rho_0$, relatively close to the densities in nuclei, particularly when strong shell effects are present.  Once an region instability is present, the system may choose different paths to reach it, not necessarily following~Eq.~\eqref{eq:phi} along the way.  In~\cite{Lesinski:2006cu}, Lesinski \etal\ find that RPA instabilities may develop for the SkP at finite wavelength at densities as low as $\rho \sim \rho_0$.  For the LNS interaction with $\Psi_1^{\nabla \rho} = -16.4 \, \text{MeV} \, \text{fm}^5$, see Table~\ref{tab:skystab}, the critical density from~\eqref{eq:rhocDr} is $\rho_c^{\nabla \rho}  \simeq 3.9 \, \rho_0$, but the RPA instabilities may develop at finite $k$ at densities as low as $\rho \sim 1.5 \, \rho_0$.

Aside from the inability to complete some SHF calculations, we encountered other problems for the interactions violating stability conditions, such as a failure of energies to follow the expectations from the volume-surface separation, cf.~Fig.~\ref{fig:Bed_smple}, whether for the symmetry coefficient or the symmetry coefficient and the energy of symmetric matter together, no matter how large~$A$ was, while the Coulomb interactions were switched off.  The~latter was particularly the case when any of the Landau conditions \eqref{eq:LandauStab} was violated (in practice either $F_0'$ or $G_0$).  On the other hand, we encountered no problems in the calculations for the interactions that were nominally unstable in the short-wavelength limit, but at sufficiently high densities compared to the normal.  Insisting on the lack of any instability under any circumstances, in the Skyrme parametrizations employed in comparisons to data, would clearly eliminate, see Table~\ref{tab:skystab}, the vast majority of the parametrization and disallow exploration of interesting physics scenarios such as with respect to different effective masses.  With this in mind, we adopt a compromise that allows to retain most of the interactions in our comparisons to IAS data.  Specifically, we accept results from those Skyrme interactions that are stable in the long-wavelength limit, meeting the Landau stability criteria \eqref{eq:LandauStab}, and for which the region of absolute short-wavelength instability, according to Eqs.~\eqref{eq:rhoct} and~\eqref{eq:rhocDr}, lies above $5 \rho_0$, given that the path towards instability may start already at a lower density.  The latter allows for interactions for which $m^*/m < 1.25$ and $\Psi_1^{\nabla \rho} > -12.9 \, \text{MeV} \, \text{fm}^5$ (rather than the strict $m^*/m \le 1$ and $\Psi_1^{\nabla \rho} > 0$).  At this stage we disregard the short-wavelength stability criteria following from the extension of the functional~\eqref{eq:eTDelta}, in accepting the interactions.  Given that the terms in \eqref{eq:eTDelta} are physically expected to be low compared to those in \eqref{eq:ESkyrme}, the associated instabilities may have a harder time to develop -- when disregarding these criteria we observed no associated troubling effects in the results of calculations.

\begin{figure}
\centerline{\includegraphics[width=.5\linewidth]{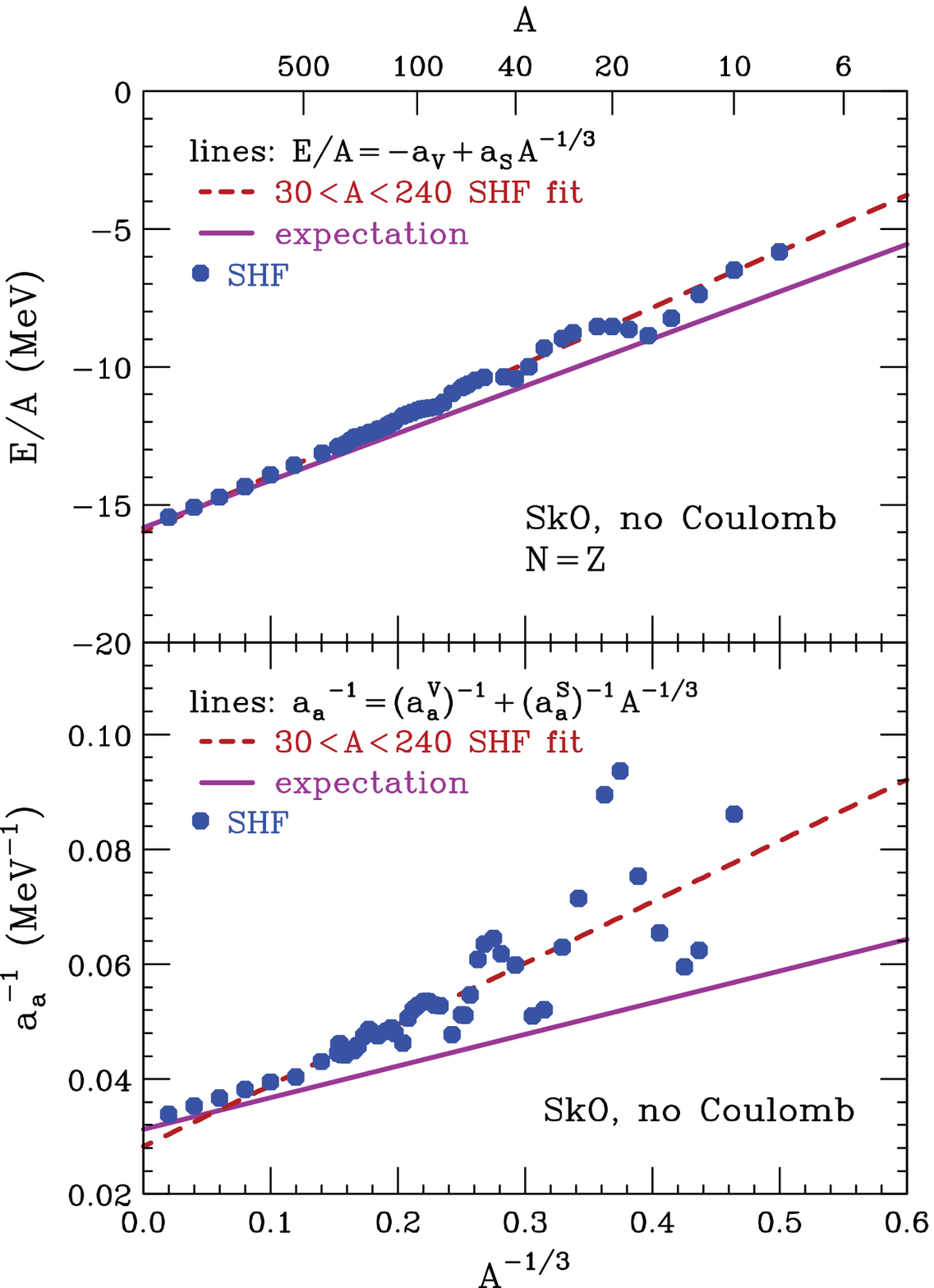}}
\caption{Energy per nucleon for $N=Z$ nuclei (top panel) and inverse symmetry coefficient, $1/a_a$, for $\eta \sim 0.2 $ nuclei (bottom), both displayed as a function of $A^-{1/3}$, in the calculations with the SkO interaction parametrization~\cite{PhysRevC.60.014316} with Coulomb interactions switched off.  The symbols represent spherical SHF calculations, solid lines represent expectations based on the volume-surface decomposition and the half-infinite matter results of~I, while dashed lines represent volume-surface fits to the spherical results in the mass region of $30 < A <240$. SkO is one of the interactions that we exclude from considerations when inferring on nuclear symmetry energy, on the basis of stability considerations.  This figure is analogous to Fig.~\ref{fig:Bed_smple}.
}
\label{fig:Bed_pat}
\end{figure}

\newpage

\pagestyle{empty}
\include{table_stability}
\pagestyle{plain}

\newpage

\bibliography{ss11,skyrme,skin}

\bibliographystyle{my}  

\end{document}

%% file: table_skin.tex
\afterpage{
\clearpage
\setlength{\LTcapwidth}{6.1in}
\begin{longtable}
               {|| c || l | c | c || c ||}

\caption[]
{Asymmetry skins from various measurements tied to nuclear rms radii.  From left to right, columns list the nucleus, reference for the skin value, type of measurements which are the source of critical information in the skin determination, and the skin value with error.  When skin values from independent determinations are available, the combined result is also shown.  The '*' over an error for the combined value indicates that error renormalization was employed.  The last column shows the average and rms deviation of skins for the Skyrme interactions and their interpolations meeting the skin ensemble and IAS constraints.}
\label{tab:skin}\\

\hline
\hline
Nucleus & Reference & Data Source & $\Delta r_{np}$ [fm] & $\Delta r_{np}^\text{GF}$ [fm]  \\
\hline
\hline
\endfirsthead

\multicolumn{5}{l}{{\hspace*{7em}\tablename} \thetable{} -- Continued}\\
\hline
\hline
Nucleus & Reference & Data Source & $\Delta r_{np}$ [fm] & $\Delta r_{np}^\text{GF}$ [fm]  \\
\hline
\hline
\endhead

\multicolumn{5}{l}{\hspace*{7em}Table Continued on Next Page\ldots}
\endfoot

\endlastfoot

$^{48}$Ca & Friedman \cite{Friedman:2012pa} & pionic atoms & $0.13 \pm 0.06$ & \\
          & Gils \etal~\cite{PhysRevC.29.1295} & elastic $\alpha$ scattering & $0.175 \pm 0.050$ & \\
          & Ray \cite{PhysRevC.19.1855} & elastic $\vec{p}$ scattering & $0.229 \pm 0.050$ & \\
          & Clark \etal~\cite{PhysRevC.67.054605} & elastic $p$ scattering & $0.103 \pm 0.040$ & \\
          & Shlomo \etal~\cite{Shlomo19795} & elastic $p$ scattering & $0.10 \pm 0.03$ & \\
          & Gibbs \etal~\cite{PhysRevC.46.1825} & elastic $\pi$ scattering & $0.11 \pm 0.04$ & \\
            \cline{2-5}
          & & combined results & $0.129 \pm 0.053^*$ & $0.218 \pm 0.015 $\\
          \hline
          \hline
$^{50}$Ti & Gils \etal~\cite{PhysRevC.29.1295} & elastic $\alpha$ scattering & $0.031 \pm 0.040$ & $0.133 \pm 0.011 $ \\
          \hline
          \hline
$^{64}$Ni & Ray \cite{PhysRevC.19.1855} & elastic $\vec{p}$ scattering & $0.167 \pm 0.050$ & $0.102 \pm 0.015 $ \\
          \hline
          \hline
$^{116}$Sn & Ray \cite{PhysRevC.19.1855} & elastic $\vec{p}$ scattering & $0.146 \pm 0.050$ & $0.103 \pm 0.015 $ \\
          \hline
          \hline
$^{124}$Sn & Ray \cite{PhysRevC.19.1855} & elastic $\vec{p}$ scattering & $0.252 \pm 0.050$ & $0.184 \pm 0.021 $ \\
          \hline
          \hline
$^{204}$Pb & Zenihiro \etal~\cite{PhysRevC.82.044611} & elastic $p$ scattering & $0.178 \pm 0.059$ & $0.161 \pm 0.024 $\\
          \hline
          \hline
$^{206}$Pb & Zenihiro \etal~\cite{PhysRevC.82.044611} & elastic $p$ scattering & $0.180 \pm 0.064$ &  \\
           & Starodubsky \etal~\cite{PhysRevC.49.2118} & elastic $p$ scattering & $0.181 \pm 0.045$ & \\
           \hline
           & & combined results & $0.181 \pm 0.037$ & $0.172 \pm 0.024 $ \\
          \hline
          \hline
\pagebreak[1]
$^{207}$Pb & Starodubsky \etal~\cite{PhysRevC.49.2118} & elastic $p$ scattering & $0.186 \pm 0.041$ & $0.178 \pm 0.024 $ \\
          \hline
          \hline
\pagebreak[4]
$^{208}$Pb & Starodubsky \etal~\cite{PhysRevC.49.2118} & elastic $p$ scattering & $0.197 \pm 0.042$ & \\
           & Ray \cite{PhysRevC.19.1855} & elastic $\vec{p}$ scattering & $0.16 \pm 0.05$ & \\
           & Clark \etal~\cite{PhysRevC.67.054605} & elastic $p$ scattering & $0.119 \pm 0.045$ & \\
           & Zenihiro \etal~\cite{PhysRevC.82.044611} & elastic $p$ scattering & $0.211 \pm 0.063$ & \\
           & Friedman \cite{Friedman:2012pa} & elastic $\pi^+$ scattering & $0.11 \pm 0.06$ & \\
           & Friedman \cite{Friedman:2012pa} & pionic atoms & $0.15 \pm 0.08$ &   \\
           \hline
           & & combined results & $0.159 \pm 0.041^* $ & $0.179 \pm 0.023 $ \\
           \hline
           \hline

\end{longtable}
\setlength{\LTcapwidth}{4in}
}

%% file: table_stability.tex
\begin{center}
\setlength{\LTcapwidth}{6.1in}
\begin{longtable}
               {l||c c c c c c|c c c c c | c}

\caption[]
{Long- and short-wavelength stability parameters for different Skyrme interactions with force constants given in the indicated references. The left side of the Table provides parameters associated with the basic functional employing time-even densities, while the right side -- parameters associated with the complement of the functional, employing time-odd densities and incorporating the tensor terms.  The Landau parameters are provided for symmetric matter at a~normal density for a~given interaction.  The short-wavelength stability parameters $\Psi$ are in units of $ \text{MeV} \, \text{fm}^5 $.  The effective masses are in neutron matter at normal density. Stability range is indicated underneath each parameter symbol.  Parameter values that fall outside of the associated stability range are represented in cursive.   Cursive is further used for the names of interactions in violation of any of the stability conditions for the basic functional.  Names of interactions that pass all the considered stability conditions are printed in bold.}
\label{tab:skystab}\\

\hline
Name & $F_0$ & $F_0'$ & $m_n^*/m$ & $m_p^*/m$  & $\Psi_0^{\nabla \rho}$ & $\Psi_1^{\nabla \rho}$ & $G_0$ & $G_0'$ & $\big(m/m^{*} \big)_\text{min} $ & $\Psi_0^{\nabla \sigma}$ & $\Psi_1^{\nabla \sigma}$ &  Ref. \\
Stability & $ >$$-1$ & $ >$$-1$ & $\le$1 & $\le$1 & $\ge$$0$ & $\ge$$0$ & $ >$$-1$ & $ >$$-1$ &  $\ge$1  &  $\ge$$0$ & $\ge$$0$  &    \\
\hline
\endfirsthead

\multicolumn{13}{l}{{\hspace*{7em}\tablename} \thetable{} -- Continued}\\
\hline
Name & $F_0$ & $F_0'$ & $m_n^*/m$ & $m_p^*/m$  & $\Psi_0^{\nabla \rho}$ & $\Psi_1^{\nabla \rho}$ & $G_0$ & $G_0'$ & $\big(m/m^{*} \big)_\text{min}$ & $\Psi_0^{\nabla \sigma}$ & $\Psi_1^{\nabla \sigma}$ &  Ref. \\
Stability & $ >$$-1$ & $ >$$-1$ & $\le$1 & $\le$1 & $\ge$$0$ & $\ge$$0$ & $ >$$-1$ & $ >$$-1$ & $\ge$1  & $\ge$$0$  & $\ge$$0$  &    \\
\hline
\endhead

\multicolumn{13}{l}{\hspace*{7em}Table Continued on Next Page\ldots}
\endfoot

\hline
\endlastfoot

\input{sympar.tex}

\end{longtable}
\setlength{\LTcapwidth}{4in}
\end{center}

%% file: sympar.tex
{\em SI      } &    0.56      &    1.21      & {\em 1.064}  &  0.797       &   44.2       &    2.2       &   -0.07      &    0.57      & {\em   0.625} &    2.2       &    2.2       & \cite{PhysRevC.5.626}          \\
SII            &   -0.06      &    0.70      &  0.689       &  0.500       &  110.0       &   33.3       &    0.08      &    0.39      & {\em   0.903} &   33.3       &   33.3       & \cite{PhysRevC.5.626}          \\
SIII           &    0.31      &    0.87      &  0.912       &  0.656       &   74.1       &   12.8       &    0.05      &    0.46      & {\em   0.667} &   12.8       &   12.8       & \cite{Beiner:1974gc}           \\
{\em SIIIs   } &    0.38      &    1.18      &  0.822       &  0.759       &   75.0       & {\em -20.7}  &    1.12      &   -0.09      & {\em   0.667} & {\em  -3.2}  &   24.1       & \cite{PhysRevC.21.2076}        \\
{\bf SIV     } &   -0.28      &    0.24      &  0.558       &  0.407       &  143.4       &   52.2       &    0.44      &    0.34      &    1.127      &   52.2       &   52.2       & \cite{Beiner:1974gc}           \\
{\bf SV      } &   -0.46      &    0.04      &  0.453       &  0.332       &  182.0       &   74.1       &    0.57      &    0.31      &    1.401      &   74.1       &   74.1       & \cite{Beiner:1974gc}           \\
{\em SVI     } &    0.68      &    1.23      & {\em 1.142}  &  0.813       &   50.9       & {\em  -0.3}  &   -0.17      &    0.53      & {\em   0.521} & {\em  -0.3}  & {\em  -0.3}  & \cite{Beiner:1974gc}           \\
{\em SVII    } &    0.78      &    1.36      & {\em 1.206}  &  0.855       &   46.2       & {\em  -3.1}  &   -0.26      &    0.55      & {\em   0.488} & {\em  -3.1}  & {\em  -3.1}  & \cite{PhysRevC.21.2076}        \\
{\bf SkT     } &   -0.04      &    0.29      &  0.663       &  0.551       &   75.4       &   32.1       &    0.89      &    0.59      &    1.200      &   32.1       &   32.1       & \cite{1974NuPhA.236..269K}     \\
SkT1           &    0.06      &    1.60      &  1.000       &  1.000       &   55.9       &    0         &   -0.40      &    0.16      & {\em   0.422} & {\em -18.6}  &    0         & \cite{1984NuPhA.420..297T}     \\
SkT2           &    0.06      &    1.59      &  1.000       &  1.000       &   56.3       &    0         &   -0.41      &    0.16      & {\em   0.418} & {\em -18.8}  &    0         & \cite{1984NuPhA.420..297T}     \\
SkT3           &    0.06      &    1.55      &  1.000       &  1.000       &   56.0       &   18.7       & {\em  -1.23} &    0.45      & {\em   0.227} & {\em -18.7}  & {\em  -6.2}  & \cite{1984NuPhA.420..297T}     \\
SkT4           &    0.07      &    1.90      &  1.000       &  1.000       &   56.9       &    0         &   -0.70      &    0.16      & {\em   0.418} & {\em -19.0}  &    0         & \cite{1984NuPhA.420..297T}     \\
SkT5           &   -0.10      &    1.96      &  1.000       &  1.000       &   61.5       &    0         &   -0.88      &    0.05      & {\em   0.351} & {\em -20.5}  &    0         & \cite{1984NuPhA.420..297T}     \\
SkT6           &    0.06      &    1.43      &  1.000       &  1.000       &   55.1       &    0         &   -0.22      &    0.18      & {\em   0.430} & {\em -18.4}  &    0         & \cite{1984NuPhA.420..297T}     \\
SkT7           &   -0.11      &    1.00      &  1.000       &  0.714       &   68.7       &   19.5       &   -0.72      &    0.49      & {\em   0.358} &    3.0       &    2.2       & \cite{1984NuPhA.420..297T}     \\
SkT8           &   -0.11      &    1.02      &  0.833       &  0.833       &   68.8       &   13.0       &    0.01      &    0.24      & {\em   0.557} & {\em -10.0}  &    8.6       & \cite{1984NuPhA.420..297T}     \\
SkT9           &   -0.12      &    1.02      &  0.833       &  0.833       &   70.8       &   13.0       &   -0.01      &    0.21      & {\em   0.538} & {\em -10.6}  &    8.6       & \cite{1984NuPhA.420..297T}     \\
SkTK           &   -0.32      &    0.71      &  0.762       &  0.509       &  104.8       &   24.5       &   -0.17      &    0.27      & {\em   0.661} &   24.5       &   24.5       & \cite{1976JPhG....2..285T}     \\
SkM            &   -0.23      &    0.97      &  0.976       &  0.661       &   72.2       &    9.1       &   -0.25      &    0.36      & {\em   0.536} &    9.1       &    9.1       & \cite{krivine:1980}            \\
SkM1           &   -0.23      &    0.61      &  0.995       &  0.653       &   76.9       &    8.8       &    0.01      &    0.31      & {\em   0.478} &    8.8       &    8.8       & \cite{gomez:1995}              \\
SkMP           &   -0.31      &    0.61      &  0.740       &  0.585       &   94.4       &   40.7       &   -0.31      &    0.47      & {\em   0.576} &   15.4       &   17.5       & \cite{PhysRevC.40.2834}        \\
SkMs           &   -0.23      &    0.93      &  0.995       &  0.653       &   76.9       &    8.8       &   -0.30      &    0.31      & {\em   0.478} &    8.8       &    8.8       & \cite{1982NuPhA.386...79B}     \\
SKa            &   -0.26      &    0.66      &  0.744       &  0.515       &  107.0       &   27.2       &   -0.02      &    0.32      & {\em   0.746} &   27.2       &   27.2       & \cite{Kohler:1976}             \\
SKb            &   -0.26      &    0.21      &  0.744       &  0.515       &  107.0       &   27.2       &    0.44      &    0.32      & {\em   0.746} &   27.2       &   27.2       & \cite{Kohler:1976}             \\
SGI            &   -0.26      &    0.44      &  0.646       &  0.574       &   96.7       &   49.8       &    0.07      &    0.50      & {\em   0.770} &   17.5       &   24.5       & \cite{1981PhLB..106..379V}     \\
SGII           &   -0.23      &    0.73      &  0.950       &  0.670       &   63.8       &   13.5       &    0.01      &    0.51      & {\em   0.612} &   11.0       &    8.5       & \cite{1981PhLB..106..379V}     \\
RATP           &   -0.28      &    0.59      &  0.819       &  0.563       &   96.2       &   41.4       &   -0.63      &    0.51      & {\em   0.398} &   18.3       &   12.6       & \cite{1982AA...116..183R}      \\
T              &    0.06      &    1.30      &  1.000       &  1.000       &   56.5       &   75.4       & {\em  -3.63} &    1.32      & {\em  -0.563} & {\em -18.8}  & {\em -25.1}  & \cite{PL123B.139}              \\
{\em SkP     } &   -0.10      &    1.42      & {\em 1.538}  &  0.741       &   60.1       & {\em -23.9}  &   -0.23      &    0.06      & {\em   0.388} &    2.3       &    0.5       & \cite{1984NuPhA.422..103D}     \\
{\bf ZR1a    } &    0.71      &   -0.24      &  1.000       &  1.000       &    0         &    0         &    2.31      &    1.04      &    1.000      &    0         &    0         & \cite{jaqaman84}               \\
{\bf ZR1b    } &    0.71      &    0.43      &  1.000       &  1.000       &    0         &    0         &    1.64      &    1.04      &    1.000      &    0         &    0         & \cite{jaqaman84}               \\
{\bf ZR1c    } &    0.71      &    1.44      &  1.000       &  1.000       &    0         &    0         &    0.64      &    1.04      &    1.000      &    0         &    0         & \cite{jaqaman84}               \\
{\bf ZR2a    } &    0.39      &   -0.87      &  1.000       &  1.000       &    0         &    0         &    2.95      &    1.04      &    1.000      &    0         &    0         & \cite{jaqaman84}               \\
{\bf ZR2b    } &    0.39      &   -0.08      &  1.000       &  1.000       &    0         &    0         &    2.15      &    1.04      &    1.000      &    0         &    0         & \cite{jaqaman84}               \\
{\bf ZR2c    } &    0.39      &    1.12      &  1.000       &  1.000       &    0         &    0         &    0.96      &    1.04      &    1.000      &    0         &    0         & \cite{jaqaman84}               \\
{\em ZR3a    } &   -0.15      & {\em -11.65} &  1.000       &  1.000       &    0         &    0         &   13.71      &    1.03      &    1.000      &    0         &    0         & \cite{jaqaman84}               \\
{\em ZR3b    } &   -0.15      & {\em  -8.70} &  1.000       &  1.000       &    0         &    0         &   10.76      &    1.03      &    1.000      &    0         &    0         & \cite{jaqaman84}               \\
{\em ZR3c    } &   -0.15      & {\em  -4.27} &  1.000       &  1.000       &    0         &    0         &    6.34      &    1.03      &    1.000      &    0         &    0         & \cite{jaqaman84}               \\
PRC45          &    0.76      &    3.42      &  1.000       &  1.000       &    0         &    0         & {\em  -1.31} &    1.05      &    1.000      &    0         &    0         & \cite{Phys.Rev.C63.044605}     \\
{\em E       } &    0.31      &    0.96      & {\em 1.058}  &  0.736       &   58.1       &    4.1       &   -0.08      &    0.44      & {\em   0.531} &    4.1       &    4.1       & \cite{PhysRevC.33.335}         \\
{\em Es      } &   -0.07      &    0.79      & {\em 1.055}  &  0.697       &   67.3       &    5.3       &   -0.09      &    0.35      & {\em   0.461} &    5.3       &    5.3       & \cite{PhysRevC.33.335}         \\
Gs             &   -0.15      &    1.02      &  0.930       &  0.678       &   63.0       &   10.3       &   -0.10      &    0.46      & {\em   0.674} &   10.3       &   10.3       & \cite{PhysRevC.33.335}         \\
Rs             &   -0.15      &    0.97      &  0.928       &  0.677       &   63.0       &   10.4       &   -0.04      &    0.46      & {\em   0.677} &   10.4       &   10.4       & \cite{PhysRevC.33.335}         \\
Z              &    0.26      &    0.85      &  0.994       &  0.731       &   53.4       &    6.2       &    0.17      &    0.51      & {\em   0.645} &    6.2       &    6.2       & \cite{PhysRevC.33.335}         \\
Zs             &   -0.18      &    0.68      &  0.954       &  0.664       &   67.9       &    9.6       &    0.11      &    0.39      & {\em   0.590} &    9.6       &    9.6       & \cite{PhysRevC.33.335}         \\
Zss            &   -0.19      &    0.80      &  0.952       &  0.653       &   71.4       &   10.1       &   -0.05      &    0.37      & {\em   0.570} &   10.1       &   10.1       & \cite{PhysRevC.33.335}         \\
SkSC1          &    0.06      &    1.28      &  1.000       &  1.000       &   53.0       &    0         &   -0.04      &    0.21      & {\em   0.453} & {\em -17.7}  &    0         & \cite{1991NuPhA.528....1P}     \\
SkSC2          &    0.06      &    1.01      &  1.000       &  1.000       &   54.5       &    0         &    0.21      &    0.19      & {\em   0.437} & {\em -18.2}  &    0         & \cite{1991NuPhA.528....1P}     \\
SkSC3          &    0.06      &    1.19      &  1.000       &  1.000       &   54.4       &   18.1       &   -0.82      &    0.47      & {\em   0.251} & {\em -18.1}  & {\em  -6.0}  & \cite{1991NuPhA.528....1P}     \\
SkSC4          &    0.06      &    1.34      &  1.000       &  1.000       &   53.2       &    0         &   -0.10      &    0.21      & {\em   0.451} & {\em -17.7}  &    0         & \cite{1992NuPhA.549..155A}     \\
SkSC4o         &    0.06      &    1.19      &  1.000       &  1.000       &   53.1       &    0         &    0.04      &    0.21      & {\em   0.452} & {\em -17.7}  &    0         & \cite{pearson:2000}            \\
SkSC5          &    0.06      &    1.52      &  1.000       &  1.000       &   52.9       &    0         &   -0.28      &    0.21      & {\em   0.454} & {\em -17.6}  &    0         & \cite{PhysRevC.50.460}         \\
SkSC6          &    0.06      &    0.99      &  1.000       &  1.000       &   54.7       &    0         &    0.22      &    0.18      & {\em   0.434} & {\em -18.2}  &    0         & \cite{PhysRevC.50.460}         \\
SkSC10         &    0.06      &    0.85      &  1.000       &  1.000       &   56.1       &    0         &    0.34      &    0.16      & {\em   0.421} & {\em -18.7}  &    0         & \cite{PhysRevC.50.460}         \\
SkSC11         &    0.06      &    1.34      &  1.000       &  1.000       &   53.2       &    0         &   -0.10      &    0.21      & {\em   0.451} & {\em -17.7}  &    0         & \cite{PhysRevC.52.2254}        \\
SkSC14         &    0.06      &    1.44      &  1.000       &  1.000       &   54.6       &    0         &   -0.22      &    0.18      & {\em   0.435} & {\em -18.2}  &    0         & \cite{pearson:2000}            \\
SkSC15         &    0.06      &    1.27      &  1.000       &  1.000       &   53.6       &    0         &   -0.04      &    0.20      & {\em   0.447} & {\em -17.9}  &    0         & \cite{pearson:2000}            \\
{\em Skyrme1p} &    0.56      &    1.22      & {\em 1.064}  &  0.797       &   44.2       &    2.2       &   -0.07      &    0.57      & {\em   0.625} &    2.2       &    2.2       & \cite{1995NuPhA.584..675P}     \\
MSkA           &    0.16      &    1.02      &  0.970       &  0.671       &   69.4       &    9.2       &   -0.17      &    0.43      & {\em   0.591} &    9.9       &    9.3       & \cite{PhysRevLett.74.3744}     \\
SkI1           &   -0.24      &    1.11      &  0.611       &  0.800       &   82.5       &  306.6       & {\em  -8.74} &    3.17      & {\em  -4.566} & {\em -11.3}  & {\em -69.4}  & \cite{1995NuPhA.584..467R}     \\
SkI2           &   -0.25      &    0.88      &  0.597       &  0.804       &   82.2       &   83.9       & {\em  -1.18} &    0.77      & {\em   0.076} & {\em -11.3}  &    7.0       & \cite{1995NuPhA.584..467R}     \\
SkI3           &   -0.32      &    0.65      &  0.451       &  0.803       &  105.3       &   63.3       &    0.57      &    0.20      & {\em   0.855} & {\em -19.0}  &   37.7       & \cite{1995NuPhA.584..467R}     \\
SkI4           &   -0.27      &    0.56      &  0.546       &  0.801       &   88.8       &  157.6       & {\em  -2.81} &    1.38      & {\em  -1.360} & {\em -13.5}  & {\em -11.3}  & \cite{1995NuPhA.584..467R}     \\
SkI5           &   -0.32      &    0.76      &  0.452       &  0.804       &  103.3       &   72.0       &    0.28      &    0.30      & {\em   0.692} & {\em -18.2}  &   35.2       & \cite{1995NuPhA.584..467R}     \\
SkI6           &   -0.28      &    0.57      &  0.533       &  0.800       &   90.7       &  115.3       & {\em  -1.42} &    0.92      & {\em  -0.442} & {\em -14.0}  &    5.1       & \cite{PhysRevC.53.740}         \\
SLy0           &   -0.28      &    0.82      &  0.619       &  0.800       &   91.0       &   16.0       &    0.78      &   -0.03      & {\em   0.885} & {\em -14.2}  &   26.6       & \cite{ChabanatPhd1995}         \\
SLy1           &   -0.28      &    0.82      &  0.619       &  0.800       &   91.6       &    4.3       &    1.15      &   -0.17      & {\em   0.844} & {\em -14.4}  &   30.5       & \cite{ChabanatPhd1995}         \\
SLy2           &   -0.28      &    0.81      &  0.618       &  0.800       &   90.4       &   29.9       &    0.34      &    0.12      & {\em   0.748} & {\em -14.0}  &   22.0       & \cite{ChabanatPhd1995}         \\
SLy230a        &   -0.28      &    0.82      &  0.535       &  1.000       &   91.8       &   21.0       &    1.15      &   -0.16      & {\em   0.851} & {\em -30.6}  &   30.6       & \cite{1997NuPhA.627..710C}     \\
SLy3           &   -0.28      &    0.81      &  0.616       &  0.800       &   90.2       &    6.4       &    1.11      &   -0.13      & {\em   0.884} & {\em -13.9}  &   30.1       & \cite{ChabanatPhd1995}         \\
SLy4           &   -0.28      &    0.81      &  0.614       &  0.800       &   91.3       &    6.7       &    1.11      &   -0.13      & {\em   0.885} & {\em -14.2}  &   30.4       & \cite{1998NuPhA.635..231C}     \\
SLy5           &   -0.28      &    0.81      &  0.618       &  0.800       &   90.8       &    5.1       &    1.14      &   -0.15      & {\em   0.860} & {\em -14.1}  &   30.3       & \cite{1998NuPhA.635..231C}     \\
SLy6           &   -0.28      &    0.80      &  0.606       &  0.800       &   86.7       &   14.3       &    0.97      &    0.00      &    1.000      & {\em -12.6}  &   28.9       & \cite{1998NuPhA.635..231C}     \\
SLy7           &   -0.28      &    0.80      &  0.604       &  0.800       &   86.5       &   15.9       &    0.94      &    0.02      &    1.000      & {\em -12.5}  &   28.8       & \cite{1998NuPhA.635..231C}     \\
SLy8           &   -0.28      &    0.81      &  0.616       &  0.800       &   90.1       &    6.7       &    1.11      &   -0.13      & {\em   0.889} & {\em -13.9}  &   30.0       & \cite{ChabanatPhd1995}         \\
SLy9           &   -0.28      &    0.80      &  0.570       &  0.800       &   95.7       &   24.9       &    0.85      &    0.04      &    1.000      & {\em -14.8}  &   31.9       & \cite{ChabanatPhd1995}         \\
SLy10          &   -0.28      &    0.81      &  0.596       &  0.800       &   80.8       &   26.0       &    0.75      &    0.19      &    1.000      & {\em -10.3}  &   26.9       & \cite{ChabanatPhd1995}         \\
{\em SkX     } &    0.24      &    1.56      & {\em 1.471}  &  0.750       &   46.3       & {\em -11.0}  &   -0.63      &    0.51      & {\em   0.443} &    6.9       & {\em  -3.1}  & \cite{PhysRevC.58.220}         \\
{\em SkXce   } &    0.25      &    1.52      & {\em 1.518}  &  0.753       &   45.8       & {\em -12.0}  &   -0.62      &    0.51      & {\em   0.426} &    6.7       & {\em  -3.9}  & \cite{PhysRevC.58.220}         \\
{\em SkXm    } &    0.05      &    1.47      & {\em 1.360}  &  0.749       &   51.3       & {\em  -2.9}  &   -0.82      &    0.51      & {\em   0.376} &    4.8       & {\em  -3.3}  & \cite{PhysRevC.58.220}         \\
SkO            &   -0.10      &    1.33      &  0.945       &  0.852       &   56.9       &   98.9       & {\em  -4.11} &    1.62      & {\em  -0.770} & {\em  -7.7}  & {\em -26.7}  & \cite{PhysRevC.60.014316}      \\
SkOp           &   -0.10      &    1.33      &  0.923       &  0.871       &   56.5       &   40.7       & {\em  -1.61} &    0.79      & {\em   0.209} & {\em  -9.3}  & {\em  -6.8}  & \cite{PhysRevC.60.014316}      \\
SKRA           &   -0.26      &    0.91      &  0.925       &  0.628       &   76.0       &   13.1       &   -0.21      &    0.39      & {\em   0.589} &   13.1       &   12.0       & \cite{2000MPLA...15.1287R}     \\
MSk1           &    0.07      &    1.47      &  1.000       &  1.000       &   51.5       &    0         &   -0.18      &    0.25      & {\em   0.478} & {\em -17.2}  &    0         & \cite{PhysRevC.62.024308}      \\
{\em MSk2    } &    0.11      &    1.59      & {\em 1.050}  & {\em 1.050}  &   48.8       & {\em  -3.1}  &   -0.26      &    0.23      & {\em   0.442} & {\em -19.4}  & {\em  -2.1}  & \cite{PhysRevC.62.024308}      \\
{\em MSk3    } &    0.07      &    1.30      & {\em 1.001}  & {\em 1.001}  &   50.5       & {\em  -0.1}  &    0.00      &    0.26      & {\em   0.487} & {\em -16.9}  & {\em  -0.1}  & \cite{PhysRevC.62.024308}      \\
{\em MSk4    } &    0.11      &    1.42      & {\em 1.050}  & {\em 1.050}  &   47.6       & {\em  -3.1}  &   -0.07      &    0.24      & {\em   0.454} & {\em -19.0}  & {\em  -2.1}  & \cite{PhysRevC.62.024308}      \\
{\em MSk5    } &    0.11      &    1.42      & {\em 1.050}  & {\em 1.050}  &   47.7       & {\em  -3.1}  &   -0.07      &    0.24      & {\em   0.454} & {\em -19.0}  & {\em  -2.1}  & \cite{PhysRevC.62.024308}      \\
MSk5s          &   -0.10      &    0.85      &  0.933       &  0.700       &   68.0       &   10.3       &    0.02      &    0.40      & {\em   0.595} &    5.8       &    9.2       & \cite{PhysRevC.64.027301}      \\
{\em MSk6    } &    0.11      &    1.42      & {\em 1.050}  & {\em 1.050}  &   48.5       & {\em  -3.1}  &   -0.08      &    0.23      & {\em   0.446} & {\em -19.3}  & {\em  -2.1}  & \cite{PhysRevC.62.024308}      \\
{\em MSk7    } &    0.11      &    1.41      & {\em 1.050}  & {\em 1.050}  &   48.6       & {\em  -3.1}  &   -0.08      &    0.23      & {\em   0.444} & {\em -19.3}  & {\em  -2.1}  & \cite{2001ADNDT..77..311G}     \\
{\em MSk8    } &    0.15      &    1.53      & {\em 1.100}  & {\em 1.100}  &   45.5       & {\em  -6.0}  &   -0.14      &    0.22      & {\em   0.418} & {\em -21.2}  & {\em  -4.0}  & \cite{2001NuPhA.688..349G}     \\
MSk9           &    0.07      &    1.30      &  1.000       &  1.000       &   51.3       &    0         &   -0.02      &    0.25      & {\em   0.480} & {\em -17.1}  &    0         & \cite{2001NuPhA.688..349G}     \\
{\em v070    } &    0.11      &    1.42      & {\em 2.100}  &  0.700       &   46.6       & {\em -24.6}  &   -0.53      &    0.42      & {\em   0.364} &   12.7       & {\em  -5.4}  & \cite{PhysRevC.64.027301}      \\
{\em v075    } &    0.11      &    1.42      & {\em 1.750}  &  0.750       &   47.3       & {\em -17.5}  &   -0.58      &    0.42      & {\em   0.349} &    6.2       & {\em  -5.7}  & \cite{PhysRevC.64.027301}      \\
{\em v080    } &    0.11      &    1.42      & {\em 1.527}  &  0.800       &   47.1       & {\em -11.8}  &   -0.58      &    0.43      & {\em   0.349} &    0.8       & {\em  -5.7}  & \cite{PhysRevC.64.027301}      \\
{\em v090    } &    0.11      &    1.42      & {\em 1.260}  &  0.900       &   47.5       & {\em  -5.4}  &   -0.46      &    0.38      & {\em   0.372} & {\em  -8.5}  & {\em  -4.8}  & \cite{PhysRevC.64.027301}      \\
{\em v100    } &    0.11      &    1.42      & {\em 1.105}  &  1.000       &   47.6       &    0         &   -0.37      &    0.35      & {\em   0.391} & {\em -15.9}  & {\em  -4.2}  & \cite{PhysRevC.64.027301}      \\
{\em v105    } &    0.11      &    1.42      & {\em 1.050}  & {\em 1.050}  &   47.5       & {\em  -3.1}  &   -0.07      &    0.25      & {\em   0.456} & {\em -19.0}  & {\em  -2.1}  & \cite{PhysRevC.64.027301}      \\
{\em v110    } &    0.11      &    1.42      & {\em 1.004}  & {\em 1.100}  &   47.4       & {\em  -6.0}  &    0.21      &    0.16      & {\em   0.514} & {\em -21.8}  & {\em  -0.2}  & \cite{PhysRevC.64.027301}      \\
{\em SKz0    } &   -0.27      &    0.82      & {\em 1.166}  &  0.500       &   82.5       & {\em -20.0}  &    0.44      &    0.20      & {\em   0.833} &   37.3       &   22.1       & \cite{PhysRevC.66.014303}      \\
{\em SKz1    } &   -0.27      &    0.82      &  0.914       &  0.567       &   82.5       & {\em  -5.1}  &    0.46      &    0.19      & {\em   0.838} &   22.0       &   22.2       & \cite{PhysRevC.66.014303}      \\
SKz2           &   -0.27      &    0.82      &  0.779       &  0.635       &   82.5       &    6.7       &    0.47      &    0.19      & {\em   0.843} &    9.7       &   22.4       & \cite{PhysRevC.66.014303}      \\
SKz3           &   -0.27      &    0.82      &  0.638       &  0.776       &   82.5       &   24.7       &    0.48      &    0.18      & {\em   0.847} & {\em  -8.7}  &   22.5       & \cite{PhysRevC.66.014303}      \\
SKz4           &   -0.27      &    0.82      &  0.553       &  0.952       &   82.5       &   39.7       &    0.50      &    0.18      & {\em   0.787} & {\em -24.2}  &   22.7       & \cite{PhysRevC.66.014303}      \\
{\em SKzm1   } &   -0.27      &    0.82      & {\em 1.284}  &  0.481       &   82.5       & {\em -28.9}  &    0.57      &    0.16      & {\em   0.685} &   42.4       &   23.4       & \cite{PhysRevC.66.014303}      \\
{\em BSk1    } &    0.11      &    1.40      & {\em 1.050}  & {\em 1.050}  &   49.3       & {\em  -3.1}  &   -0.08      &    0.22      & {\em   0.438} & {\em -19.6}  & {\em  -2.1}  & \cite{2002NuPhA.700..142S}     \\
{\em BSk2    } &    0.11      &    1.40      & {\em 1.321}  &  0.860       &   48.9       & {\em  -2.7}  &   -0.71      &    0.45      & {\em   0.316} & {\em  -5.6}  & {\em  -6.2}  & \cite{PhysRevC.66.024326}      \\
{\em BSk2p   } &    0.12      &    1.42      & {\em 1.336}  &  0.864       &   48.6       & {\em  -2.9}  &   -0.73      &    0.45      & {\em   0.307} & {\em  -5.8}  & {\em  -6.6}  & \cite{PhysRevC.66.024326}      \\
{\em BSk3    } &    0.21      &    1.58      & {\em 1.510}  &  0.894       &   43.7       & {\em  -5.8}  &   -0.99      &    0.50      & {\em   0.245} & {\em  -6.8}  & {\em -10.3}  & \cite{PhysRevC.68.054325}      \\
{\em BSk4    } &    0.00      &    1.12      & {\em 1.003}  &  0.850       &   57.5       &   13.0       &   -0.48      &    0.46      & {\em   0.400} & {\em  -7.6}  & {\em  -0.6}  & \cite{PhysRevC.68.054325}      \\
{\em BSk5    } &    0.00      &    1.17      & {\em 1.023}  &  0.836       &   58.6       &   12.5       &   -0.58      &    0.45      & {\em   0.381} & {\em  -6.6}  & {\em  -0.8}  & \cite{PhysRevC.68.054325}      \\
BSk6           &   -0.16      &    0.84      &  0.749       &  0.859       &   71.6       &   25.1       &    0.07      &    0.31      & {\em   0.577} & {\em -13.0}  &   10.0       & \cite{PhysRevC.68.054325}      \\
BSk7           &   -0.16      &    0.84      &  0.740       &  0.871       &   72.4       &   30.6       &   -0.10      &    0.36      & {\em   0.526} & {\em -14.4}  &    8.5       & \cite{PhysRevC.68.054325}      \\
BSk8           &   -0.16      &    0.83      &  0.739       &  0.871       &   74.8       &   25.7       &    0.04      &    0.26      & {\em   0.542} & {\em -15.3}  &   10.0       & \cite{samyn:044309}            \\
BSk9           &   -0.16      &    0.96      &  0.715       &  0.907       &   77.2       &   30.0       &   -0.16      &    0.25      & {\em   0.503} & {\em -19.1}  &    9.5       & \cite{samyn:044309}            \\
{\em BSk10   } &    0.00      &    1.25      & {\em 1.059}  &  0.813       &   60.5       &   13.4       &   -0.83      &    0.46      & {\em   0.323} & {\em  -5.2}  & {\em  -1.9}  & \cite{goriely:2006}            \\
{\em BSk11   } &    0.00      &    1.26      & {\em 1.049}  &  0.819       &   58.2       &   13.0       &   -0.76      &    0.49      & {\em   0.359} & {\em  -5.0}  & {\em  -1.6}  & \cite{goriely:2006}            \\
{\em BSk12   } &    0.00      &    1.26      & {\em 1.051}  &  0.818       &   57.9       &   13.1       &   -0.76      &    0.49      & {\em   0.360} & {\em  -4.8}  & {\em  -1.6}  & \cite{goriely:2006}            \\
{\em BSk13   } &    0.00      &    1.26      & {\em 1.048}  &  0.820       &   57.7       &   13.3       &   -0.76      &    0.50      & {\em   0.362} & {\em  -4.9}  & {\em  -1.6}  & \cite{goriely:2006}            \\
BSk14          &   -0.13      &    0.97      &  0.818       &  0.783       &   70.8       &   33.4       &   -0.63      &    0.51      & {\em   0.420} & {\em  -5.5}  &    4.6       & \cite{goriely:064312}          \\
BSk15          &   -0.12      &    0.96      &  0.836       &  0.767       &   69.9       &   33.1       &   -0.67      &    0.54      & {\em   0.412} & {\em  -3.5}  &    4.1       & \cite{goriely:2008}            \\
BSk16          &   -0.12      &    0.97      &  0.819       &  0.782       &   71.9       &   33.8       &   -0.65      &    0.51      & {\em   0.403} & {\em  -5.7}  &    4.5       & \cite{chamel:2008}             \\
BSk17          &   -0.12      &    0.97      &  0.818       &  0.782       &   73.0       &   34.3       &   -0.69      &    0.50      & {\em   0.387} & {\em  -6.1}  &    4.3       & \cite{goriely:2009}             \\
{\em SK255   } &   -0.07      &    1.46      & {\em 1.034}  &  0.649       &   73.0       &    5.7       &   -0.71      &    0.37      & {\em   0.521} &   11.4       &    8.6       & \cite{PhysRevC.68.031304}      \\
SK272          &   -0.03      &    1.40      &  0.951       &  0.651       &   74.6       &   10.5       &   -0.61      &    0.39      & {\em   0.577} &   11.0       &   10.7       & \cite{PhysRevC.68.031304}      \\
{\em QMC1    } &    0.52      &    1.48      & {\em 1.504}  &  0.669       &  122.1       & {\em  -3.3}  & {\em  -1.81} &   -0.16      & {\em  -0.166} & {\em  -3.3}  & {\em  -3.3}  & \cite{guichon:132502}          \\
{\em QMC2    } &    0.36      &    1.13      & {\em 1.033}  &  0.700       &   75.0       &    6.8       &   -0.38      &    0.38      & {\em   0.506} &    6.8       &    6.8       & \cite{guichon:132502}          \\
{\em QMC3    } &    0.36      &    2.06      & {\em 1.035}  &  0.685       &   69.9       &    6.3       & {\em  -1.37} &    0.35      & {\em   0.473} &    6.3       &    6.3       & \cite{guichon:132502}          \\
KDE0v          &   -0.26      &    0.92      &  0.672       &  0.769       &   80.8       &    9.1       &    0.75      &    0.05      & {\em   0.922} & {\em  -7.5}  &   24.4       & \cite{agrawal:014310}          \\
KDE0v1         &   -0.25      &    1.06      &  0.685       &  0.814       &   77.2       &    6.5       &    0.67      &    0.00      & {\em   0.878} & {\em -11.4}  &   21.9       & \cite{agrawal:014310}          \\
{\em LNS     } &   -0.26      &    1.12      &  0.955       &  0.727       &   50.0       & {\em -16.4}  &    0.83      &    0.14      & {\em   0.852} &    5.6       &   14.7       & \cite{cao:014313}              \\
SV-min         &   -0.05      &    1.37      &  0.975       &  0.930       &   55.5       &   39.4       & {\em  -1.84} &    0.80      & {\em   0.101} & {\em -13.6}  & {\em -10.4}  & \cite{arXiv:0804.3385}         \\
{\em SV-bas  } &   -0.05      &    1.20      & {\em 1.214}  &  0.715       &   58.8       &   21.3       & {\em  -1.40} &    0.73      & {\em   0.209} &    6.3       & {\em  -6.1}  & \cite{arXiv:0804.3385}         \\
{\em SV-K241 } &   -0.01      &    1.21      & {\em 1.215}  &  0.715       &   58.2       &   10.8       &   -0.96      &    0.60      & {\em   0.325} &    6.7       & {\em  -2.6}  & \cite{arXiv:0804.3385}         \\
{\em SV-K226 } &   -0.08      &    1.19      & {\em 1.212}  &  0.715       &   59.4       &   34.4       & {\em  -1.95} &    0.90      & {\em   0.065} &    6.0       & {\em -10.4}  & \cite{arXiv:0804.3385}         \\
{\em SV-K218 } &   -0.12      &    1.18      & {\em 1.210}  &  0.715       &   60.1       &   42.7       & {\em  -2.31} &    1.00      & {\em  -0.031} &    5.6       & {\em -13.1}  & \cite{arXiv:0804.3385}         \\
{\em SV-mas10} &    0.06      &    1.45      & {\em 1.665}  &  0.715       &   50.7       &    4.8       & {\em  -1.58} &    0.73      & {\em   0.165} &    9.1       & {\em -10.2}  & \cite{arXiv:0804.3385}         \\
SV-mas08       &   -0.16      &    0.95      &  0.907       &  0.715       &   69.0       &   50.7       & {\em  -1.54} &    0.84      & {\em   0.174} &    2.9       & {\em  -3.8}  & \cite{arXiv:0804.3385}         \\
SV-mas07       &   -0.26      &    0.71      &  0.685       &  0.715       &   82.2       &   96.2       & {\em  -1.94} &    1.03      & {\em  -0.139} & {\em  -1.5}  & {\em  -3.6}  & \cite{arXiv:0804.3385}         \\
{\em SV-sym34} &   -0.04      &    1.50      & {\em 1.214}  &  0.715       &   60.7       &   44.6       & {\em  -2.68} &    1.03      & {\em  -0.048} &    5.8       & {\em -13.9}  & \cite{arXiv:0804.3385}         \\
{\em SV-sym32} &   -0.05      &    1.35      & {\em 1.214}  &  0.715       &   59.8       &   29.7       & {\em  -1.91} &    0.83      & {\em   0.113} &    6.0       & {\em  -8.9}  & \cite{arXiv:0804.3385}         \\
{\em SV-sym28} &   -0.05      &    1.06      & {\em 1.214}  &  0.715       &   57.6       &   23.3       & {\em  -1.32} &    0.77      & {\em   0.201} &    6.7       & {\em  -6.8}  & \cite{arXiv:0804.3385}         \\
{\em SV-kap60} &   -0.05      &    1.20      & {\em 1.603}  &  0.625       &   58.9       &   16.1       & {\em  -1.72} &    0.83      & {\em   0.129} &   19.3       & {\em  -8.7}  & \cite{arXiv:0804.3385}         \\
SV-kap20       &   -0.05      &    1.20      &  0.977       &  0.834       &   58.7       &   27.8       & {\em  -1.13} &    0.64      & {\em   0.276} & {\em  -6.6}  & {\em  -3.9}  & \cite{arXiv:0804.3385}         \\
{\em SV-kap00} &   -0.05      &    1.20      &  0.817       & {\em 1.001}  &   58.6       &   37.3       &   -0.98      &    0.59      & {\em   0.313} & {\em -19.6}  & {\em  -2.7}  & \cite{arXiv:0804.3385}         \\
{\em SV-tls  } &   -0.05      &    1.20      & {\em 1.214}  &  0.715       &   59.6       &   13.9       & {\em  -1.10} &    0.62      & {\em   0.277} &    6.1       & {\em  -3.6}  & \cite{arXiv:0804.3385}         \\
ska25s20       &   -0.02      &    1.69      &  0.976       &  0.984       &   52.8       &   11.6       &   -0.89      &    0.40      & {\em   0.379} & {\em -16.5}  & {\em  -2.5}  & \cite{brown08}                 \\
{\em ska35s15} &    0.11      &    1.54      & {\em 1.017}  & {\em 1.011}  &   46.8       &    8.6       &   -0.63      &    0.46      & {\em   0.407} & {\em -16.3}  & {\em  -3.9}  & \cite{brown08}                 \\
ska35s20       &    0.10      &    1.76      &  1.000       &  1.000       &   49.5       &    9.9       &   -0.88      &    0.44      & {\em   0.397} & {\em -16.5}  & {\em  -3.3}  & \cite{brown08}                 \\
ska35s25       &    0.09      &    2.01      &  0.989       &  0.993       &   51.9       &   10.9       & {\em  -1.18} &    0.42      & {\em   0.383} & {\em -16.8}  & {\em  -3.0}  & \cite{brown08}                 \\
{\em ska45s20} &    0.23      &    1.83      & {\em 1.029}  & {\em 1.019}  &   46.0       &    8.0       &   -0.89      &    0.47      & {\em   0.409} & {\em -16.6}  & {\em  -4.3}  & \cite{brown08}                 \\